\documentclass[12pt]{article}

\usepackage[T1]{fontenc}
\usepackage{amsmath}
\usepackage{amssymb}
\usepackage{mathrsfs}
\usepackage{graphicx}
\usepackage{subfigure}
\usepackage{color}
\usepackage{fullpage}
\usepackage{multirow}
\usepackage{hyperref}
\hypersetup{hidelinks}
\usepackage{amscd}
\usepackage{xcolor}
\usepackage{braket}
\usepackage{amsfonts}
\usepackage{tikz}
\usepackage{pgfplots}
\usepackage{cite}
\usepackage{booktabs}
\usepackage{placeins}
\usepackage{makecell}
\usepackage{enumitem}
\usepackage[font=small, labelfont=bf, textfont={small,it}]{caption}

\newcommand{\rmd}{{\rm d}}

\def\thefootnote{\fnsymbol{footnote}} 
\catcode`\@=11 
\def\numberbysection{\@addtoreset{equation}{section} 
        \def\theequation{\thesection.\arabic{equation}}}
    
\renewcommand{\thefootnote}{\arabic{footnote}}

\newcommand{\printdoi}[1]{\href{https://doi.org/#1}{[\texttt{#1}]}}

\def\be{\begin{equation}} 
\def\ee{\end{equation}} 
\def\ba{\begin{eqnarray}} 
\def\ea{\end{eqnarray}} 
\def\bali{\begin{align}}
\def\eali{\end{align}}
 
\def\ov{\overline}

\def\Z{\mathbb{Z}}

\def\nl{\nonumber \\}

\def\wt{\widetilde}

\def\b{\beta} 
\def\g{\gamma} 
\def\G{\Gamma} 
\def\D{\Delta} 
 
\def\e{\epsilon} 
\def\eps{\varepsilon}

\def\l{\lambda}

\def\n{\nu}

\def\s{\sigma}

\def\w{\omega}

\begin{document} 

\begin{titlepage} 
\begin{center} 
\vskip .6 in 
{\LARGE \bf Benchmarking the Ising Universality Class}\\
\bigskip
{\LARGE \bf in $3 \le d < 4$ dimensions} \\
\vskip 0.3in 
\bf Claudio BONANNO${}^{\,1,a}$, Andrea CAPPELLI${}^{\,1,b}$, \\
Mikhail KOMPANIETS${}^{\,2,3,c}$, Satoshi OKUDA${}^{\,4,d}$,
Kay J\"org~WIESE${}^{\,5,e}$\vspace{0.2cm}

\let\thefootnote\relax\footnotetext{
\\[-1em]
${}^a$ \href{mailto:claudio.bonanno@fi.infn.it}{\texttt{claudio.bonanno@fi.infn.it}}\\
${}^b$ \href{mailto:andrea.cappelli@fi.infn.it}{\texttt{andrea.cappelli@fi.infn.it}}\\
${}^c$ \href{mailto:m.kompaniets@spbu.ru}{\texttt{m.kompaniets@spbu.ru}}\\
${}^d$ \href{mailto:okudas@rikkyo.ac.jp}{\texttt{okudas@rikkyo.ac.jp}}\\
${}^e$ \href{mailto:wiese@lpt.ens.fr}{\texttt{wiese@lpt.ens.fr}}}

{\it ${}^1$ INFN, Sezione di Firenze, Via G. Sansone 1,
50019 Sesto Fiorentino (FI), Italy\\\vspace{0.1cm}
${}^2$ Saint Petersburg State University,
7/9 Universitetskaya Embankment, St. Petersburg, 199034, Russia\\\vspace{0.1cm}
${}^3$ Bogoliubov Laboratory of Theoretical Physics, JINR,
6 Joliot-Curie, Dubna, 141980, Russia\\\vspace{0.1cm}
${}^4$ Department of Physics, Rikkyo University Toshima,
Tokyo 171-8501, Japan\\\vspace{0.1cm}
${}^5$ Laboratoire de Physique de l'E\'cole Normale Sup\'erieure,
Universit\'e PSL, CNRS, Sorbonne Universit\'e, Universit\'e
Paris-Diderot, Sorbonne Paris Cit\'e, 24 rue Lhomond, 75005 Paris,
France}

\end{center} 
\vskip .2 in 

\begin{abstract}
The Ising critical exponents $\eta$, $\nu$ and $\omega$ are
determined up to one-per-thousand relative error in the whole range
of dimensions $3 \le d < 4$, using numerical conformal-bootstrap
techniques. A detailed comparison is made with results by the
resummed epsilon expansion in varying dimension, the analytic
bootstrap, Monte Carlo and non-perturbative renormalization-group
methods, finding very good overall agreement. Precise conformal
field theory data of scaling dimensions and structure constants are
obtained as functions of dimension, improving on earlier findings,
and providing benchmarks in $3 \le d < 4$.
\end{abstract} 
 
\vfill 
\end{titlepage} 
\pagenumbering{arabic} 
\numberbysection

%-1--------------------------------------------- 
\section{Introduction}
\vspace{-\baselineskip}\indent

Many approaches to critical phenomena obtain results in continuous
space dimension, although physically relevant dimensions are
integer. Most notable is the perturbative renormalization group in
$d=4-\e$ dimensions~\cite{wf,zinn_ising,zinn_on,mc-rev}. This is not merely
a technical issue: quantities as functions of real $d$ can clarify
features that are harder to see at discrete values. E.g., one can
follow the topology of the renormalization-group (RG) flow as
a function of dimension and find instances where the universality class
changes at non-integer values.
This proved particularly useful for systems with long-range
interactions~\cite{PrudnikovPrudnikovFedorenko2000,slava-long,tromb}
or disorder~\cite{Ludwig1987,JacobsenLeDoussalPiccoSantachiaraWiese2008,
  Komargodski:2016auf,ParisiSourlas1979,Kaviraj:2021qii,Wiese2021}.

The recent very precise  numerical conformal bootstrap~\cite{slava-I,slava-IM,slava-rev}
has been formulated in continuous dimension~\cite{slava-d,Sirois:2022vth}, in particular for the Ising model in its whole range $4>d\ge 2$~\cite{Behan:2016dtz,CMO,Henriksson:2022gpa}.
The interest lies in understanding how the strongly interacting Ising
conformal field theory connects to a free scalar in $d=4$ and to the
integrable fully-solvable model in $d=2$~\cite{BPZ,cft}.
Analytic bootstrap approaches which use the dimension as a tunable
parameter were also developed~\cite{null-ref1,null-ref2,alday,4th-1,Bissi:2022mrs,Hartman:2022zik,Bertucci:2022ptt,Henriksson:2022rnm,4th-2}.
Initially, the non-unitarity of the theory in non-integer dimensions~\cite{slava-uni}
was thought to hamper the numerical methods involving positive
quantities. These concerns have been overcome
by \emph{de facto} never observing problems for the quantities of interest, as explained later.

In this paper, we extend the numerical approach of Ref.~\cite{CMO}
using a single correlator, the SDPB~\cite{SDPB} routine for
determining the unitarity domain, and the Extremal Functional
Method~\cite{extr1,extr2} for solving the bootstrap equations. We
obtain improved results for the scaling dimensions in $4>d \ge 3$ by a
denser scanning of the unitary region near the Ising point, i.e., the
kink. The latter gets parametrically sharper as $d$ approaches $4$,
allowing for its better identification.
The conformal spectrum in dimensions $4>d\ge 2.6$ has
  also been obtained in Ref.~\cite{Henriksson:2022gpa} via the
    advanced \emph{navigator} bootstrap technique~\cite{Reehorst:2021ykw}.  We use these very precise results in
  combination with ours to obtain a consistent description of the
  low-lying spectrum.

The achieved precision allows us to perform a detailed
comparison with state-of-the-art epsilon expansion in two regimes: for
$d$ close to $4$, the series is directly compared to bootstrap data,
using the necessary finer scale for the latter; for intermediate
values between $4$ and $3$ (included), the divergent perturbative
series is resummed using well-established methods involving the Borel
transform~\cite{Kompaniets:2016lmy,Batkovich:2016jus,panzer,Kompaniets:2019zes}.

The analysis is done on the dimensions of the conformal fields
  $\s, \e, \e'$,   corresponding to spin, energy and
  subleading energy. They determine the critical exponents
  $\eta, \nu, \w$.  The precision of our bootstrap data is
  summarized by the (mostly) $d$-independent value of the relative error
  ${\rm Err}(\g)/\g=O(10^{-3})$ for the
  anomalous dimensions $\g$ of the conformal fields $\s$ and $ \e$. As the
  anomalous dimensions are very small for $d\approx 4$, the precision
  for the conformal dimensions $\D_\s, \D_\e$ is even higher in
  this region. Regarding the subleading energy, the relative error
  ${\rm Err}(\D_{\e'})/\D_{\e'}$ stays at three digits, as explained
  later. Some of the structure constants are determined
  with     a higher $O(10^{-4})$ accuracy.

We compare our data with recent results of the
analytic bootstrap~\cite{4th-1,Bissi:2022mrs,Hartman:2022zik,
Bertucci:2022ptt,Henriksson:2022rnm,4th-2},
Monte Carlo
simulations~\cite{mc-hasen-old,FerrenbergXuLandau2018,mc-hasen-new}
and the non-perturbative
RG~\cite{Balog:2019rrg,DupuisCanetEichhornMetznerPawlowskiTissierWschebor2021}.
We find that the data by all methods agree very well. This is rather
rewarding given the achieved precision. Besides confirming the high
quality of conformal-bootstrap results, our analysis provides a
reference point for further analytic and numerical methods aiming at
exploring critical phenomena in varying dimensions.

The outline of this paper is the following. In Sec.~\ref{sec:boot_res}
we summarize our bootstrap protocol~\cite{CMO} and present the results
for the three main conformal dimensions mentioned above, together with
their polynomial fits as a function of dimension and the estimation of
errors. In Sec.~\ref{sec:eps_exp_comp} we briefly recall the
properties of the epsilon expansion and resummation techniques. We
then compare its predictions with our bootstrap data and the results
by other methods, and authors. A detailed analysis of all issues is
presented. In Sec.~\ref{sec:higher_fields}, we report the numerical
bootstrap data for scaling dimensions of structure constants
and other conformal fields, and compare them to the existing
epsilon expansion. In the conclusions in Sec.~\ref{sec:conclusions} we
discuss open questions.

%-2--------------------------------------------- 
\section{Conformal bootstrap in non-integer dimension}\label{sec:boot_res}
\vspace{-\baselineskip}\indent

The aim of this section is to summarize our procedure for deriving
conformal data of scaling dimensions and structure constants, as a
function of the space-time dimension $4 > d \ge 2$. We  
first discuss the conformal dimensions of three main fields
${\mathcal{O}}= \s, \e, \e'$. Our goal is to provide  
a polynomial description of $\D_{\mathcal{O}}$ as a function of
$y = 4-d$, by performing a \emph{best fit} of the data obtained
at several values of
$d$\footnote{Note that $\e$ is the energy field, the
next-to-lowest scalar primary field, not to be confused with the
deviation from four dimensions denoted by $y$.}.
Our results are finally compared
to those obtained from the resummed epsilon expansion in
Section~\ref{sec:eps_exp_comp}.

%-2.1---------------------------------------------
\subsection{Summary of numerical methods}
\vspace{-\baselineskip}\indent

The conformal dimensions and structure constants of the critical Ising
model as a function of $d$ are computed in the setup of
Ref.~\cite{CMO}, which we shortly summarize for the reader's
convenience. We consider a single 4-point correlator
$\braket{\s(x_1)\s(x_2)\s(x_3)\s(x_4)}$, where $\s(x)$ is the primary
scalar field with lowest dimension, denoted $\D_\s$. We truncate the
functional bootstrap equation to $190$ components\footnote{
This corresponds to the standard bootstrap parameter $\Lambda=18$, which counts
the number of derivatives in the approximation of the functional
basis.}. The unitarity condition for this equation is determined
through the SDPB algorithm~\cite{SDPB}, leading to a bound in the
$(\D_\s, \D_\e)$ plane; next, the Extremal Functional Method
(EFM)~\cite{extr1,extr2} is used to solve the equations on this
boundary. We use the generalization of these numerical methods to
non-integer dimensions developed in Ref.~\cite{CMO}, and detailed in
its Appendix A.

Our $1$-correlator numerical bootstrap approach has been surpassed by
more recent implementations~\cite{slava-rev,sd-old,sd,Behan:2016dtz,Henriksson:2022gpa}, but
we find it   convenient for determining the low-lying spectrum
with modest computing resources.
 The complete determination of the
conformal data for one value of $d$ requires about $20$ hours on $256$
cores, corresponding to $5000$ core hours. This simple setting allows
us to evaluate the spectrum for various dimensions $d$.

The first crucial step is to locate the Ising critical point in
parameter space. To this end, we adopt the twofold strategy
of Ref.~\cite{CMO}, consisting in searching the kink on the
unitarity boundary in the $(\D_\s,\D_\e)$ plane and, at the same time,
minimizing the central charge $c$~\cite{slava-IM}. This procedure
allow us to determine for each value of $d$ an interval of values for
$\D_\s,\D_\e$ and $c$, that we take as the Ising conformal theory, accompanied by an estimate of the uncertainty.

\begin{figure}
\centering
\hspace*{-0.4cm}
\includegraphics[scale=0.219]{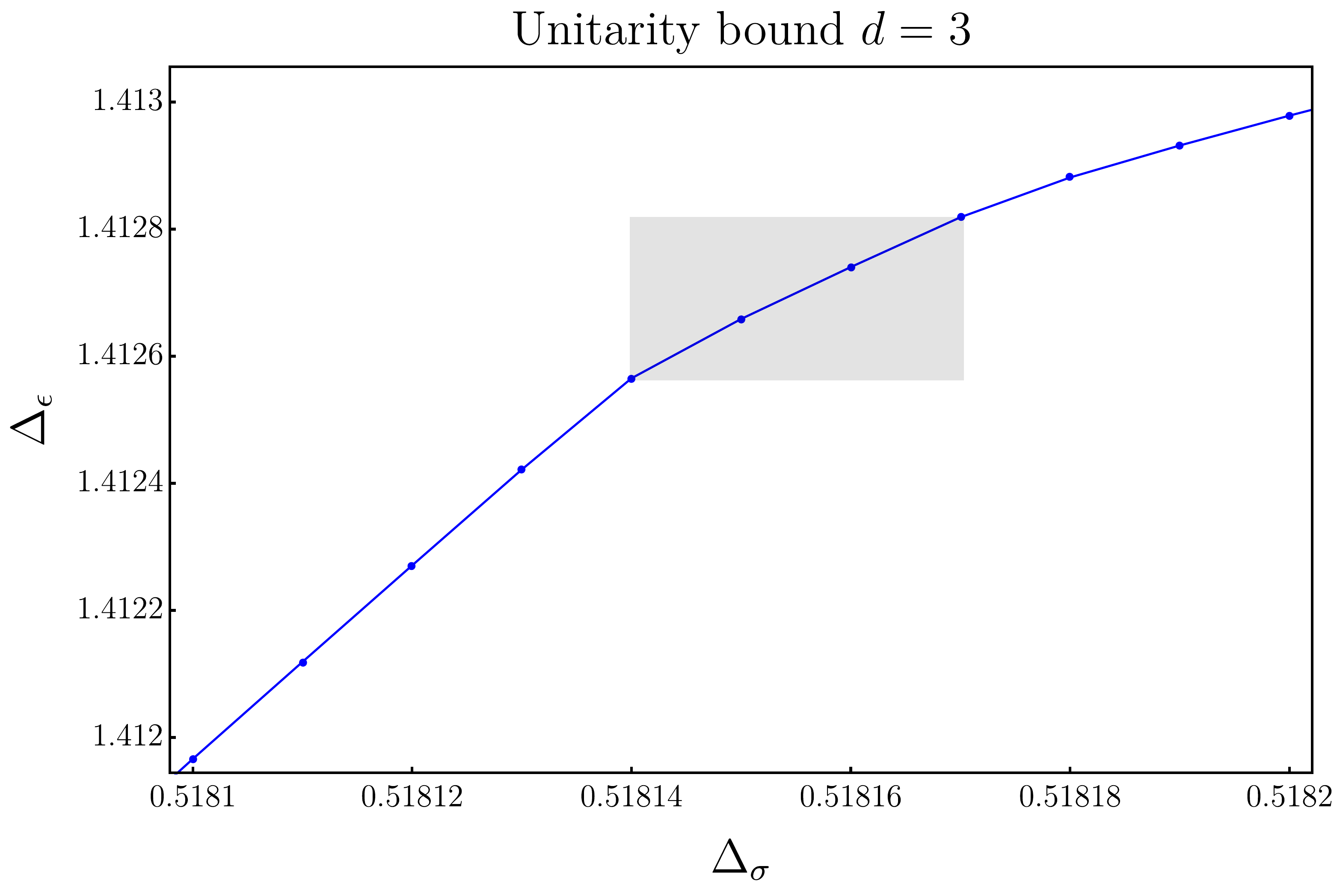} 
\includegraphics[scale=0.222]{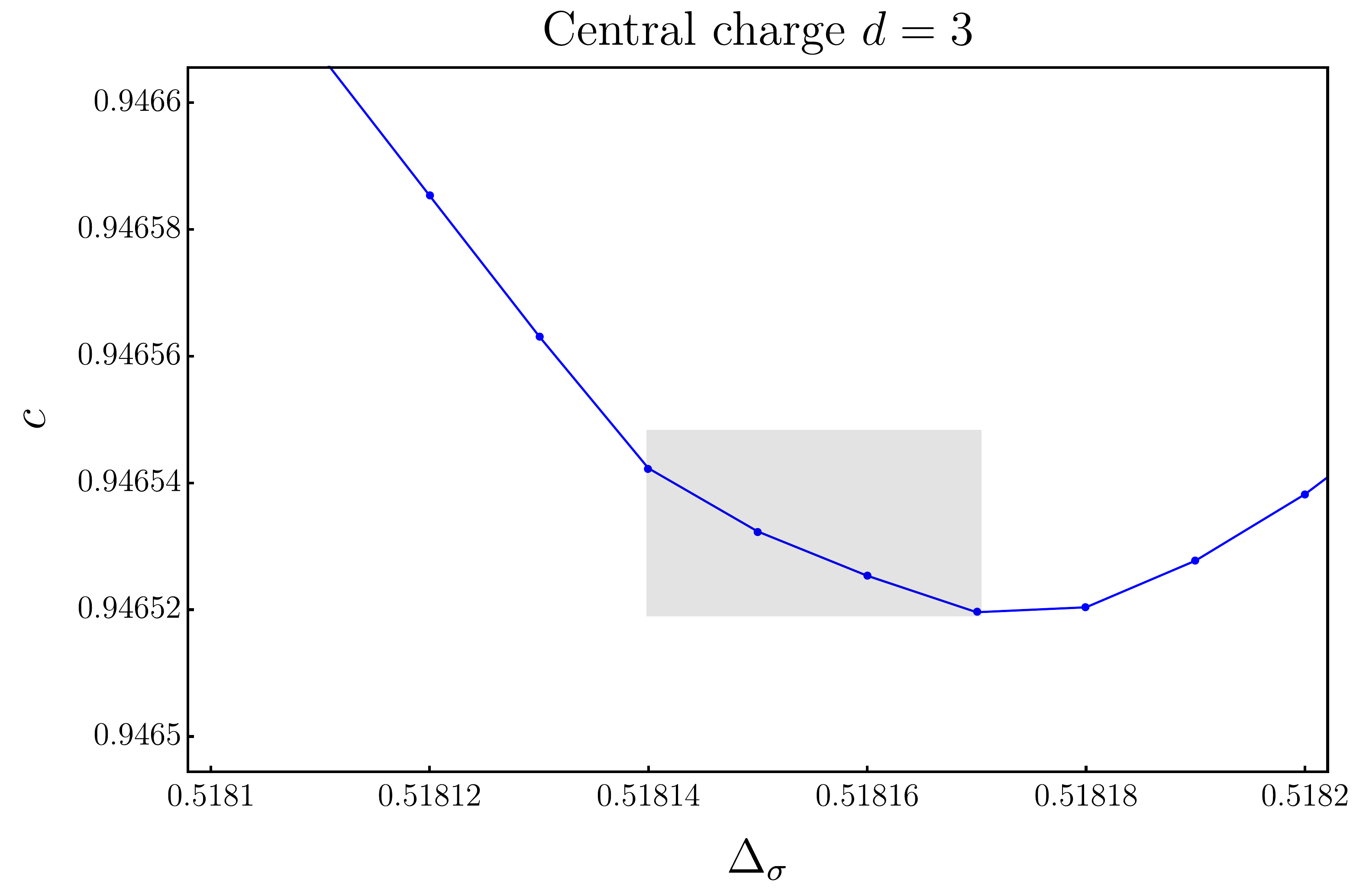}
\hspace*{-0.2cm}
\includegraphics[scale=0.235]{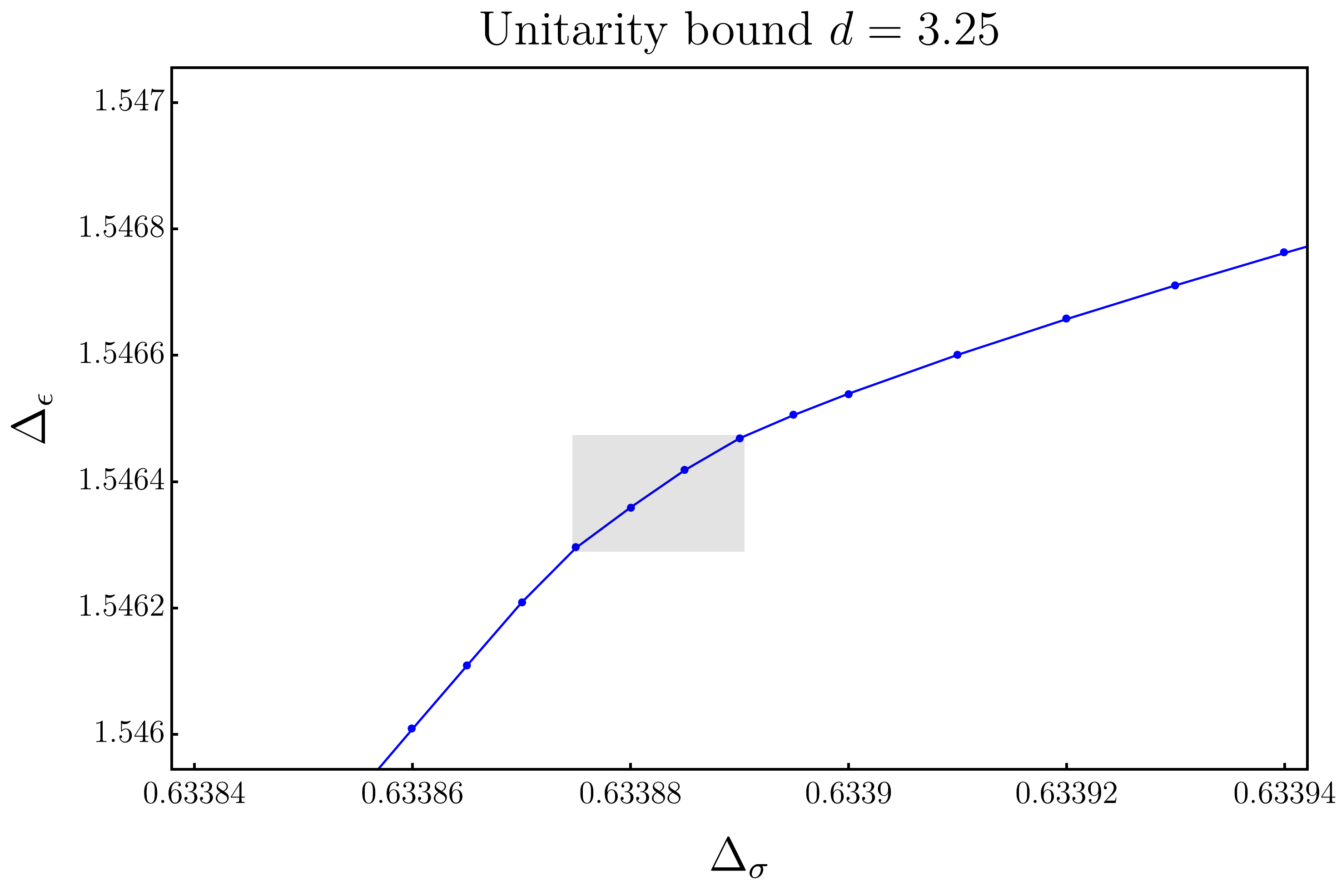} 
\includegraphics[scale=0.238]{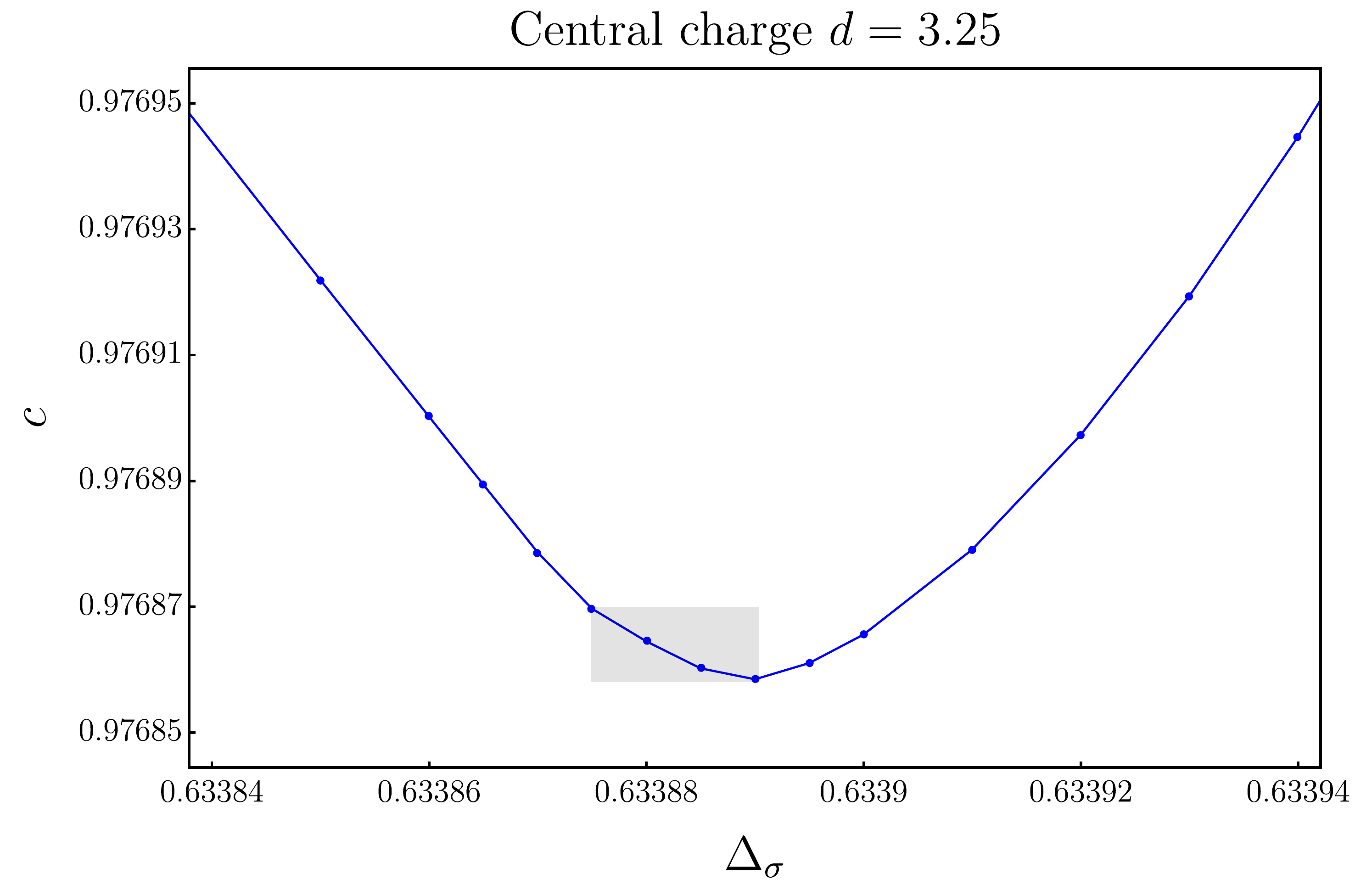}
\includegraphics[scale=0.235]{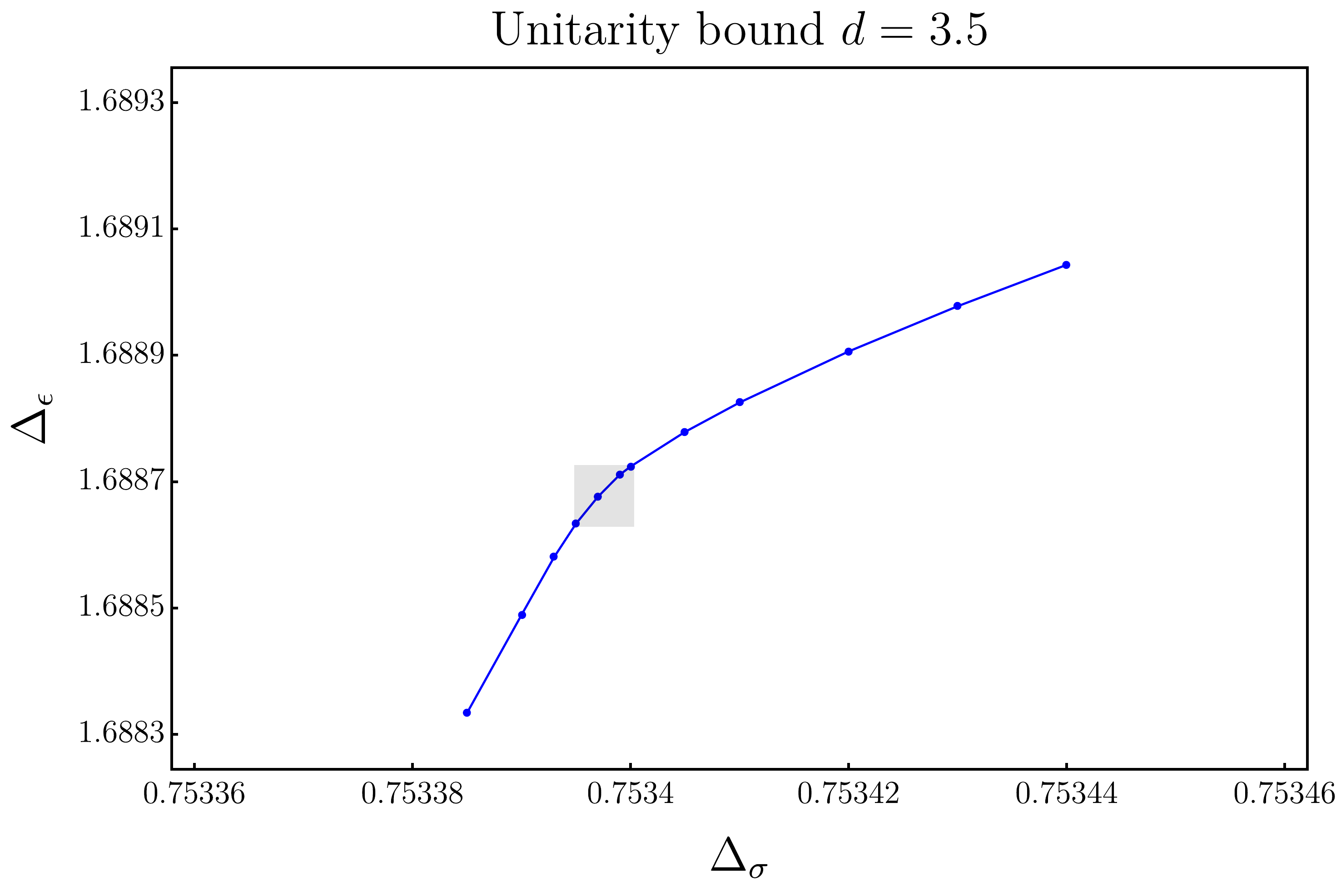} 
\includegraphics[scale=0.238]{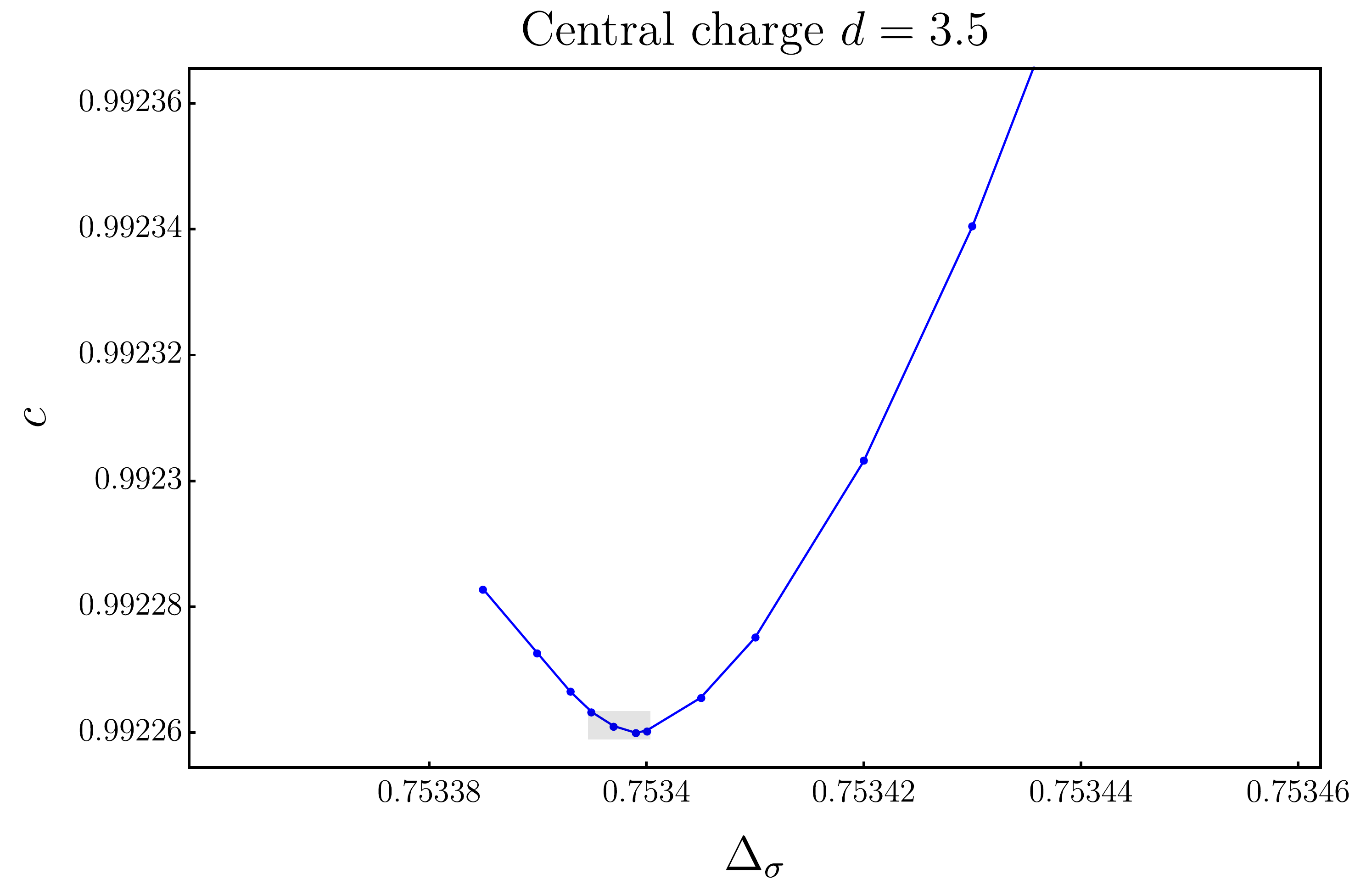} 
\includegraphics[scale=0.223]{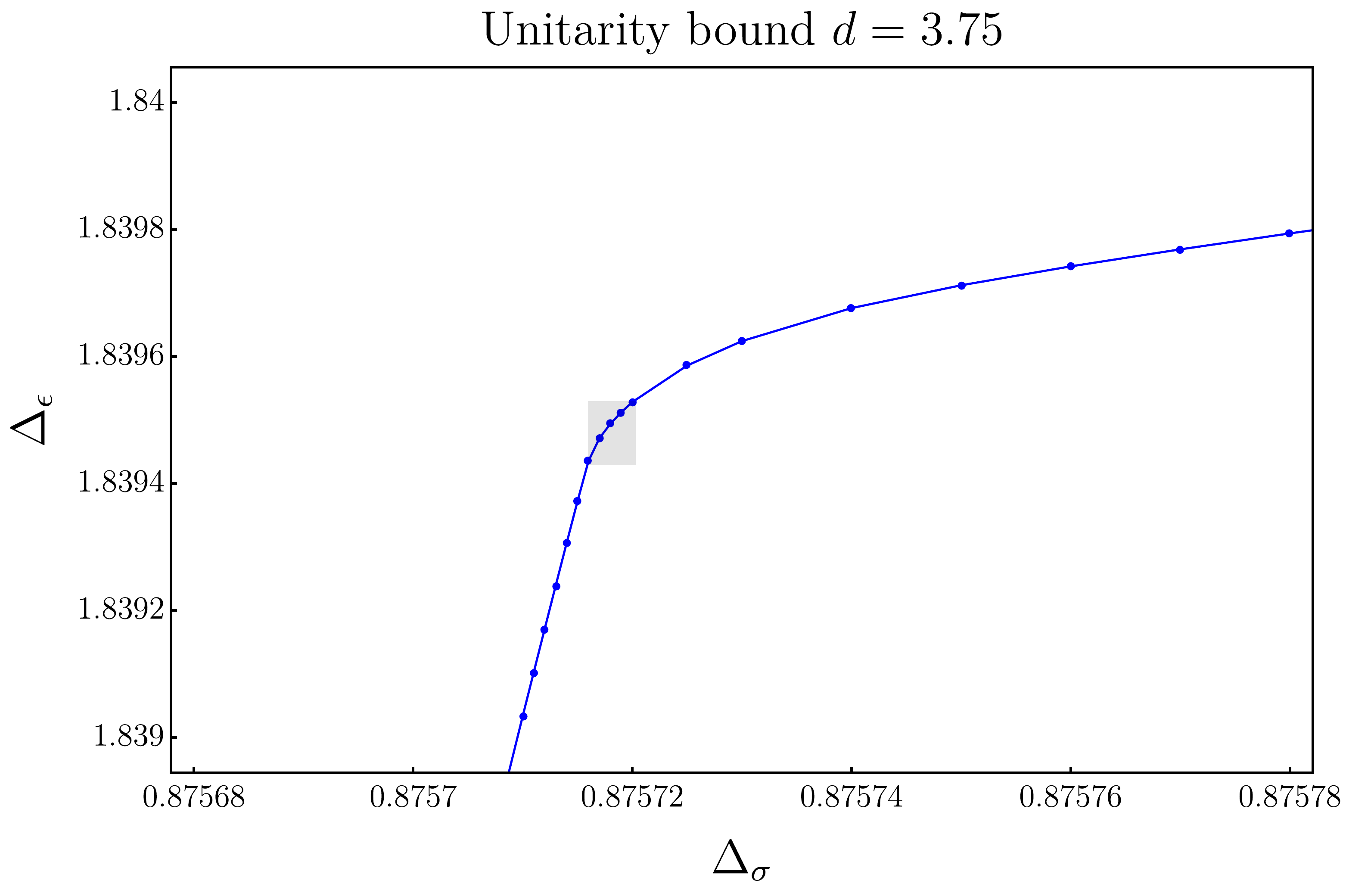}
\includegraphics[scale=0.223]{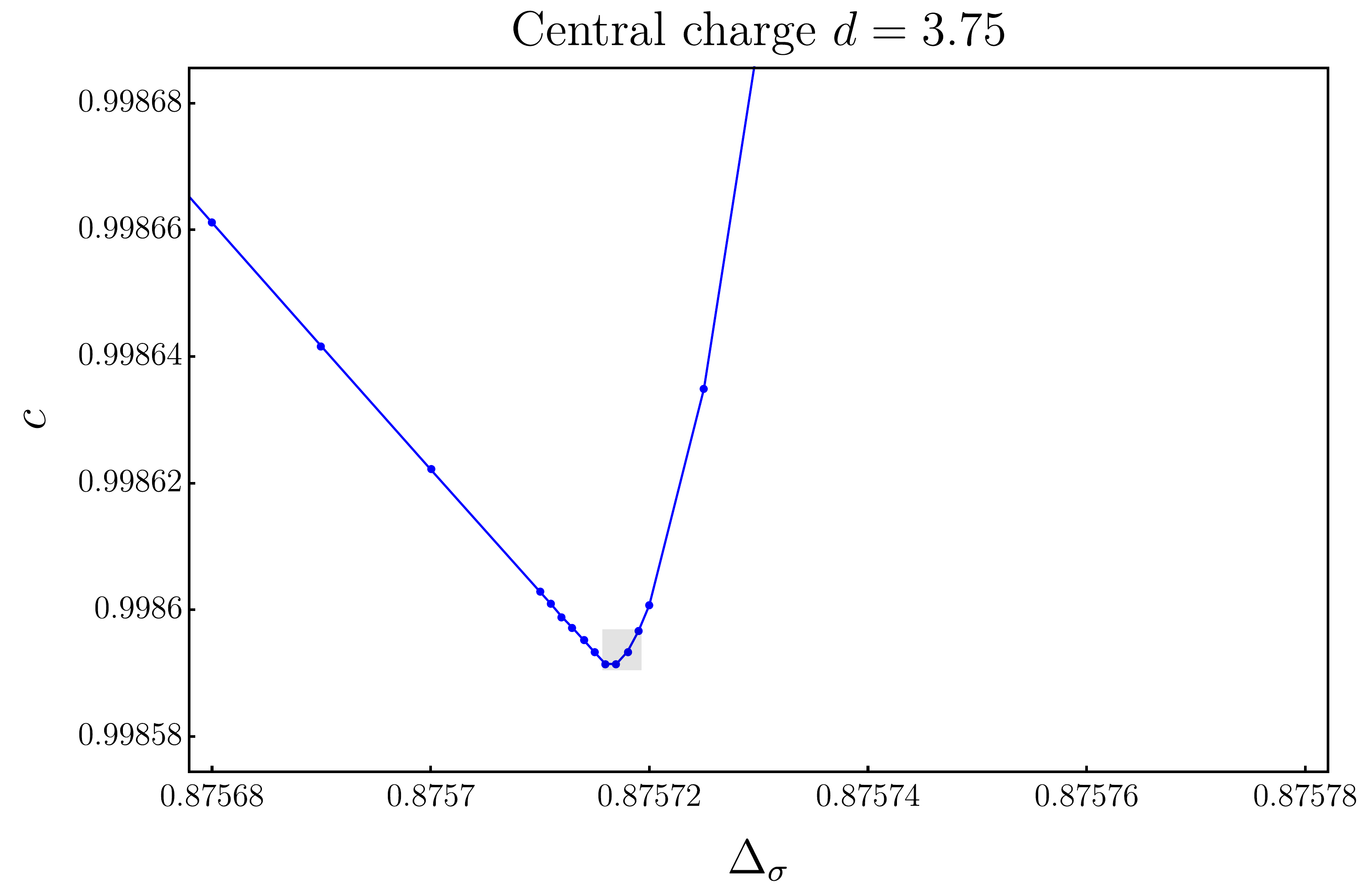} 
\caption{Determination of the Ising critical point for
$d=3, 3.25, 3.5, 3.75$ ($d=3$ data from
Ref.~\cite{CMO}). Left plots: Identification of the kink;
the blue points correspond to the solutions of the
bootstrap equations. Right plots: position of the $c$ minimum.
The grey shaded areas represent the estimated
errors on $\D_\s$, $\D_\e$ and $c$.}
\label{fig:ising_point}
\end{figure}

\begin{figure}
\centering
\includegraphics[scale=0.24]{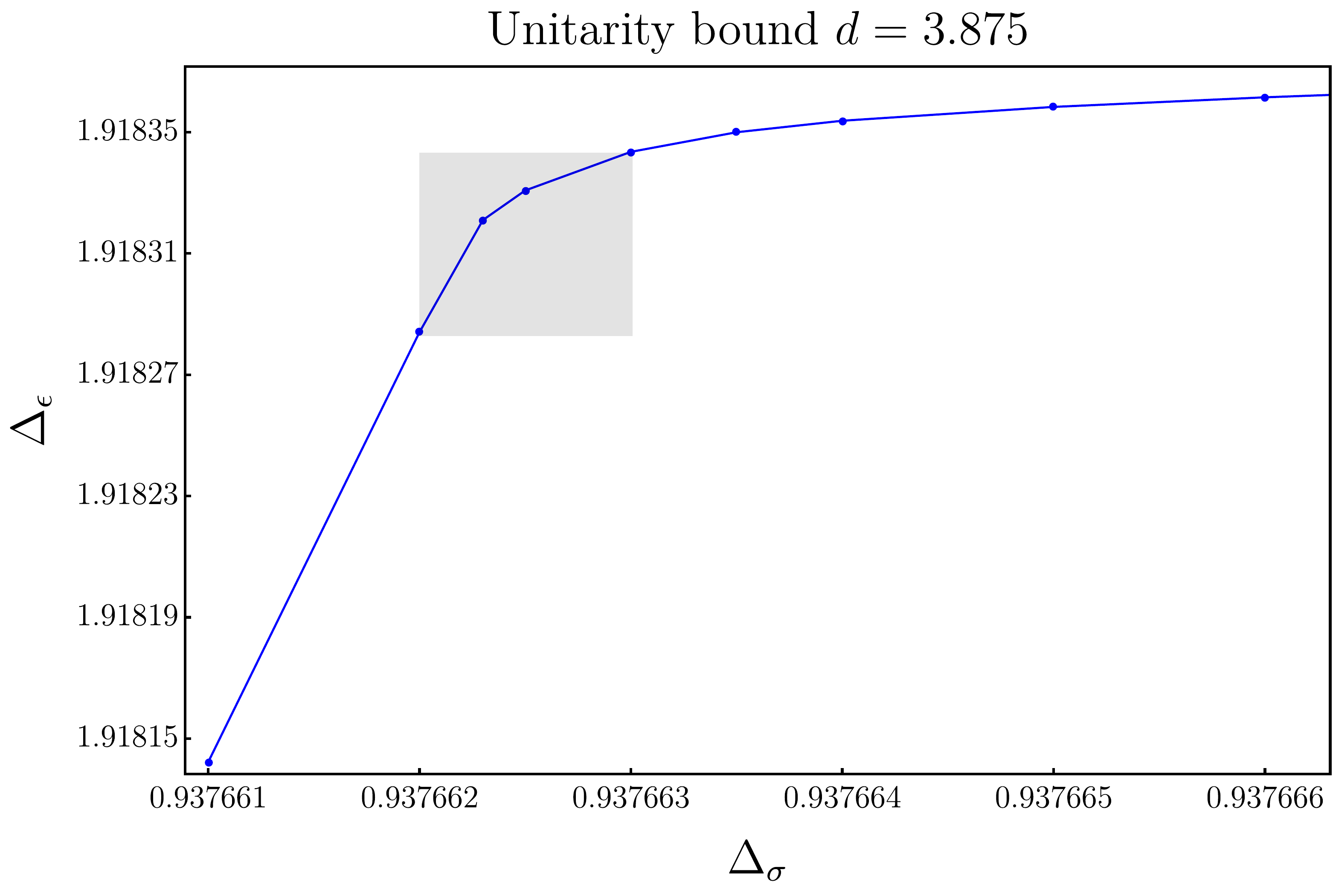}
\includegraphics[scale=0.24]{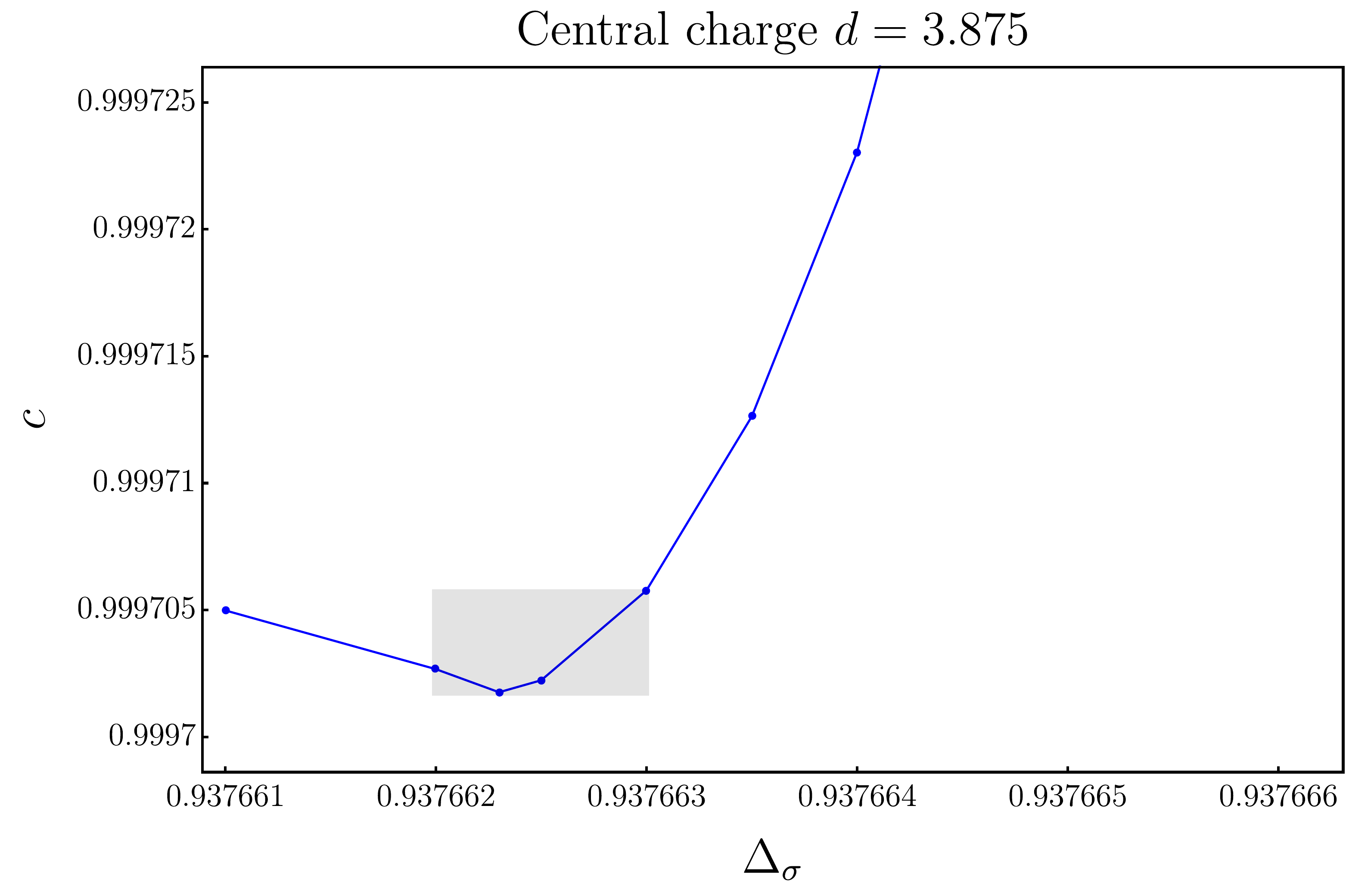} 
\caption{Determination of the Ising point for $d=3.875$, as in
Fig.~\ref{fig:ising_point}. Note the magnified scale on both axis
with respect to those of Fig.~\ref{fig:ising_point}.}
\label{fig:ising_point_new}
\end{figure}

This procedure is displayed in Fig.~\ref{fig:ising_point}, where we
show the identification of the Ising point for $d=3, 3.25, 3.5$ and
$3.75$. The gray area in the plots indicates the chosen errors for
$\D_\s, \D_\e$ and $c$, which are roughly determined by the mismatch
between the positions of the minimum and the kink. As a conservative
choice, we consider an interval of four data points for each value of $d$.

The precision is greater than in 
Ref.~\cite{CMO}, because we perform a finer scan of the $\D_\s$ values
around the kink. We observe that the kink and the minimum
get sharper for $d\to 4$, as shown by the four pairs of plots
drawn on the same scale in Fig.~\ref{fig:ising_point}; this is 
convenient in our approach, since it leads to an increased
precision when anomalous dimensions are smaller.
In Fig.~\ref{fig:ising_point_new}, we show the point $d=3.875$, not considered in the earlier work. It is necessary for
studying the region of $d\to 4$. Here the curves are so steep that
magnified scales are needed.

Once the Ising point is determined, we obtain the rest of the
conformal data as follows. The solution of the bootstrap equations
gives a spectrum of conformal dimensions $\D_{\mathcal{O}}$ and
structure constants $f_{\s\s\mathcal{O}}$ as a function of $\D_\s$; they 
are divided into different sets characterized by the spin
$\ell=0,2,4,\dots$ of the operator $\mathcal{O}$. The 
estimation of $\D_{\mathcal{O}}$ and $f_{\s\s\mathcal{O}}$ is obtained
by taking the central value of such quantities for $\D_\s$ varying in
the interval previously identified as the Ising point (grey areas
in Figs.~\ref{fig:ising_point} and~\ref{fig:ising_point_new}). The
error is obtained from their dispersion.

It is interesting to point out that, although we largely improved the
precision of our results for $4>d>3$ with respect to Ref.~\cite{CMO},
we observe no signs of trouble associated to non-unitarity in our bootstrap
spectrum. On general
grounds, non-unitarity contributions are expected to appear for
non-integer values of $d$ due to the presence of negative-norm
states~\cite{slava-uni}. However, these occur at very high order in
the OPE expansion of the correlator $\langle \s\s\s\s\rangle$, thus we may
argue that they have numerically negligible structure constants. 
As a matter of fact,
their presence does not seem to yield problems in solving the
bootstrap equations with our method. This conclusion was also reached
by recent $3$-correlator bootstrap studies of the critical
$\mathrm{O}(N)$ models~\cite{Sirois:2022vth} and the Ising
model~\cite{Henriksson:2022gpa} in non-integer space dimensions using
the navigator method~\cite{Reehorst:2021ykw}.

\FloatBarrier

%-2.2---------------------------------------------
\subsection{Analysis of conformal dimensions of the three leading fields for $4>d \ge 3$}\label{sec:3_lead_fields}
\vspace{-\baselineskip}\indent

In Tab.~\ref{tab:conf_dim_results} we present our results for the
conformal dimensions $\D_{\mathcal{O}}$ in $4>d>3$ along with those
of Ref.~\cite{CMO} for $3 \ge d > 2$, also
employed in the following. Our implementation of the
bootstrap determines with high precision the conformal
dimensions and structure constants for the first few low-lying
operators with $\ell=0,2$ and $4$: 
$\mathcal{O}_{\ell=0}=\s,\e,\e'$, $\mathcal{O}_{\ell=2}=T'$ and
$\mathcal{O}_{\ell=4}= C$~\cite{CMO}.

\begin{table}[!htb]
\centering
\be 
\resizebox{\hsize}{!}{$
\begin{array}{|c|l|l|l|l|l|l|l|}
\hline 
d & \Delta_\sigma & \Delta_\e & \Delta_ {\e'} & \Delta_ {\e''} & \Delta_{ T'} & \Delta_{C}  & \Delta_{C'} \\
\hline 
\mathbf{4} & \mathbf{1} & \mathbf{2} & \mathbf{4} & \mathbf{6} & \mathbf{6} & \mathbf{6} & \mathbf{8} \\
\hline
3.875   & 0.9376625(5)  & 1.91831(3)  & 3.992(2)    & 7.0(3)  & 5.9307(6)  & 5.8752253(9) & 7.903(3) \\
3.75    & 0.8757175(15) & 1.83948(4)  & 3.9771(12)  & 6.8(2)  & 5.8616(12) & 5.75111(13)  & 7.81(3) \\
3.5     & 0.753398(3)   & 1.68868(5)  & 3.9296(8)   & 6.82(7) & 5.734(7)   & 5.5053(5)    & 7.55(6) \\ 
3.25    & 0.633883(8)   & 1.54639(9)  & 3.8776(11)  & 6.92(6) & 5.59(2)    & 5.264(2)     & 7.25(10) \\
\hline
\hline
3       & 0.518155(15) & 1.41270(15) & 3.8305(15) & 7.01(5)  & 5.505(10)  & 5.026(4)     & 6.7(2) \\ 
2.75    & 0.40747(4)   & 1.2887(2)   & 3.800(2)   & 7.12(8)  & 5.445(15)  & 4.790(5)     & 6.3(2) \\ 
2.5     & 0.30341(1)   & 1.17625(15) & 3.7970(10) & 7.32(2)  & 5.46(3)    & 4.574(9)     & 5.78(13) \\
2.25    & 0.20822(3)   & 1.0784(2)   & 3.847(1)   & 7.53(2)  & 5.58(5)    & 4.344(14)    & 5.36(6) \\
2.2     & 0.19053(8)   & 1.0610(5)   & 3.864(4)   & 7.64(3)  & 5.69(4)    & 4.325(15)    & 5.29(4) \\
2.15    & 0.17333(8)   & 1.0444(4)   & 3.891(6)   & 7.73(3)  & 5.64(13)   & 4.28(3)      & 5.19(1) \\
2.1     & 0.15663(8)   & 1.0286(5)   & 3.9215(5)  & 7.82(3)  & 5.820(10)  & 4.17(4)      & 5.12(4) \\
2.05    & 0.14048(8)   & 1.0134(7)   & 3.9565(5)  & 7.93(3)  & 5.9050(10) & 4.13(6)      & 5.065(15) \\
2.01    & 0.12803(8)   & 1.001(2)    & 3.9900(10) & 8.035(5) & 5.9815(5)  & 4.01440(10)  & 5.0115(15) \\
2.00001 & 0.125000(10) & 0.99989(14) & 4.0002(2)  & 7.99(10) & 6.0006(2)  & 4.000055(10) & 5.00048(8) \\
\hline
\mathbf{2} & \mathbf{0.125} & \mathbf{1} & \mathbf{4} & \mathbf{8} & \mathbf{6} & \mathbf{4} & \mathbf{5}\\
\hline
\end{array}
\nonumber
$}
\ee
\caption{Conformal dimensions of the first few low-lying states for $4>d>2$. Exact values for $d=2,4$ are given in bold, results for $3\ge d >2$ are taken from Ref.~\cite{CMO}.}
\label{tab:conf_dim_results}
\end{table}

\noindent
The goal of this section is to determine the behavior of
$\D_{\mathcal{O}}$ as a function of the variable $y = 4-d$, by finding
the best fitting polynomial that describes the data in
Tab.~\ref{tab:conf_dim_results}. We use all available values, but
focus on the range $4>d\ge 3$ where results are more precise and
allow for a comparison with other approaches. The points for
$3>d\ge 2$ are mainly used for stabilizing the higher powers of the
fitting polynomials\footnote{Note that the lower quality of
$3>d> 2$ data is due to the coarse scanning of $\D_\s$
values, not to an intrinsic limitation of the numerical bootstrap
approach~\cite{CMO}.}.

We employ an improved fit method for $\D_{\cal O}(y)$ that uses
orthogonal polynomials~\cite{ortho_poly}: the idea is to expresses the
$n^{\text{th}}$-order polynomial fit function $f_n(y)$ in terms of
orthogonal polynomials $P_k(y)$ of degree $k=0,1,\dots,n$, instead of
a parameterization in terms of monomials,
$1,y,y^2,\dots,y^n$. To this aim we write
\be\label{poly-def}
f_n(y) = \sum_{k=0}^{n} \alpha_k P_k(y), \qquad
\braket{P_r(y) P_s(y)} \propto \sum_{i=1}^{14} P_r(y_i)P_s(y_i) \propto
\delta_{rs},
\ee
where $y_i$ are the values in Tab.~\ref{tab:conf_dim_results}. This
method is  equivalent to the naive one, but is numerically
more stable and the fit parameters $\alpha_k$ can be determined with
improved precision and less statistical noise.

The optimal degree $n$ for the fitting polynomial is not known \emph{a
priori} and is determined in the following way: The fit with weights
proportional to the inverse square of errors is done for several
values of $n$, and the least chi-square
$\chi^2_{\mathrm{min}}$ is found as a function of $n$.
At a given order $\ov n$, adding a further term
$\alpha_{\ov n+1}P_{\ov n+1}$   
results in a negligible change of $\chi^2_{\mathrm{min}}$ and
the best fit yields a result for $\alpha_{\ov n+1}$ which is
compatible with zero within errors. This identifies $\ov n$ as the degree of
the optimal polynomial. Finally, we use the
results of our best fit for $\{\alpha_k\}$ to assign an error to
$f_n(y)$ in the whole range of $4>d\ge 3$. Details on the  
fitting procedure and the computation of errors
can be found in App.~\ref{appendix:ortho_poly}.

In this section we focus on the three leading operators
$\s, \e$ and $\e'$ (corresponding to $\phi,\phi^2$ and $\phi^4$
in the $\phi^4$ field theory), which are determined with
very good precision. The analysis of
higher-dimensional operators is postponed to
Sec.~\ref{sec:higher_op_dim}. Instead of working with conformal
dimensions, we consider the anomalous dimensions
\be
\label{def_g}
\g_\s    = \D_\s - \frac{d-2}{2},\qquad\qquad
\g_\e    = \D_\e - (d-2), \qquad\qquad
\g_{\e'} = \D_{\e'} - 2(d-2).
\ee
They are related to the Ising critical exponents $\eta$, $\nu$ and $\omega$
by
\be
\label{crit-exp}
\eta   = 2\g_\s, \qquad\qquad
\frac{1}{\nu}    = 2-\g_\e,\qquad\qquad
\omega = d-4+ \g_{\e'}\; .
\ee
The vanishing of anomalous dimensions in the free theory ($d=4$)
is assumed in the following fits.

\begin{figure}[!htb]
\centering
\includegraphics[scale=0.48]{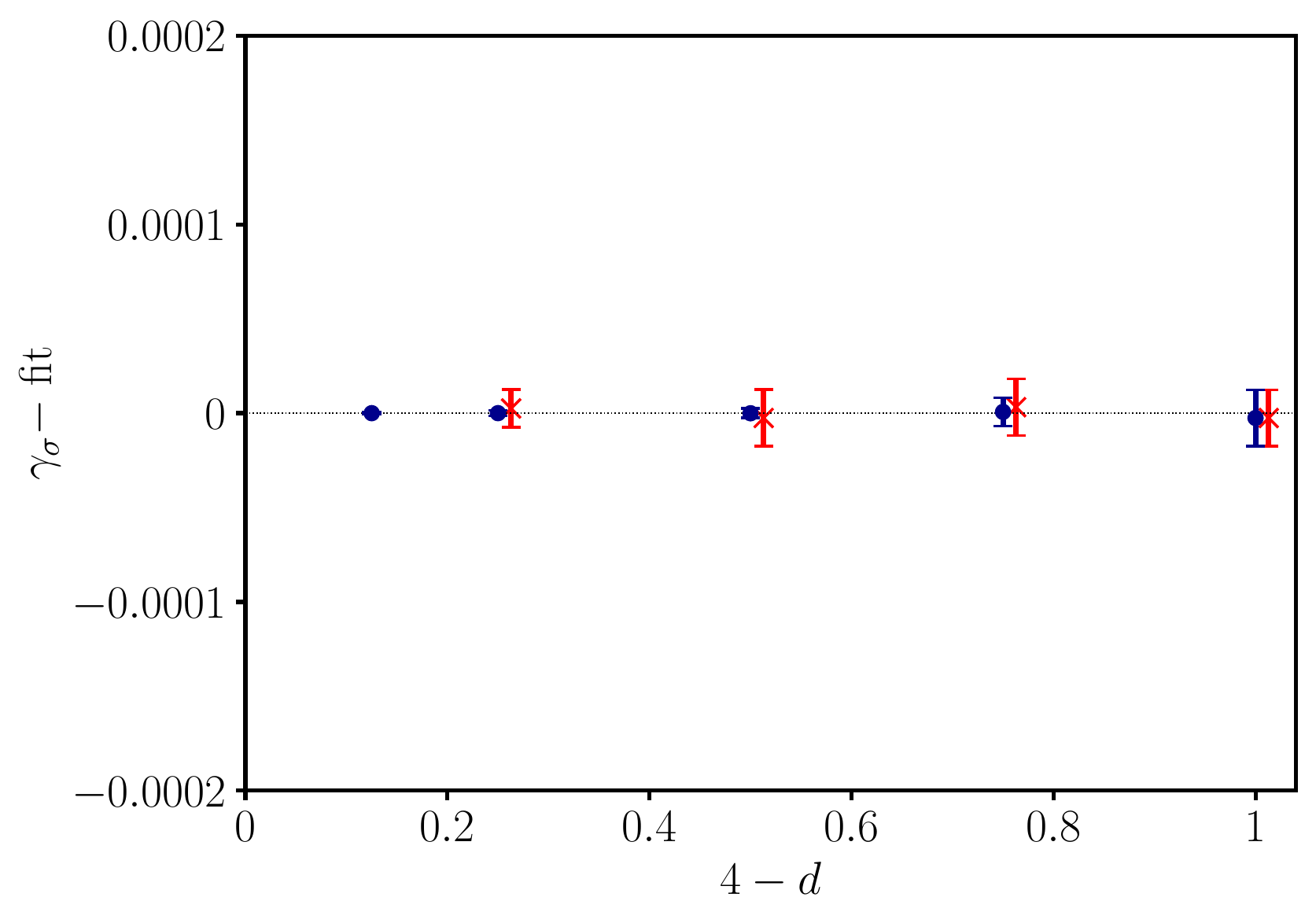}
\includegraphics[scale=0.48]{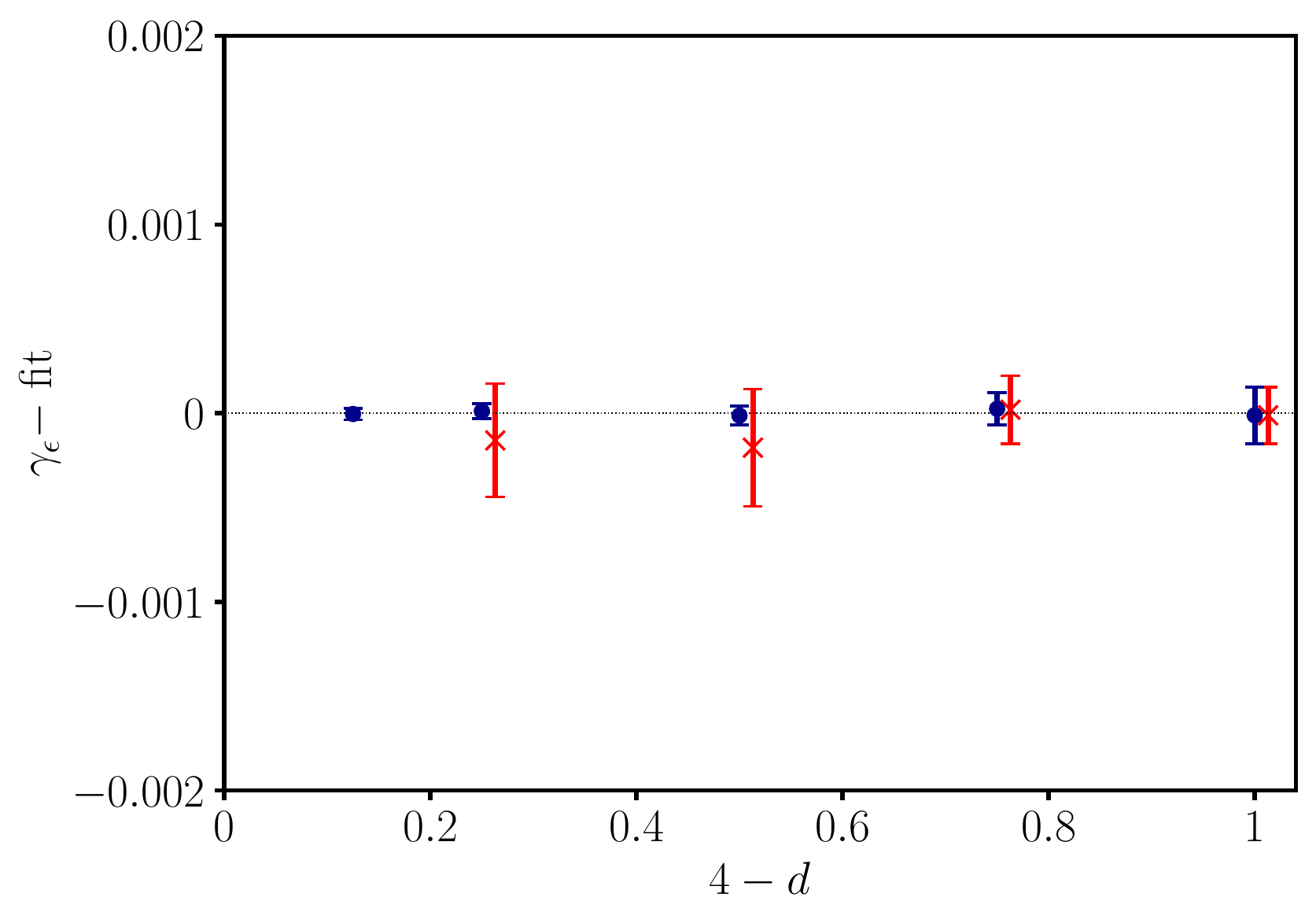}
\includegraphics[scale=0.48]{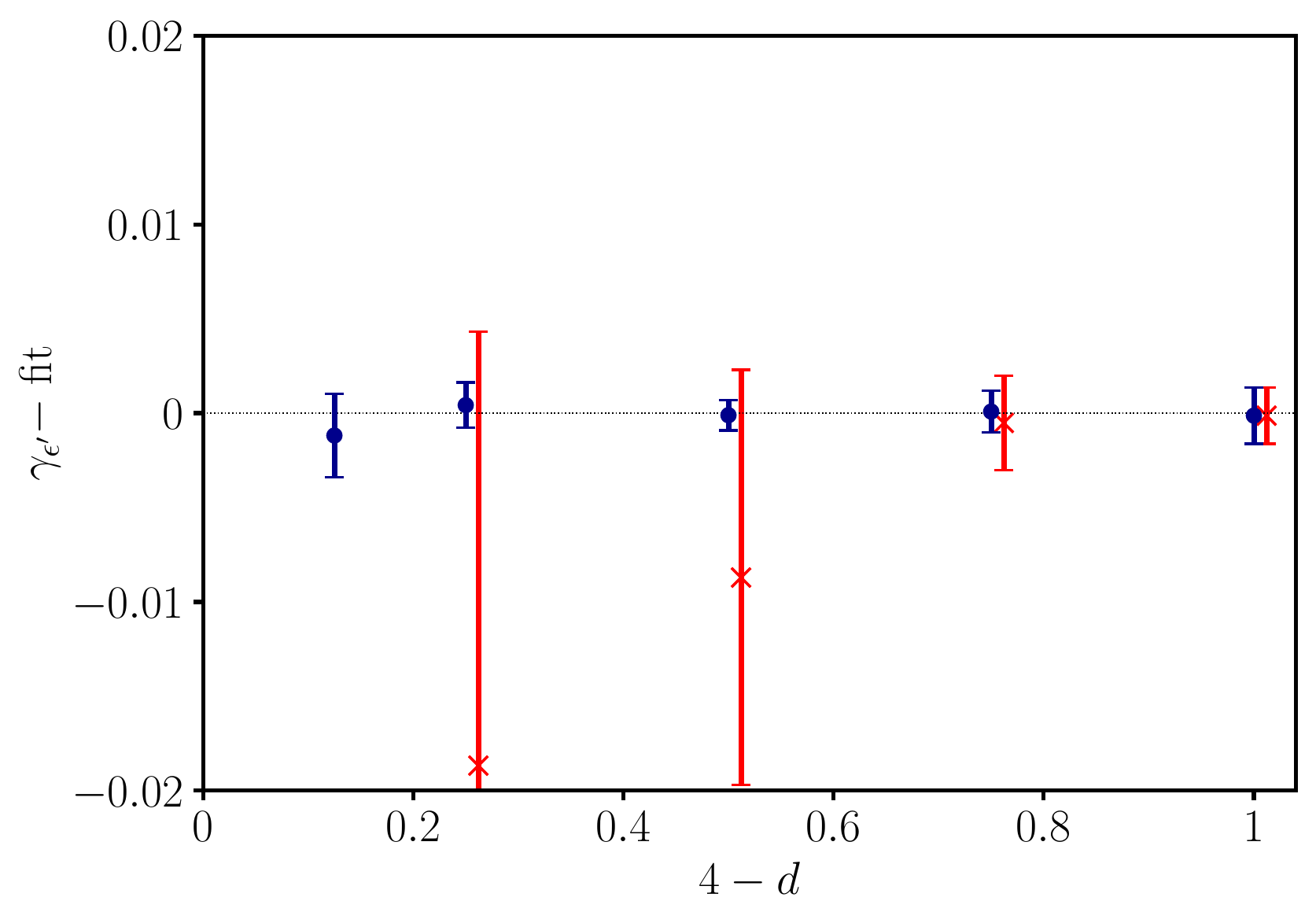}
\caption{Old~\cite{CMO} (red crosses) and new (blue circles) bootstrap
data for $\g_\s, \g_\e, \g_{\e'}$, minus the corresponding
best fits. The plots use the same scales as in Ref.~\cite{CMO}.}
\label{fig:old_vs_new_comp}
\end{figure}

Our analysis starts by comparing the old~\cite{CMO} and new data for
$4>d>3$. In Fig.~\ref{fig:old_vs_new_comp} the new results (blue
circles) show much smaller errors than the earlier findings (red crosses),
due to a more accurate localization of the Ising point, as explained
above. In these and later figures we report the differences
$(\g_{\cal O}- {\rm fit})$ between data and fitting polynomial, because
simpler plots would not capture the small errors involved (note that the abscissas of the three plots differ by factors of ten). The explicit form of the best fitting polynomials are provided in Sec.~\ref{sec:eps_exp_comp}.

\begin{figure}[!htb]
\centering
\includegraphics[scale=0.45]{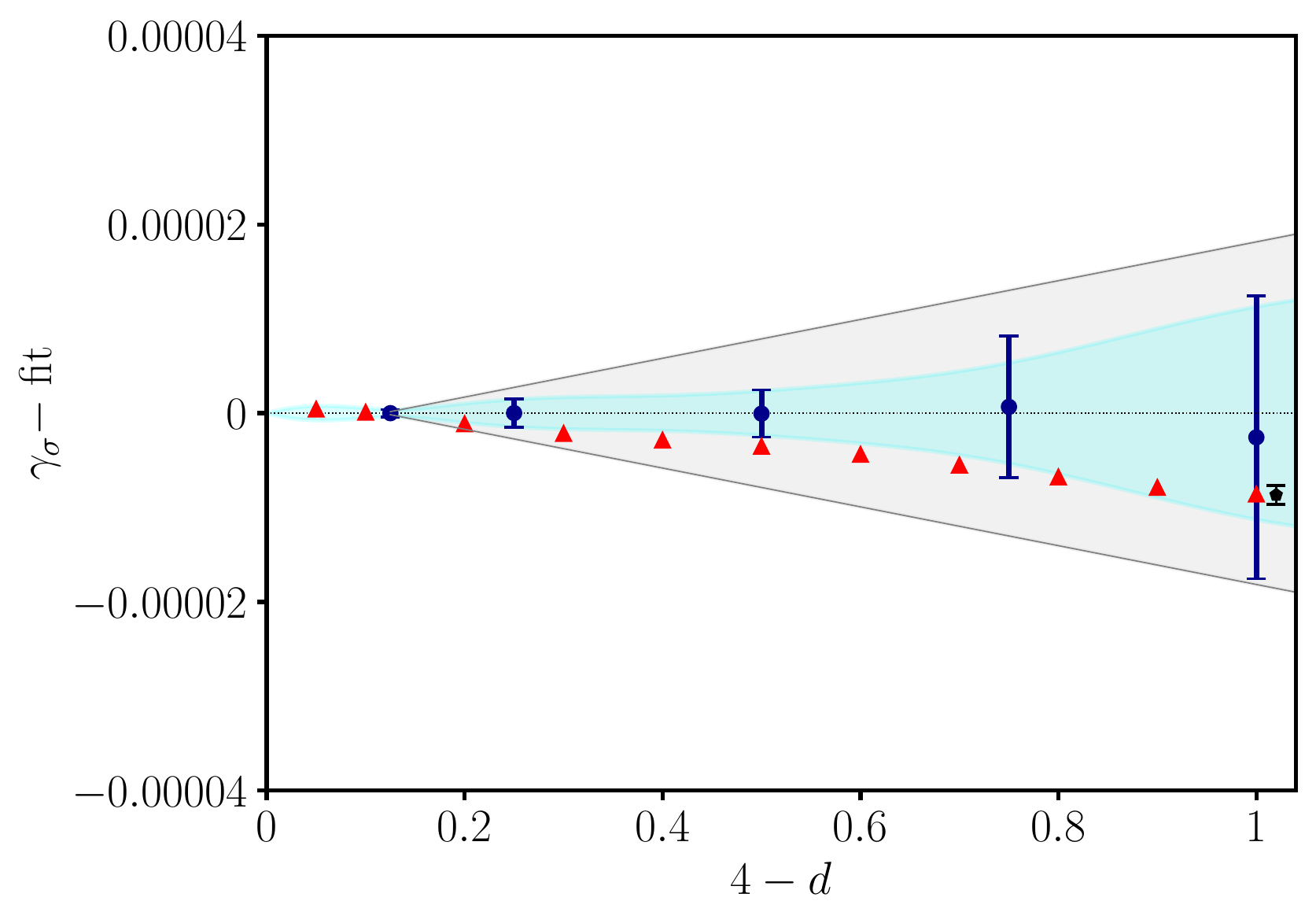}
\includegraphics[scale=0.45]{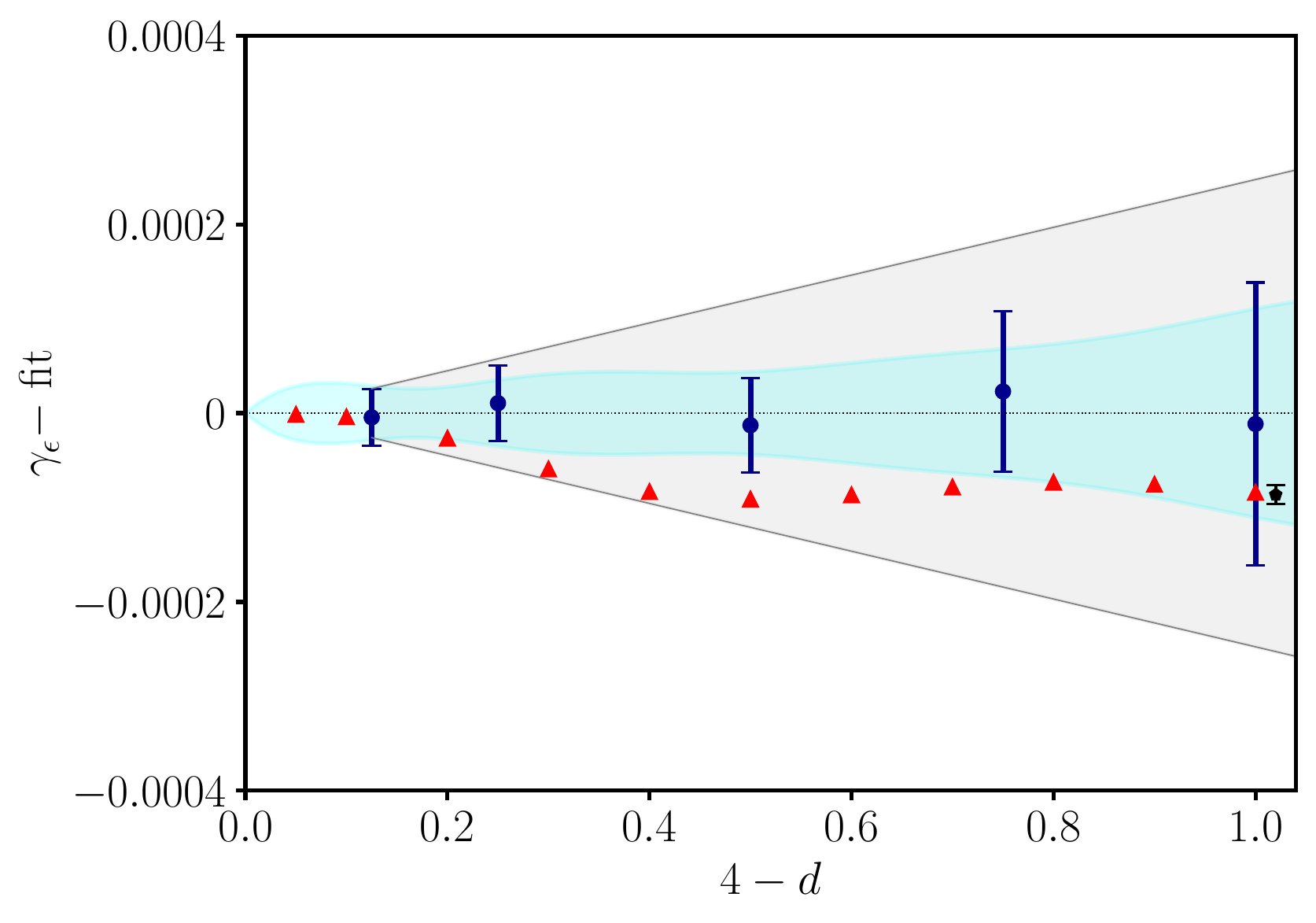}
\includegraphics[scale=0.45]{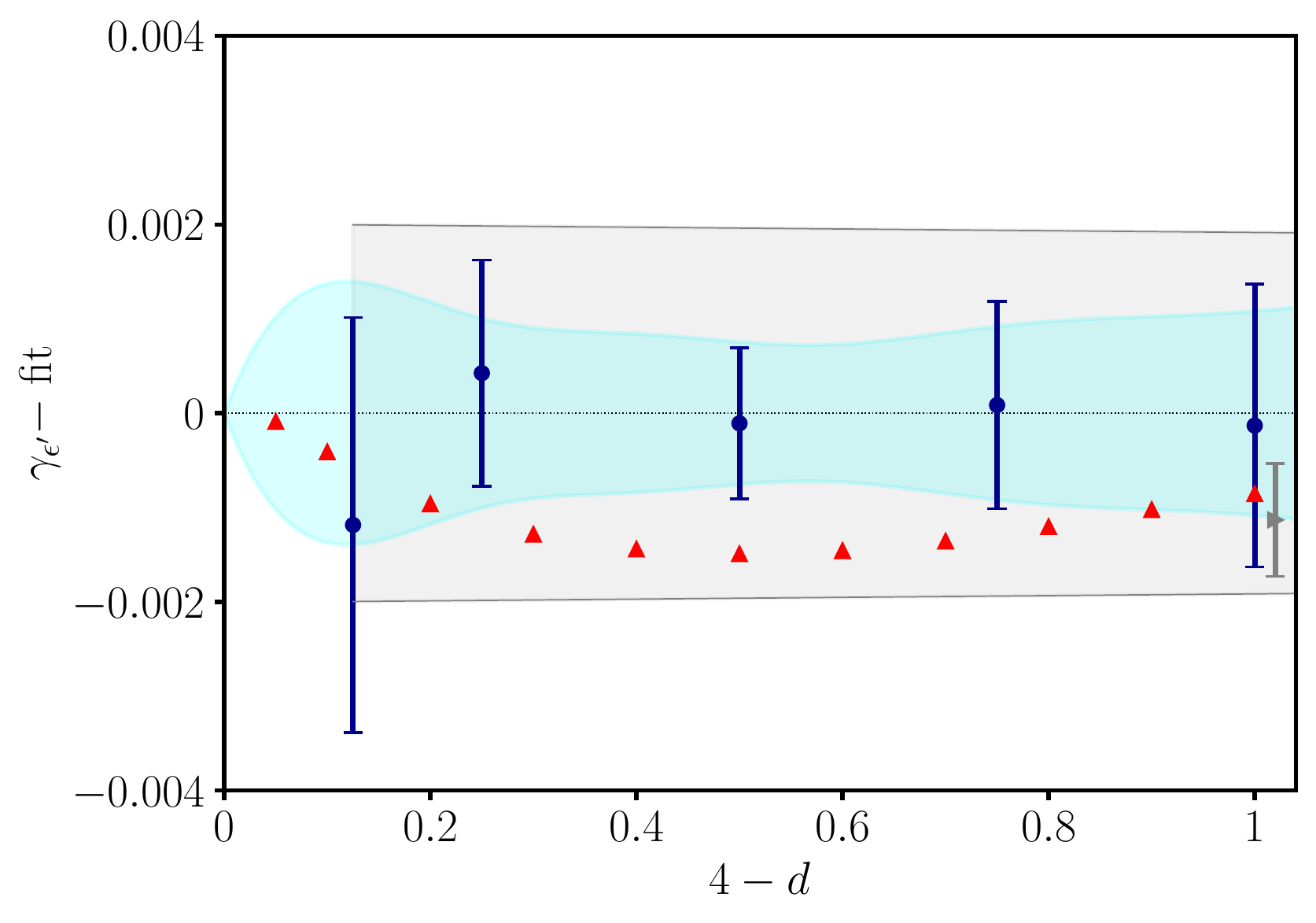}
\caption{Plot of bootstrap data for $\g_\s, \g_\e , \g_{\e'}$ minus
  the best fit values.The shaded area represents the error obtained
  from the $\chi^2$ minimization of the fitting polynomial. The red
  triangles are results from Ref.~\cite{Henriksson:2022gpa} using the
  navigator method in a $3$-correlator bootstrap setup (no error
  bars). Black diamonds and grey rightward triangle for $d=3$ represent
  respectively results by Ref.~\cite{sd} ($\g_\s$ and
  $\g_\e$) and Ref.~\cite{Reehorst:2021hmp} ($\g_{\e'}$); these data
  points are slightly displaced around $d=3$ to improve
  readability. The gray shaded bands represents
    the error bounds reported in Eq.~\eqref{error-size}.}
\label{fig:henriksson_comp}
\end{figure}

Next, we compare these results with those recently obtained
  by solving the
$3$-correlator bootstrap with the navigator method \cite{Henriksson:2022gpa}.
In Fig.~\ref{fig:henriksson_comp} our data, given in earlier figures
  (blue circles), are shown on a finer scale, together with the
  estimated error of the fit (cyan shaded area).
The red triangles are the navigator values: they come with no errors
and thus cannot be directly used for the fits\footnote{
  Earlier results of Ref.~\cite{Behan:2016dtz} are not considered
here due to their large errors.}.
A first observation is the fairly good agreement between the
  two different bootstrap approaches at our level of precision.

We propose to estimate the error of navigator data  
  as follows. We suppose that they are roughly of the same size as those
  found in other 3-correlator studies at $d=3$
(rigorous bounds) \cite{sd,Reehorst:2021hmp}, which are  plotted in
  Fig.~\ref{fig:henriksson_comp} as black diamonds ($\g_\s$ and
  $\g_\e$), and a grey rightward triangle ($\g_{\e'}$).
  Assuming these very small uncertainties for each value of $d$, there seems to be a negative offset with respect to our data,
  in particular for $\eps'$.
This could be a systematic error due to our approximate identification of the Ising point within the unitarity region
(Section 2.1), while the navigator method rigorously determines it within
a unitarity island \cite{Reehorst:2021ykw}.
However, other explanations are possible.

In conclusion, taking into account these considerations,
  we  enlarge the error estimate of our fits
  to the shaded gray bands in Figs.~\ref{fig:henriksson_comp},
which correspond to the following bounds:
  \be\label{error-size}
\frac{\rm{Err}(\g_\s)}{\g_\s} 
\approx \frac{{\rm Err}(\g_\e)}{\g_\e} \lesssim 1\times 10^{-3}, \qquad
\frac{\rm{Err}(\D_{\e'})}{\D_{\e'}}
\lesssim 0.5 \times 10^{-3}, 
\qquad\quad 3.875\ge d\ge 3.
\ee
 Given the small value of anomalous dimensions for $d\to 4$, these
imply extremely low absolute errors,
${\rm Err}(\g_\s)=O(10^{-6})$ and ${\rm Err}(\g_\e)=O(10^{-5})$ in
this range, as spelled out in the following sections.
This allows us to give a precise
comparison to other methods, as a benchmark  
for the Ising universality class in non-integer dimensions.

\FloatBarrier

%--3--------------------------------------------- 
\section{Comparison with the epsilon expansion in
  \hbox{$4>d \ge 3$}}\label{sec:eps_exp_comp}

In this section, we recall some features of the epsilon expansion
and the resummation methods employed for it. We compare unresummed and
resummed series with the bootstrap results for $\g_\s$.
Then, the analysis is extended to $\g_\e$ and $\g_{\e'}$.

%--3.1--------------------------------------------- 
\subsection{Warm-up analysis of the anomalous dimensions $\g_\s$}\label{sec:g_s_analysis}
\vspace{-\baselineskip}\indent

We start with a brief summary of the properties of the perturbative
expansion of the $\phi^4$ field theory in $d=4-y$, which describes the
Ising universality class. This is a textbook
subject~\cite{zinn-justin-book} but we would like to single out a few
aspects that are important in the following comparison with bootstrap
results in varying dimensions\footnote{An up-to-date discussion of
epsilon expansion can be found in
Refs.~\cite{Kompaniets:2016lmy,Batkovich:2016jus,panzer,Kompaniets:2019zes}. We
refer to these works for a proof of the following statements and
appropriate referencing.}.

The $\beta$-function $\b(g,y)$ and the anomalous dimensions
$\g_{\cal O}(g)$, where ${\cal O}=\phi,\phi^2,\phi^4$, take the
following form, in the Minimal Subtraction (MS)~\cite{zinn-justin-book,vasiliev-book} renormalization scheme,
\be
\b(g,y)=- y g + \sum_{k=2}^{n+1} \b_k\; g^k,
\qquad\quad \g_{\cal O} (g)=\sum_{k=1}^n \g_{{\cal O},k}\; g^k.
\label{pert_series_beta}
\ee
The numerical coefficients $\b_k, \g_{{\cal O},k}$ were computed up to
order $n=6$ in Ref.~\cite{panzer}, and $n=7$ in
Ref.~\cite{Schnetz:2016fhy}. While results up to
order $n=15$ are known for a subclass of Feynman diagrams believed to
give the dominant contribution, they are not used
here~\cite{panzer,Dunne:2021lie}.

The coefficients of the $\beta$-function~\eqref{pert_series_beta} grow
exponentially with $k$, and their asymptotic behavior can be estimated
from the contribution of instanton field configurations~\cite{zinn-justin-book}
\be
\b_k \underset{k\to\infty}{\sim} C\; (-a)^k\; k^b\; k! \,.
\label{asympt}
\ee
Similar behaviors are found for the coefficients $\g_{{\cal O},k}$.
The parameters $a, b, C$ depend on the quantity considered.
One finds that the known values of the coefficients up to
order $n=7$ grow very fast with $n$ but have not yet reached their
asymptotic values~\eqref{asympt}~\cite{panzer,Dunne:2021lie}.

The behavior~\eqref{asympt} can be understood as follows: The
perturbative series has a vanishing radius of convergence in the
complex $g$ plane, because real negative values of $g$ correspond to
an upside-down potential and an action not bounded from below.
This fact can be exemplified by the simple \emph{zero-dimension path integral}
(see App.~\ref{appendix:baby_integral}):
\be\label{toy}
{\cal I}(g)=
\int_{-\infty}^\infty \frac{{\rm d} x}{\sqrt{2\pi}}\, {\rm e}^{-\frac
  {x^2}2 - g x^4} = \sum_{k=0}^\infty a_k (-g)^k, \qquad a_k =
\frac{(4k)!}{2^{2k}(2k)! k! }\underset{k\to\infty}{\sim}
\frac{2^{4k}}{\sqrt 2 \pi k
}\times k! \ .
\ee
This is the generating function counting the
number of vacuum Feynman diagrams. The asymptotic
behavior of $a_k$ can be found by a saddle-point analysis of the
integral. In field theory the corresponding saddle point is given by instantons~\cite{zinn-justin-book}\footnote{There is growing consensus that the
large-order behavior is governed by an instanton rather than a
renormalon~\cite{Dunne:2021lie}. If one could go to much higher
orders in the series expansion (e.g., 20-loop order) one could apply
methods of resurgence and trans-series~\cite{Aniceto:2018bis}.}.

The solution of the fixed-point equation $\b(g,y)=0$
gives $g=g(y)$ by perturbative inversion around $g=y=0$; this is used
to rewrite the anomalous dimensions as a series in $y$,
\be
\g_{\cal O}(y) =\sum_{k=1}^n \ov{\g}_{{\cal O},k}\; y^k.
\label{pert_gamma}
\ee
This is again a divergent series of asymptotic form~\eqref{asympt}, with suitable parameters $a$, $b$ and $C$.

The ratio of two consecutive terms in the
series~\eqref{pert_gamma} can be estimated from \eqref{asympt} as,
$ \ov{\g}_{{\cal O},k}\; y/\ov{\g}_{{\cal O},k-1}\approx -a k y$, which is
larger than one for $y> 1/|ak|$. A
simple conclusion is that the more terms are present in the
perturbative series~\eqref{pert_gamma}, the sooner it diverges as a series
in $y$. We can draw two main conclusions:

\begin{enumerate}[label=\textit{\roman*})]
\item\label{item:pertub_theo_arg_1} As it stands, the perturbative
  series~\eqref{pert_gamma} is basically useless for physical
  dimension $y=1$, apart from the first couple of terms, and
  resummation methods are necessary for extracting precise values of anomalous
  dimensions.  The resummation is based on the Borel transform,
  followed by a conformal mapping, as will be explained later, and
  further discussed in App.~\ref{appendix:baby_integral}. This
  procedure gives resummed finite expressions $\wt{\g}_{\cal O}(y)$.
\item\label{item:pertub_theo_arg_2}
For dimensions close to $d=4$, i.e., $y\ll 1$, there is an optimal
number of terms $n_{\text{opt}}(y)$, for each $y$ value, for which the distance
between the series and the resummed function
$\wt{\g}_{{\cal O}}(y)$, 
$\vert \wt{\g}_{{\cal O}}(y)-\sum_1^{n_{\rm opt}}\ov{\g}_{{\cal O},k}y^k\vert$,
is minimal before growing again.
\end{enumerate}

The perturbative anomalous dimensions $\wt{\g}_{{\cal O}}$ may differ
from results obtained by other methods, such as the lattice
formulation of the path-integral for the Ising model, or by the
bootstrap. These differences are non-analytic, e.g.,
$\delta \g_{\cal O}(y)\sim\exp(-A/y)$. Within the resummation
procedure, these terms may change according to how the inverse Borel
transform is performed~\cite{Aniceto:2018bis}.
  
Before discussing the resummation methods in the next section, a first comparison of the perturbative expansion and the bootstrap data for $\g_\s$ clarifies the issues at stake.

The perturbative series is~\cite{panzer,Schnetz:2016fhy}
\ba
\g_\s(y)&=&0.00925926 y^2 + 0.00934499 y^3 - 0.00416439 y^4 +0.0128282 y^5
\nl
&&-0.0406363 y^6 +0.15738 y^7,
\qquad\qquad  \mbox{(epsilon expansion)}.
\label{g_s-ee}
\ea
The best polynomial fit of bootstrap data in
Tab.~\ref{tab:conf_dim_results} using the methods outlined in
Sec.~\ref{sec:3_lead_fields} is\footnote{Note that the best-fit
polynomial~\eqref{g_s-CB} starts with an $O(y^2)$ term, because
the linear term vanishes within errors. If a linear term is
included in the fit procedure, it leads to a coefficient three
orders of magnitude smaller than the quadratic term.
Therefore, the conformal bootstrap implies $\g_\s(y)=O(y^2)$ close to $d=4$, in agreement with perturbation theory.}
\ba
\g_\s(y) &=& 
0.009306473 y^2 + 0.008899908 y^3 -0.001435107 y^4 + 0.001788710 y^5\nl
&&-0.000533980 y^6 + 0.000128667 y^7,
\qquad \mbox{(conformal\ bootstrap)}.
\label{g_s-CB}
\ea
The two polynomials~\eqref{g_s-ee} and~\eqref{g_s-CB} have
different meanings, although their first two coefficients are  
close. On one hand the Feynman-diagram series is exact, but has a vanishing radius of convergence. On the other hand, the numerical bootstrap data in Tab.~\ref{tab:conf_dim_results} should converge to exact non-perturbative results upon increasing the numerical precision. The collection of these values for any dimension $d=4-y$ gives the exact function $\g^{\rm ex}_\s(y)$, which however cannot be expressed in terms of a simple polynomial. Therefore, the fit~\eqref{g_s-CB} gives approximated values around $\g^{\rm ex}_\s(y)$, whose precision is \emph {a priori} limited. Nonetheless, this description is sufficient at the present level of numerical accuracy.

\begin{figure}[!htb]
\centering
\includegraphics[scale=0.63]{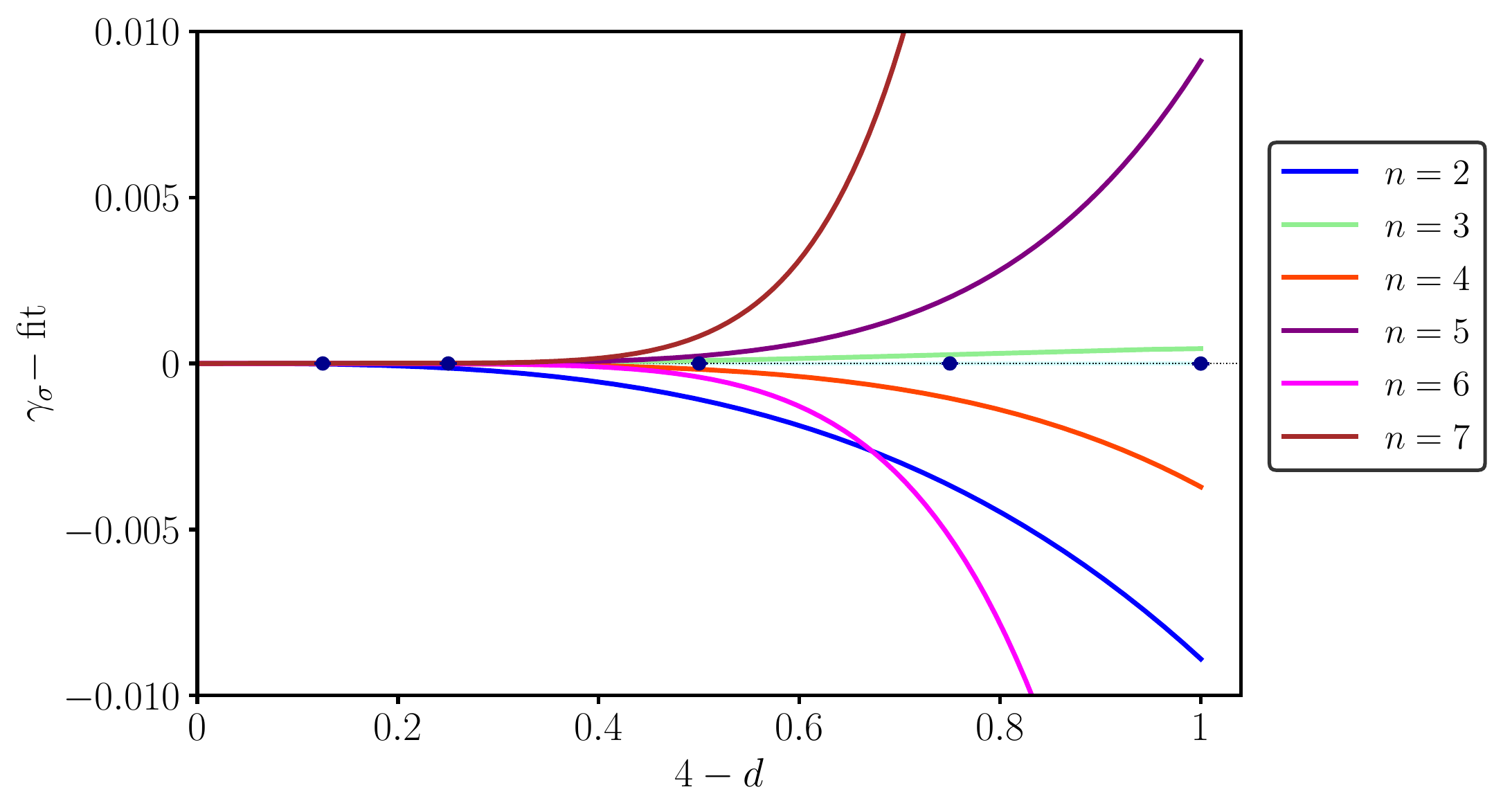} 
\caption{Comparison of $\g_\s$ bootstrap data with unresummed  epsilon expansion~\eqref{g_s-ee} in the region $4>d>3$ for 
truncations of the series to order $n=2,\dots,7$ (see color legend).
All quantities have been
subtracted by the best fit values (see~\eqref{g_s-CB}).}
\label{fig:s_ex1}
\end{figure}

In Fig.~\ref{fig:s_ex1} we show the difference between the
perturbative series~\eqref{g_s-ee} and the bootstrap fit~\eqref{g_s-CB} for $4> d\ge 3$. Color lines correspond to the series~\eqref{g_s-ee} truncated at different orders $n=2,3,\dots,7$ (cf.~color legend in the plot). One sees that, the higher the order
$n \ge 4$, the sooner the perturbative series diverges from the
bootstrap data (corresponding to the zero horizontal line in
Fig.~\ref{fig:s_ex1}). The tiny errors of bootstrap points cannot be
seen at this scale, thus showing that the unresummed perturbative
series cannot be used for a precise determination of critical
exponents in $d=3$,  as stated in point~\ref{item:pertub_theo_arg_2} above.
Yet, the lower terms $n=2,3$ may provide crude estimates.

\begin{figure}[!htb]
\centering
\includegraphics[scale=0.65]{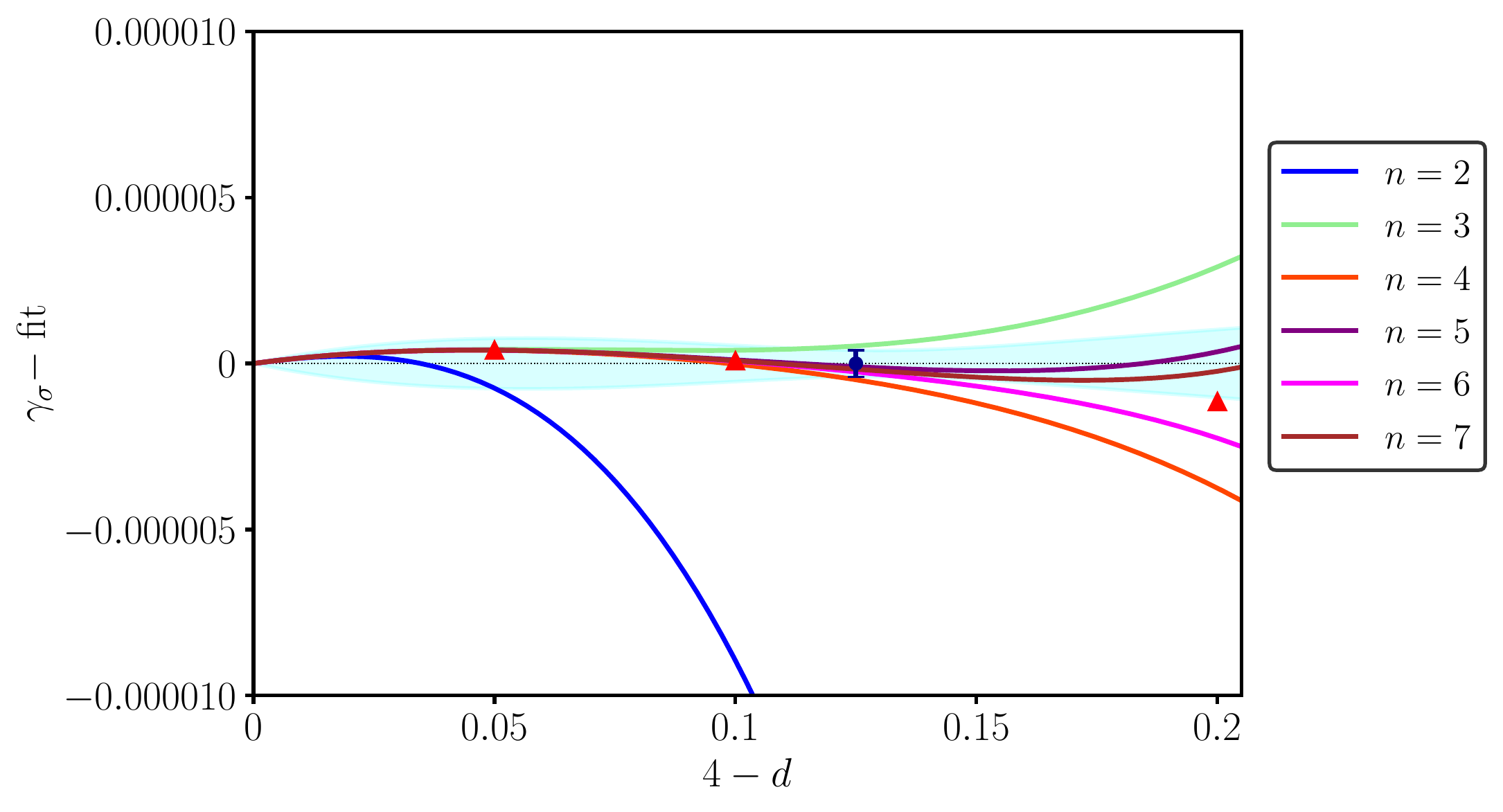} 
\caption{Comparison of $\g_\s$ data minus best fit in the region
$4>d>3.8$, between bootstrap (blue circle) and unresummed epsilon
expansion~\eqref{g_s-ee} with different truncations of the
perturbative series (cf.~Fig.~\ref{fig:s_ex1}). The red triangles
are the results of the bootstrap navigator
method~\cite{Henriksson:2022gpa}. The cyan shaded area is the fit
error.}
\label{fig:s_ex2}
\end{figure}

Fig.~\ref{fig:s_ex2} shows the other regime, close to four
dimensions. Only the bootstrap point for $d=3.875$ is present in this
range, but we also show results of Ref.~\cite{Henriksson:2022gpa} for
$d\ge 3.8$, which match very well while lacking error
bars, as discussed earlier\footnote{Note that the red triangles
  are not used in our fit of
bootstrap data.}. In contrast to the $d\approx 3$ region, we
observe that the truncated perturbative series shows a different
behavior. At any given $y$ value, upon increasing the perturbative
order up to an optimal value $n_{\rm opt}\sim 1/y$, the perturbative
series approaches the zero horizontal line (with a cyan error
band), before starting to diverge. Namely, it matches the exact bootstrap
value $\g_\s^{\rm ex}(y)$, within numerical errors.

Therefore, the comparison between
non-perturbative bootstrap results and unresummed epsilon expansion
for $\g_\s(y)$ is extremely good in the region $4>d>3.8$, with precision
${\rm Err}( \g_\s) \approx 1 \times 10^{-6}$, i.e.,
${\rm Err}( \g_\s )/\g_\s<1\times 10^{-3}$.
According to the previous discussion, we conclude that we do not see any
non-perturbative difference for $d\to 4$.

We remark that the epsilon expansion can also be obtained by
analytic solution of the bootstrap equations around $d=4$, assuming a
perturbative expansion near the free
theory~\cite{null-ref1,null-ref2,Henriksson:2022rnm,Bertucci:2022ptt,Bissi:2022mrs,4th-1,4th-2}. Thus,
is our comparison in Fig.~\ref{fig:s_ex2} tautological? It is not,
because the bootstrap identity is a set of consistency
conditions that depends on the kind of quantities they act on. Our
numerical solution does not assume any perturbative expansion,
i.e., it is an independent solution of the bootstrap constraints. That without any perturbative input, our conformal bootstrap results accurately reproduce
perturbative predictions close to $d=4$ is  non-trivial.

A natural question is how our numerical bootstrap approach can
reproduce the perturbative series, i.e., in which regime
the two polynomials~\eqref{g_s-ee} and~\eqref{g_s-CB} may agree
beyond the $O(y^3)$ term. As said earlier, the bootstrap
polynomial~\eqref{g_s-CB} is approximated, it can at most describe
a band of values around $\g_\s^{\rm ex}(y)$. While the size
${\rm Err(\g_\s)}$ of this band stays finite in the whole range
$0<y<1$ (see plots), that of the epsilon expansion is expanding in $y$
and can be finite only for $y<y_{\rm max}\sim O(1/n)$, $n$ being the
perturbative order. We expect that,
upon running the bootstrap for several points $y_i$, with
$0<y_i<y_{\rm max}\ll 1$, and by performing best fits with polynomials
limited to such a small interval, one may find that the two
expressions~\eqref{g_s-ee} and~\eqref{g_s-CB} match order
by order, i.e., the epsilon expansion is fully recovered.

\FloatBarrier

%--3.2--------------------------------------------- 
\subsection{Bootstrap data versus resummed perturbative results}\label{sec:Borel_transform}
\vspace{-\baselineskip}\indent

Precise estimates of the critical exponents have been obtained
over the years by refining the resummation techniques applied to the
epsilon expansion
series~\cite{zinn_ising,zinn_on,zinn-justin-book,mc-rev,panzer,serone,Mera:2018qte,Kompaniets:2019zes}.
In this work, we use the methods of Refs.~\cite{panzer, Kompaniets:2019zes} extended to dimension $4>d\ge 3$. Let us briefly recall the main steps involved~\cite{zinn-justin-book}.
The Borel transform ${\cal B}_{\g_{\cal O}} (t)$ of the perturbative expansion for the 
anomalous dimension $\g_{\cal O}$~\eqref{pert_gamma} is defined by removing the
factorial growth from the series,
\be
{\cal B}_{\g_{\cal O}} (t) =\sum_{k=1}^n
\frac{\ov{\g}_{{\cal O},k}}{k!}\; t^k \; .
\label{borel-g}
\ee
One infers from the asymptotic behavior~\eqref{asympt} that
this function has a singularity
${\cal B}_{\g_{\cal O}}(t)\sim (1+ta)^{-b-1}$ and a corresponding
finite radius of convergence.

The resummed quantity is defined by the inverse Borel transform,
\be
\wt{\g}_{\cal O}(y)=\int_0^\infty \rmd t\; e^{-t}\;
{\cal B}_{\g_{\cal O}} (y t).
\label{inv-borel}
\ee
By definition $\g_{\cal O}(y)$ in~\eqref{pert_gamma} and $\wt{\g}_{\cal O}(y)$ in~\eqref{inv-borel} have the same
perturbative expansion; however, the latter should be better
behaved if ${\cal B}_{\g_{\cal O}} (t)$ is suitably
continued analytically outside the original disc
$|t|<1/|a|$ to a region including the real positive axis\footnote{In particular, a real negative value of the parameter $a$ in~\eqref{asympt}, i.e., a perturbative series~\eqref{pert_gamma} of definite sign, is problematic.}.  
Such analytic continuation in principle requires the knowledge of all  
singularities of ${\cal B}_{\g_{\cal O}}(t)$ in the complex $t$-plane.
At this point, one can only make educated guesses
on these singularities, that
translate into (physical) ansatzes for $\wt{\g}_{\cal O}(y)$.

In practice, one assumes that the only singularity of
${\cal B}_{\g_{\cal O}}(t)$ lies at $t=-1/a$ real and negative, and
that it is a branch cut extending to $t=-\infty$. Using a conformal
mapping $t(z)$, this branch cut is mapped onto the unit
circle, with the start of the branch cut mapped onto $z=-1$, and $t=-\infty$ to $z=1$,
preserving the origin $z=t=0$. As long as there are no other
singularities, ${\cal B}(t(z))$ has a radius of convergence one in
$z$. As $t=\infty$ corresponds to $z=1$, this allows one to perform
the inverse Borel transform~\eqref{inv-borel}.
Details on this procedure can be
found in App.~\ref{appendix:baby_integral}.

This general idea can be improved in several ways, allowing one to
introduce a set of free parameters. The latter are determined
such that the final result is the least sensible to their variation.
Apart from providing a robust resummation scheme, the parameter uncertainty
implies an estimate of the resummation error. These methods have
been improved over the years by taking into account the phenomenology
of critical phenomena \cite{zinn-justin-book}.
In our work, the resummed data are obtained by extending the
setup of Refs.~\cite{Kompaniets:2016lmy,panzer,Kompaniets:2019zes} from $d=3$
to non-integer dimensions.
A complete account of these methods is too long
to be presented here; nonetheless, we provide some introductory
material that will allow the reader to assess the
original works. In App.~\ref{appendix:baby_integral}, the resummation
is worked out in a toy model, where one can compare it with the exact result.
In App.~\ref{app:KP17_details}, instead,
a ``reader's guide'' to Ref.~\cite{panzer} is presented, together with
the values of the resummation parameters used here.

Let us also mention that another option for the analytic
continuation is to use Hypergeometric functions, for which the inverse
Borel transform can be written as a Meijer-G
function~\cite{Mera:2018qte}. One drawback of this approach is the
possibility for spurious poles on the integration contour.
As here we could not give justice to their influence, we exclude this resummation method.

\begin{figure}[!htb]
\centering
\includegraphics[scale=0.65]{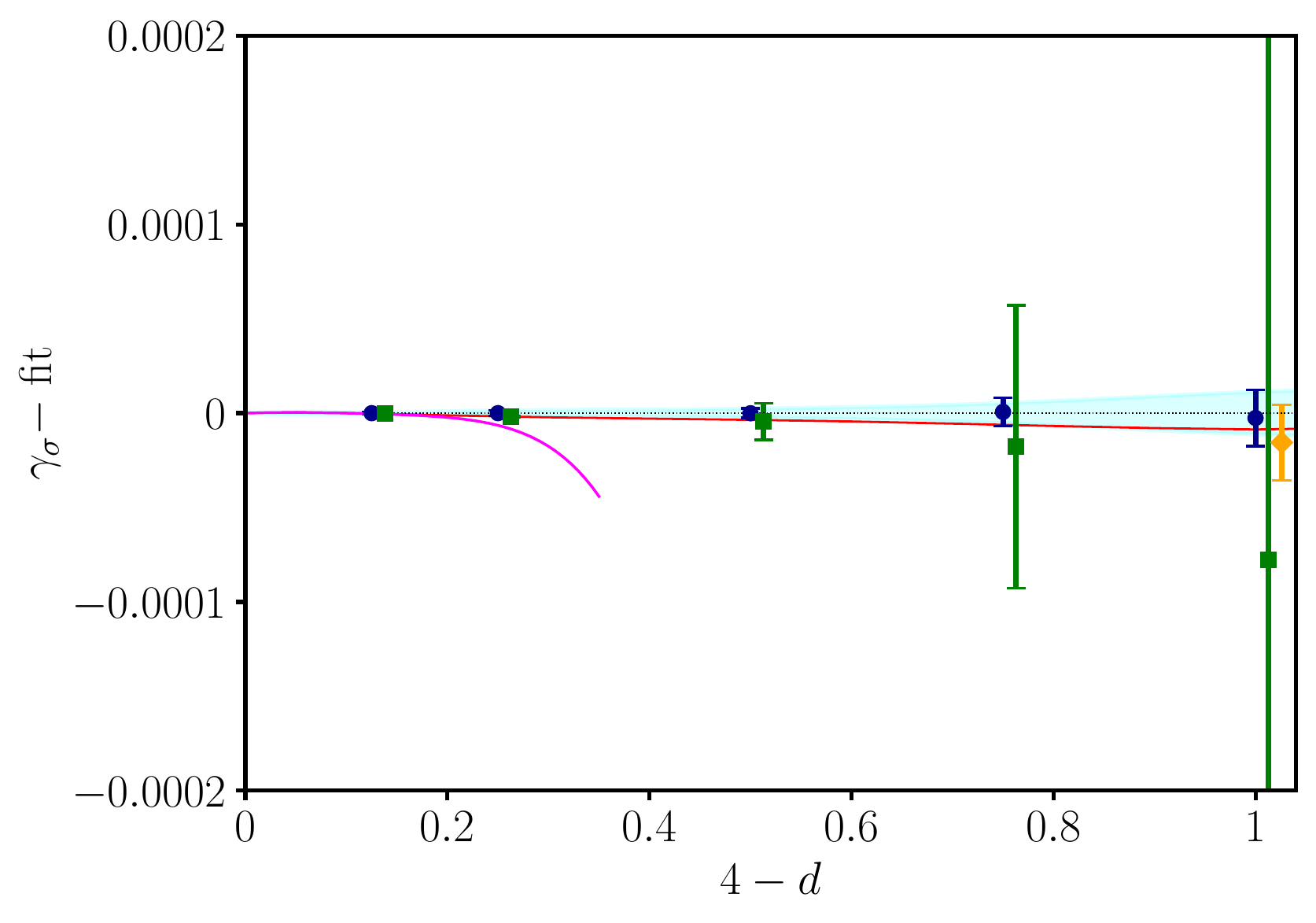}
\caption{Comparison of $\g_\s$ data minus best-fit values:
bootstrap (blue circles), Borel-resummed epsilon expansion~\cite{panzer}
(green squares), unresummed high-order epsilon expansion (magenta
solid curve), $d=3$ Monte Carlo~\cite{mc-hasen-new} (yellow rhombus). We also plot a solid red line linearly interpolating results of Ref.~\cite{Henriksson:2022gpa} for $4>d\ge3$. Note that data points are slightly displaced around the same $d$ values ($d=3.875$, $d=3.75$, $d=3.5$, $d=3.25$ and $d=3$) to improve readability. Results from earlier work~\cite{zinn_on} have been omitted due to their large error bars.}
\label{fig:s-exp}
\end{figure}

Figure~\ref{fig:s-exp} shows the fitted bootstrap data (blue points)
of $\g_\s(y)$ already reported in Fig.~\ref{fig:henriksson_comp}, now
compared to the resummed epsilon-expansion values of
Tab.~\ref{tab:resummed_KP17} (green
squares)\footnote{Resummations in this section use the
    $6$-loop  results, that were verified in several independent
    works~\cite{panzer,Kompaniets:2019zes,Schnetz:2016fhy}. We do not
    use the $7$-loop results of Ref.~\cite{Schnetz:2016fhy}, since they
    were not yet checked independently. Past experience, e.g., with the $5$-loop
    results, teaches us that involved perturbative calculations require
    confirmation.}. The
agreement between these two results is very good, especially for
$d\ge 3.5$, where the unresummed series (magenta line) is already
diverging, and greatly improves on earlier
studies~\cite{zinn_ising,zinn_on} analyzed in~\cite{CMO}. Let us
remark that resummed $\wt{\g}_\s(y)$ values have been obtained for
non-integer dimensions down to $d=2$, still finding agreement with
bootstrap data, although with larger uncertainties. Finally,
Fig.~\ref{fig:s-exp} shows the latest Monte Carlo results in $d=3$
(yellow rhombus), that match extremely well the bootstrap
points. Further $d=3$ results by these and other methods are
summarized in a later figure.  Finally, Fig.~\ref{fig:s-exp} and later
plots for the dimensions $\g_\eps$ and $ \g_{\eps'}$ also report a
solid red curve linearly interpolating the navigator points of
Ref.~\cite{Henriksson:2022gpa} obtained for $4>d\ge3$. This allows one
to assess the negligible difference between the two sets of bootstrap
data in the comparison to the epsilon-expansion.

\begin{table}[!htb]
\centering
\begin{center}
\begin{tabular}{|c|l|l|l|}
\hline
&&&\\[-1em]
$d$ & $\D_\s$ & $\D_\e$ & $\D_\e'$ \\
\hline
&&&\\[-1em]
3.875 & 0.937662197(7) & 1.91831086(14) & 3.9924550(11) \\
3.75  & 0.8757158(3)   & 1.839419(4)    & 3.97529(3) \\
3.5   & 0.753393(10)   & 1.68854(7)     & 3.9276(5) \\
3.25  & 0.63386(8)     & 1.5458(4)      & 3.873(2) \\
3     & 0.5181(3)      & 1.4108(12)     & 3.820(7) \\
\hline
\end{tabular}
\end{center}
\caption{Conformal dimensions of $\s$,$\e$ and $\e'$ field from resummed perturbative expansion, obtained according to the methods of~\cite{panzer}.}
\label{tab:resummed_KP17}
\end{table}

\FloatBarrier

We now extend the previous analysis to the energy 
field $\e$. The best fit of the conformal bootstrap data is
\ba\label{g_e-CB}
\g_\e(y) &=& 0.333441601 y + 0.114095325 y^2 -0.083458310 y^3 \nl
&& + 0.081381007 y^4 -0.045296977 y^5 + 0.014290102 y^6\nl
&& -0.001741325 y^7, \qquad\qquad\qquad\qquad \mbox{(conformal\ bootstrap)}.
\ea
The epsilon-expansion series reads~\cite{panzer,Schnetz:2016fhy}
\ba\label{g_e-ee}
\g_\e(y)
&=&0.333333 y + 0.117284 y^2 - 0.124527 y^3 + 0.30685 y^4 - 0.95124 y^5
\nl && 
+3.57258 y^6 -15.2869 y^7, \qquad\qquad\qquad \mbox{(epsilon expansion)}.
\ea
One remarks the agreement, within errors, of the first two
coefficients of this series; this corrects less precise
results of~\cite{CMO} (cf.~Fig.~6\textit{b} there).

\begin{figure}[!htb]
\centering
\includegraphics[scale=0.65]{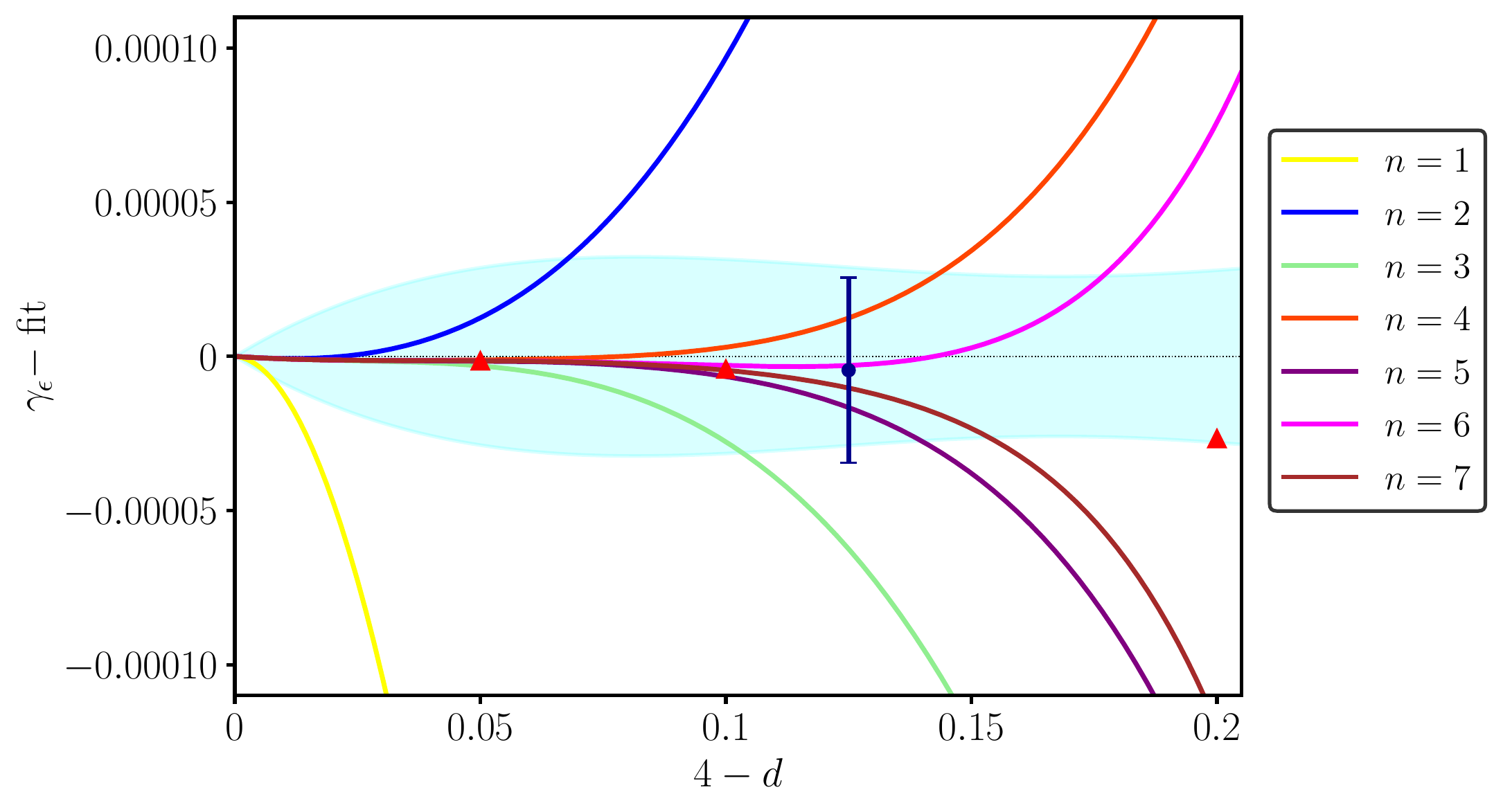}
\caption{Comparison of the $\g_\e$ data minus the best fit in the region $4>d>3.8$.
Our bootstrap point is the blue circle with error bar; the triangles are obtained by the navigator method~\cite{Henriksson:2022gpa}; the different
truncations of the perturbative series are as in Fig.~\ref{fig:s_ex1}.
The cyan shaded area is the fit error.}
\label{fig:e-unres}
\end{figure}

The comparison for $d\to 4$ before resummation is shown in
Fig.~\ref{fig:e-unres}. As for Fig.~\ref{fig:s-exp}, the
truncated perturbative series for $\g_\e$ are plotted. Their curves
approach the bootstrap fit (horizontal zero axis with cyan error band)
with better and better precision. Note the remarkable quality of
the navigator method (red triangles)\cite{Henriksson:2022gpa}.
Altogether, the agreement for $d\to 4$ is found with high precision,
${\rm Err}(\g_\e)=3 \times 10^{-5}$ and
${\rm Err}(\g_\e)/\g_\e= 1 \times 10^{-3}$.

\begin{figure}[!htb]
\centering
\includegraphics[scale=0.67]{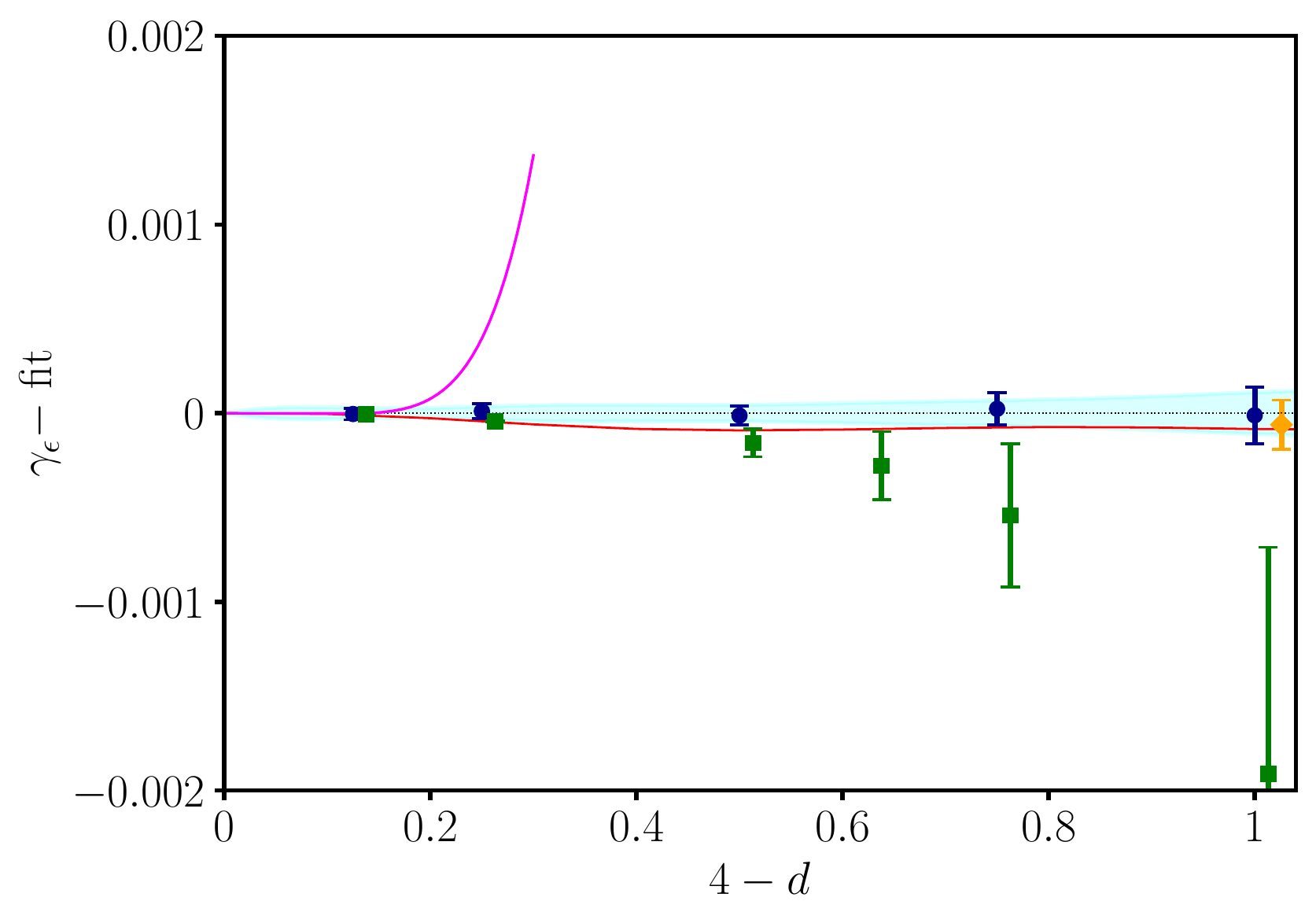}
\caption{Comparison of $\g_\e$ data minus best fit: bootstrap (blue
  circles), Borel-resummed epsilon expansion~\cite{panzer} (green
  squares), unresummed epsilon expansion (magenta solid curve), $d=3$
  Monte Carlo~\cite{mc-hasen-new} (yellow rhombus). We also plot a
  solid red line linearly interpolating results of
  Ref.~\cite{Henriksson:2022gpa} for $4>d\ge3$. The cyan shaded area
  is the fit error as in earlier plots.}
\label{fig:e-eps}
\end{figure}

\begin{figure}
\centering
\includegraphics[scale=0.67]{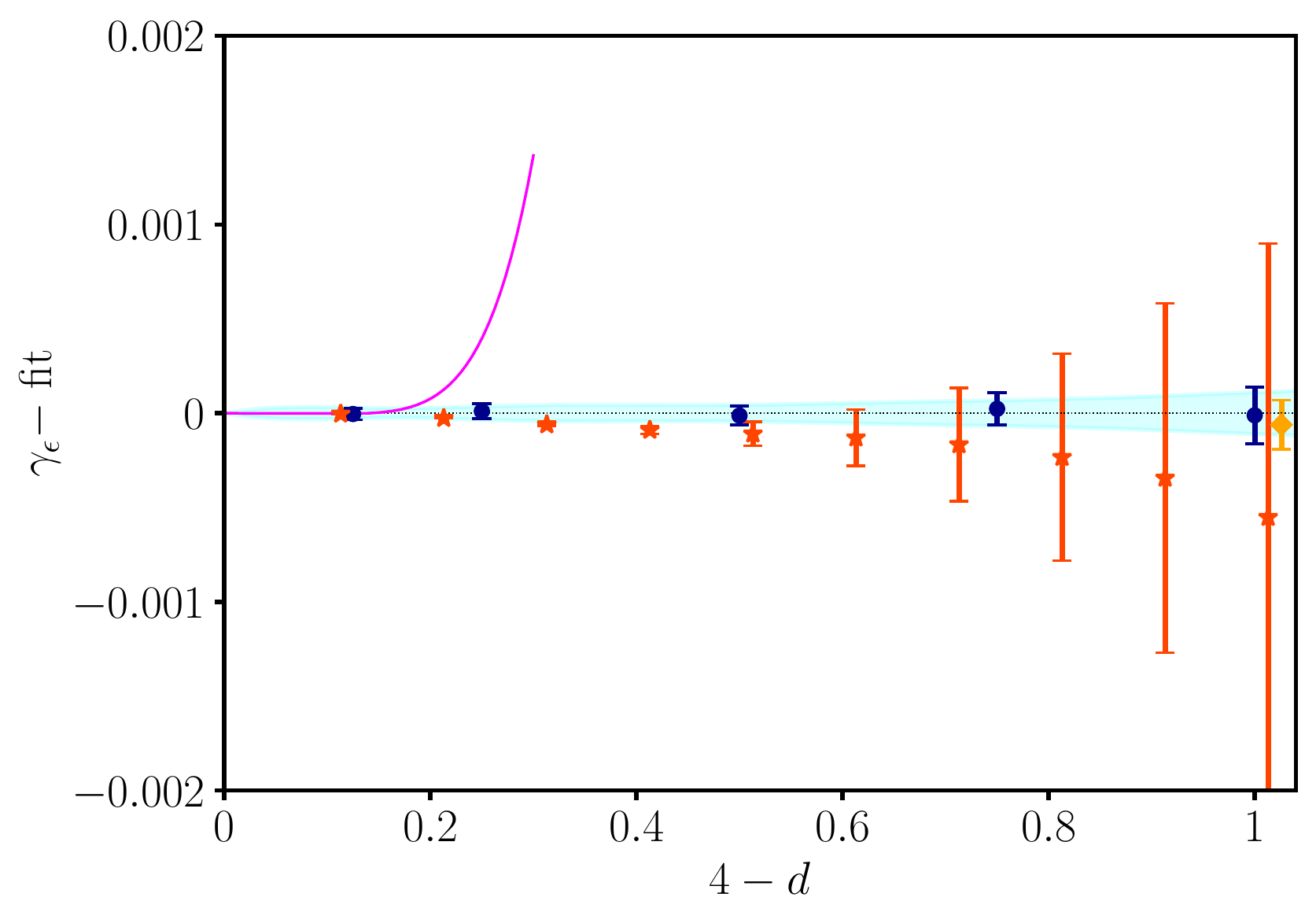}
\caption{Comparison of $\g_\e$ minus best fit: bootstrap (blue
  circles), Self-Consistent resummed epsilon
  expansion~\cite{Kompaniets:2019zes} (red stars), unresummed epsilon
  expansion (magenta solid curve), $d=3$ Monte
  Carlo~\cite{mc-hasen-new} (yellow rhombus).}
\label{fig:e-eps-SC}
\end{figure}

Figure~\ref{fig:e-eps} presents a comparison with the resummed
perturbative series (Tab.~\ref{tab:resummed_KP17}): the agreement is again very good for
$4>d\ge3.5$; there is a small $O(10^{-3})$ deviation 
from the bootstrap and Monte Carlo results~\cite{mc-hasen-new} (yellow rhombus) in $d=3$.  Probably there is a slight underestimation of the error. Let us remark
that this resummation procedure is {\em honest}, as it does not use the exact $d=2$ conformal dimension as an input, with which it could be improved. The comparison with another method, called Self-Consistent
(SC) resummation\footnote{See Ref.~\cite{Kompaniets:2019zes} for a
detailed discussion of this approach.} is presented in Fig.~\ref{fig:e-eps-SC}, where we plot data of Tab.~\ref{tab:resummed_KW19}. In this
case, the Borel transform is done on the perturbative series of $1/\nu^3$, instead of $1/\nu=2-\g_e$: this choice is
motivated by a match with the $d=2$ conformal field theory, that is achieved through comparing the $n$ dependence of the $O(n)$-symmetric $\phi^4$ theory~\cite{Kompaniets:2019zes}. We
conclude that adding information of the exact results in $d=2$
improves the resummation of the perturbative series (for this particular critical exponent). A similar constraint does not seem to be possible for the other critical exponents, as discussed in Ref.~\cite{Kompaniets:2019zes}.

\begin{table}[!htb]
\centering
\begin{center}
\begin{tabular}{|c|l|}
\hline
&\\[-1em]
$d$ & $\D_\e$ \\
\hline 
&\\[-1em]
3.9 & 1.93440534057(12) \\
3.8 & 1.8706742(6) \\
3.7 & 1.808546(5) \\
3.6 & 1.747876(2) \\
3.5 & 1.68858(6) \\
3.4 & 1.63062(15) \\
3.3 & 1.5740(3) \\
3.2 & 1.5187(5) \\
3.1 & 1.4647(9) \\
3   & 1.4122(15) \\
\hline
\end{tabular}
\end{center}
\caption{Conformal dimension of $\e$ field from resummed perturbative expansion, obtained according to the methods of~\cite{Kompaniets:2019zes}.}
\label{tab:resummed_KW19}
\end{table}

Summarizing, the bootstrap and epsilon-expansion results agree very
well: for $d\to 4$ the unresummed series fits perfectly, for
$4>d \ge3$ there is   remarkable agreement, keeping in mind that
the resummation error is roughly one order of magnitude larger than
that of bootstrap and Monte Carlo results.

\FloatBarrier

\begin{figure}[!htb]
\centering
\includegraphics[scale=0.67]{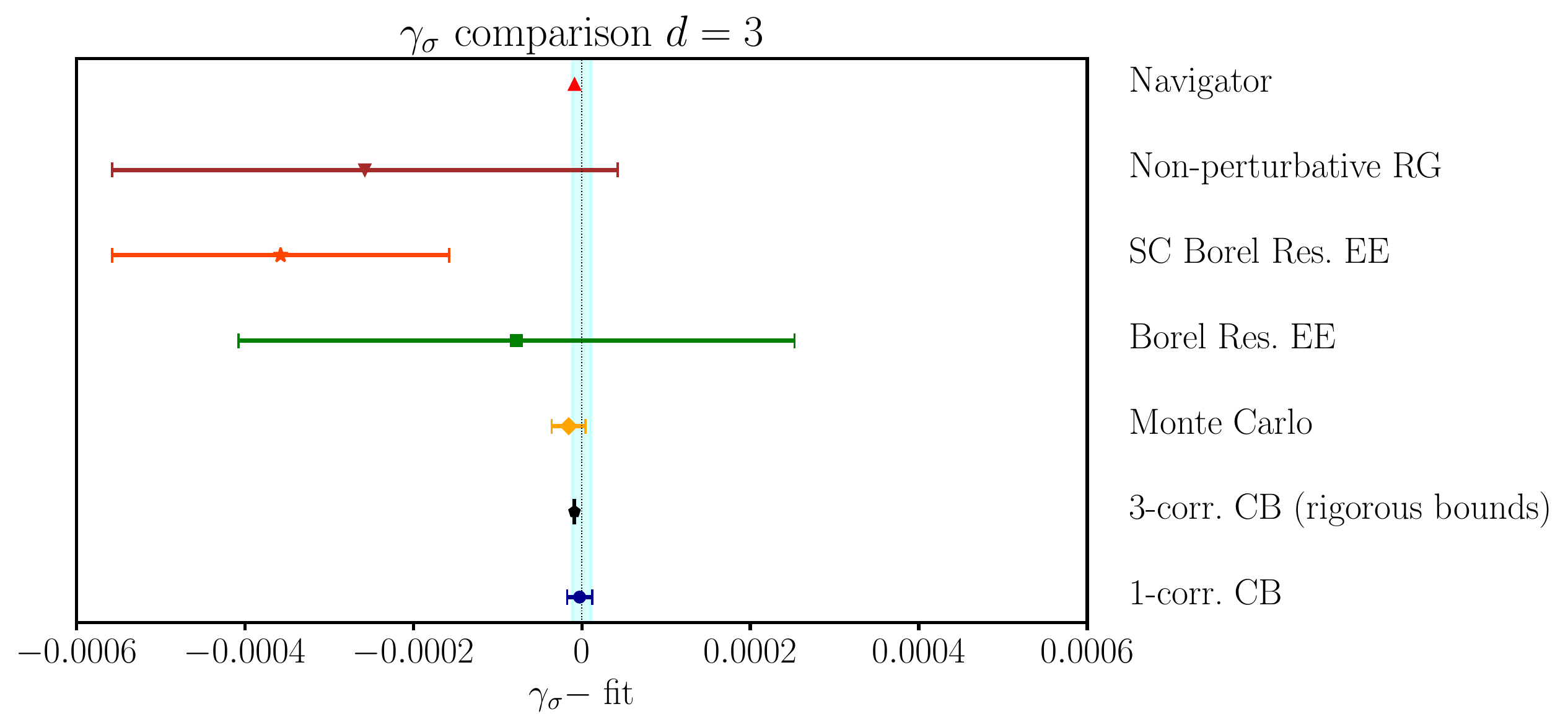}
\caption{Summary of up-to-date predictions for $\g_\s$ at $d=3$ (minus best fit): $1$-correlator bootstrap~\cite{CMO} (blue circle), $3$-correlator bootstrap with rigorous bounds~\cite{sd} (black pentagon), Monte Carlo~\cite{mc-hasen-new} (yellow rhombus), Borel-resummed epsilon expansion~\cite{panzer} (green square), Self-Consistent resummed epsilon expansion~\cite{Kompaniets:2019zes} (red star), non-perturbative renormalization group~\cite{Balog:2019rrg} (brown downward triangle), bootstrap navigator method~\cite{Henriksson:2022gpa} (red upward triangle).}
\label{fig:s-lit}
\end{figure}

\begin{figure}[!htb]
\centering
\includegraphics[scale=0.67]{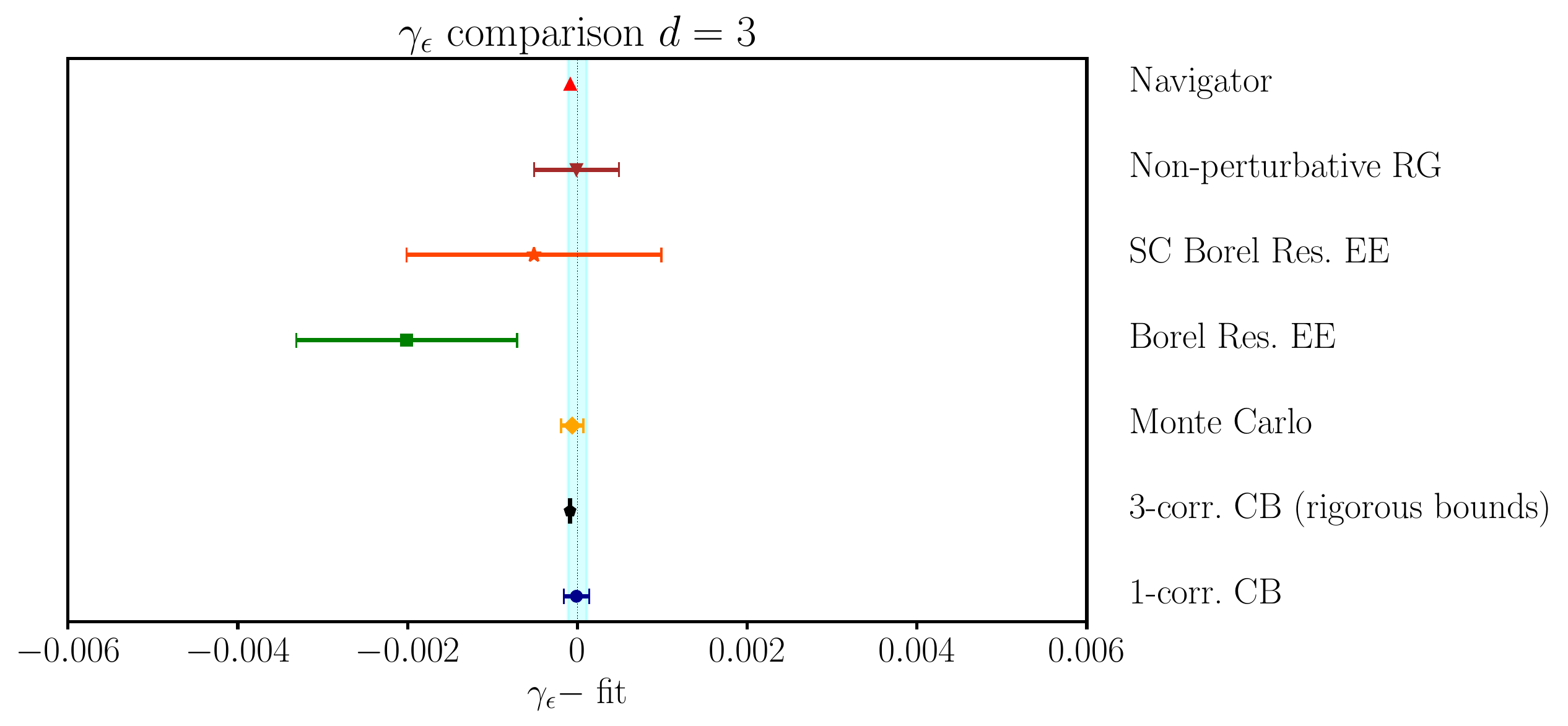}
\caption{Summary of up-to-date predictions for $\g_\e$ in $d=3$ (minus best fit): $1$-correlator bootstrap~\cite{CMO} (blue circle), $3$-correlator bootstrap with rigorous bounds~\cite{sd} (black pentagon), Monte Carlo~\cite{mc-hasen-new} (yellow rhombus), Borel-resummed epsilon expansion~\cite{panzer} (green square), Self-Consistent resummed epsilon expansion~\cite{Kompaniets:2019zes} (red star), non-perturbative renormalization group~\cite{Balog:2019rrg} (brown downward triangle), bootstrap navigator method~\cite{Henriksson:2022gpa} (red upward triangle).}
\label{fig:e-lit}
\end{figure}

A comparison of all $d=3$ results available in the literature for
$\g_\s$ and $\g_\e$ is given in Figs.~\ref{fig:s-lit}
and~\ref{fig:e-lit}. The corresponding numerical values are in
Tab.~\ref{tab:d=3_comp}. Besides data already discussed (drawn in
earlier colors), we report recent results of the non-perturbative
renormalization group~\cite{Balog:2019rrg} (brown downward
triangle). The central value is given by our fit of the bootstrap data
with error given by the cyan band, not by the mean
value of all results. The Figs.~\ref{fig:s-lit} and~\ref{fig:e-lit}
respect our convention of plotting the two anomalous dimensions on
scales differing by one order of magnitude, roughly equal to the ratio
of their actual value. Finally, Tab.~\ref{tab:d=3_comp} and Figs.~\ref{fig:s-lit},~\ref{fig:e-lit} report also the results of other $3$-correlator bootstrap approaches, using EFM~\cite{sd} and the navigator method~\cite{Reehorst:2021hmp}, and paying particular attention to error estimates (cf.~rigorous bounds). We also remark that the results obtained by perturbative expansions directly in $d=3$~\cite{mc-rev,zinn_on} are consistent with bootstrap results too, but have one order of magnitude larger errors and are therefore not plotted in Figs.~\ref{fig:s-lit} and~\ref{fig:e-lit}.

\begin{table}[!htb]
\centering
\be 
\begin{array}{|l|l|l|l|}  
\hline   
d=3 \ {\rm Ising\ critical\ indices}              &\D_\s          &\D_\e         & \D_{\e'} \\
\hline
\mbox{\rm Bootstrap\ (1 correlator)}              & 0.518155(15)  & 1.41270(15)  & 3.8305(15) \\
\mbox{\rm Bootstrap\ (3 correlators)}             & 0.5181489(10) & 1.412625(10) & 3.8297(2) \\
\mbox{\rm Borel\ resummed\ epsilon expansion}     & 0.5181(3)     & 1.4107(13)   & 3.820(7) \\
\mbox{\rm SC\ Borel\ resummed\ epsilon expansion} & 0.5178(2)     & 1.4122(15)   & 3.827(13) \\
\mbox{\rm Monte\ Carlo}                           & 0.51814(2)    & 1.41265(13)  & 3.832(6) \\
\mbox{\rm Non-perturbative RG}                    & 0.5179(3)     & 1.41270(50)  & 3.832(14) \\
\mbox{\rm Navigator\ (rigorous\ bounds)}          & 0.518157(35)  & 1.41265(36)  & 3.8295(6) \\
\hline
\end{array}
\nonumber
\ee
\caption{Comparison of $d=3$ results for the conformal dimensions of low-lying fields: $1$-correlator bootstrap~\cite{CMO}, $3$-correlator bootstrap~\cite{sd} (errors on $\D_\s$ and $\D_\e$ are rigorous bounds), Borel-resummed epsilon expansion~\cite{panzer}, Self-Consistent (SC) Borel-resummed epsilon expansion~\cite{Kompaniets:2019zes}, Monte Carlo~\cite{mc-hasen-old,mc-hasen-new}, non-perturbative renormalization group~\cite{Balog:2019rrg,DupuisCanetEichhornMetznerPawlowskiTissierWschebor2021} and bootstrap navigator method with rigorous bounds~\cite{Reehorst:2021hmp}.}
\label{tab:d=3_comp}
\end{table}

We now analyze the subleading $\Z_2$-even scalar field $\e'$, which is
related to the critical exponent
$\w=\D_{\e'}-d = d-4 + \g_{\e'}$. The best fit of our data
gives\footnote{The fit again assumes $\g_{\e'}=0$ for $d=4$.}:
\ba
\g_{\e'}(y) &=& 2.000178549 y -0.518006835 y^2 + 0.721996645 y^3\nl
&& -0.684437170 y^4 + 0.447648598 y^5 -0.162903635 y^6\nl
&& + 0.026155257 y^7, \qquad\qquad\qquad\mbox{(conformal\ bootstrap)}.
\label{ep-fit}
\ea
The large errors of the earlier analysis~\cite{CMO} have been
reduced, as explained earlier (see Fig.~\ref{fig:old_vs_new_comp}).
The epsilon-expansion series is~\cite{panzer,Schnetz:2016fhy},
\ba
\g_{\e'}(y) &=& 2y -0.62963 y^2 + 1.61822 y^3 - 5.23514 y^4 + 20.7498 y^5
\nl
&&-93.1113 y^6 + 458.7424 y^7,
\qquad\qquad \mbox{(epsilon expansion)}.
\ea

\begin{figure}[!htb]
\centering
\includegraphics[scale=0.65]{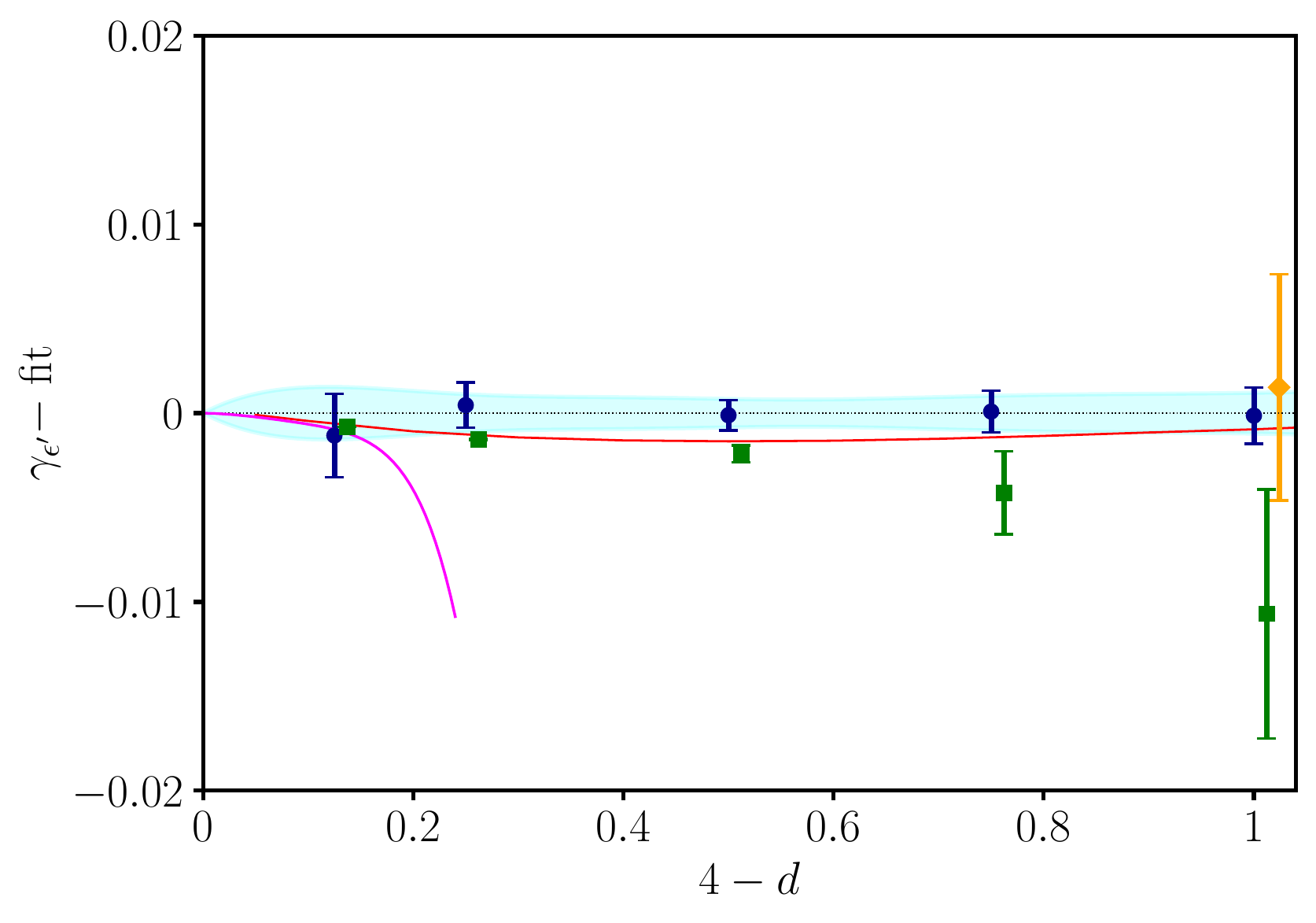}
\caption{Comparison of $\g_{\e'}$ data minus best fit: bootstrap
(blue circles), Borel-resummed epsilon expansion~\cite{panzer}
(green squares), unresummed epsilon expansion (magenta solid curve), $d=3$ Monte Carlo~\cite{mc-hasen-old} (yellow rhombus). We also plot a solid red line linearly interpolating results of Ref.~\cite{Henriksson:2022gpa} for $4>d\ge3$.}
\label{fig:ep_exp}
\end{figure}

\begin{figure}[!htb]
\centering
\includegraphics[scale=0.65]{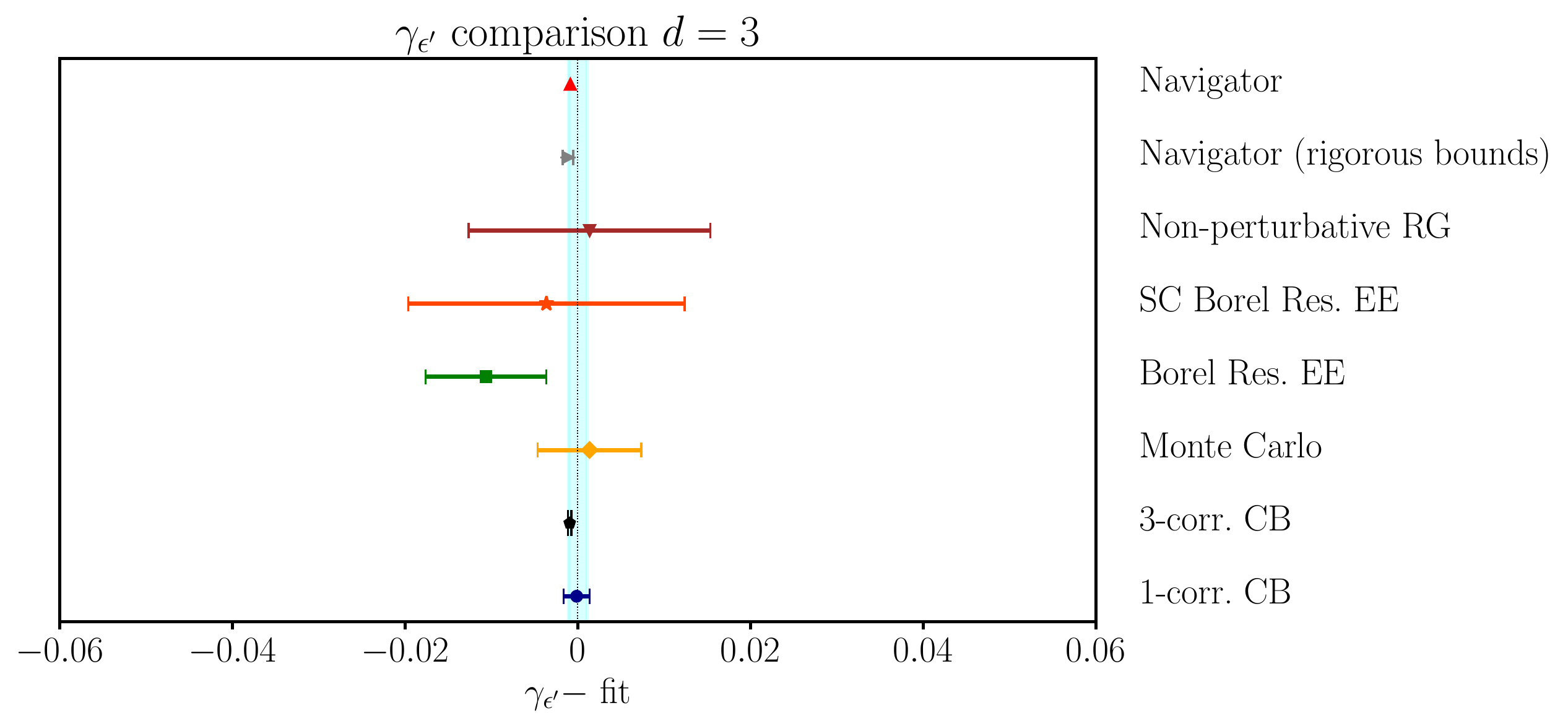}
\caption{Summary of up-to-date predictions for $\g_{\e'}$ in $d=3$ (minus our best fit, from bottom to top): $1$-correlator bootstrap~\cite{CMO} (blue circle), $3$-correlator bootstrap~\cite{sd} (black pentagon), Monte Carlo~\cite{mc-hasen-new} (yellow rhombus), Borel-resummed epsilon expansion~\cite{panzer} (green square), Self-Consistent resummed epsilon expansion~\cite{Kompaniets:2019zes} (red star), non-perturbative renormalization group~\cite{Balog:2019rrg} (brown downward triangle), bootstrap navigator method with rigorous bounds~\cite{Reehorst:2021hmp} (grey rightward triangle), bootstrap navigator method~\cite{Henriksson:2022gpa} (red upward triangles).}
\label{fig:ep-lit}
\end{figure}

\begin{figure}[!htb]
\centering
\includegraphics[scale=0.65]{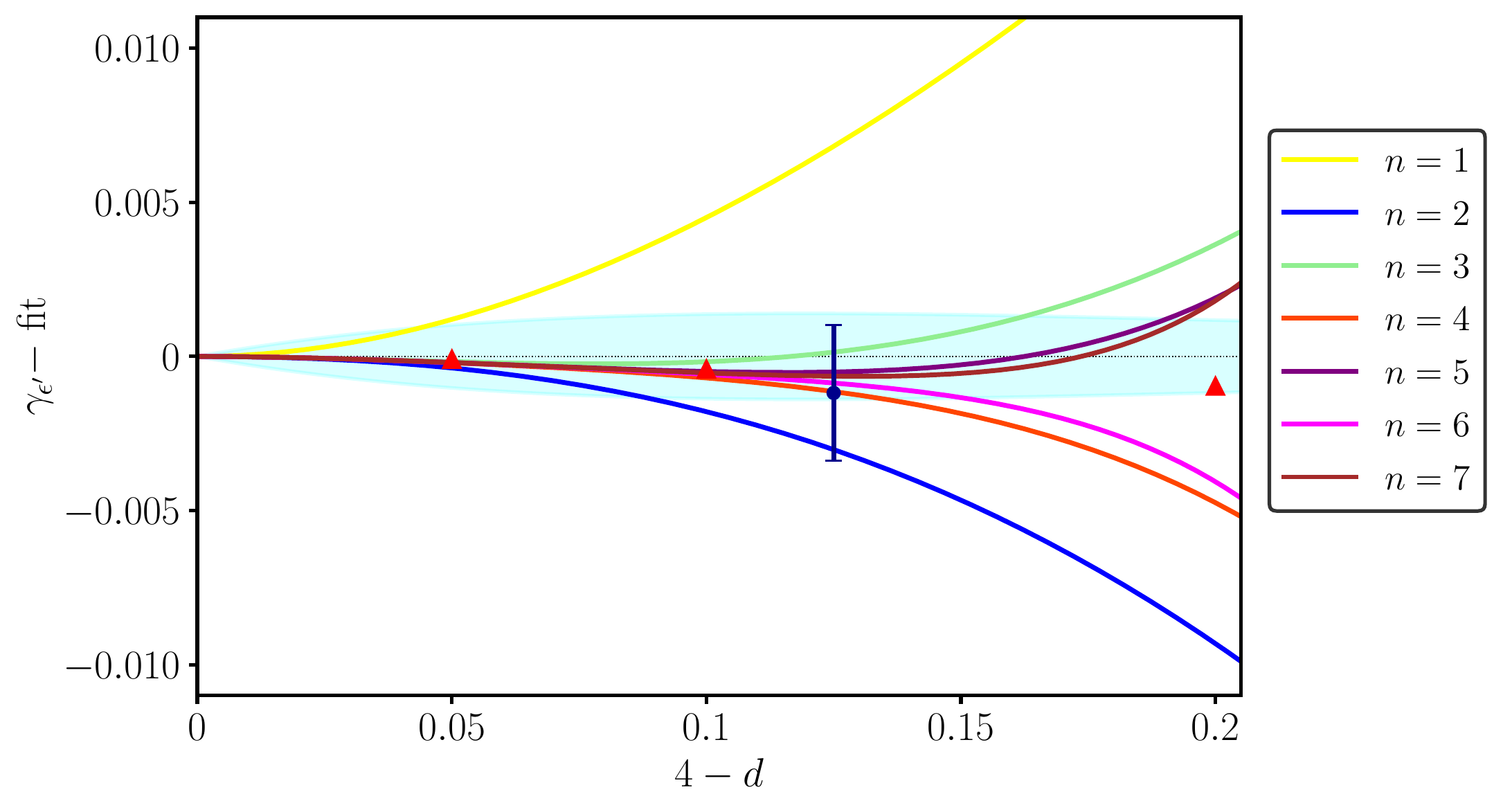}
\caption{Comparison of the $\g_{\e'}$ data minus the best fit in the region $4>d>3.8$.
Our bootstrap point is the blue circle with error bar; the triangles are obtained by the navigator method~\cite{Henriksson:2022gpa}; the different truncations of the perturbative series are as in Fig.~\ref{fig:s_ex1}. The cyan shaded area is the fit error.}
\label{fig:ep-unres}
\end{figure}

In Fig.~\ref{fig:ep_exp} we show the difference between the data and
the bootstrap best fit~\eqref{ep-fit}. The overall error of the fit
for $\g_{\e'}$ is estimated to be less than $2.0 \times 10^{-3}$ in
the whole range. The relative error is
${\rm Err}(\g_{\e'})/\g_{\e'}=1 \times 10^{-3}$ for $d=3$ but goes up
to\footnote{The growth of the
error when passing from $d=3.75$ to $d=3.875$ is due to the
instability of the higher part of the spectrum when approaching
$d=4$. This issue is further discussed in Sec.~\ref{high_spectrum}.}
$1 \times 10^{-2} $ for $d=3.875$. The
comparison with Monte Carlo~\cite{mc-hasen-old,mc-hasen-new} in $d=3$,
and the resummed epsilon-expansion series are also shown, finding
again good agreement at the coarser scale (note a factor of $10$
w.r.t.~Fig.~\ref{fig:e-eps}).
A systematic difference between
  bootstrap and epsilon-expansion points is seen for $d \to 3$,
  similar to what was   found for $\g_\e$ in Fig.~\ref{fig:e-eps}.
  Such a drift is smaller for the navigator
  results~\cite{Henriksson:2022gpa} (red line) than for our data,
  for $4 >d \ge 3.5$. Further values of $\D_{\e'}$ in $d=3$ found in
the literature are reported in Tab.~\ref{tab:d=3_comp} and plotted in
Fig.~\ref{fig:ep-lit}. A zoom over the region close to $d=4$ is drawn
in Fig.~\ref{fig:ep-unres}, showing the same features as in
Figs.~\ref{fig:s_ex2} and~\ref{fig:e-unres}.

We conclude this section by stressing the very good overall agreement
of bootstrap and resummed epsilon expansion. The study in
varying dimensions clarifies the different behavior of quantities in
the perturbative and non-perturbative regimes.

\FloatBarrier
\clearpage

%-4--------------------------------------------- 
\section{Structure constants and scaling dimensions of higher fields}\label{sec:higher_fields}
\vspace{-\baselineskip}\indent

In this section we analyze further bootstrap data. The structure
constants (OPE coefficients) of low-lying fields $\s,\e,\e',T$ are
very precise, the error being on the fifth decimal, thus better than
those of the corresponding conformal dimensions presented
earlier. Next we discuss subleading and spinful fields,
$\e'', T', C, C'$, presenting results for both dimensions and
structure constants. Some of them are good, others are not completely
correct, showing the limits of our numerical bootstrap approach.

%-4.1--------------------------------------------- 
\subsection{Structure constants in $4>d \ge 3$}\label{sec:higher_op_const}
\vspace{-\baselineskip}\indent

\begin{table}[!b]
\centering
\be 
\resizebox{1\hsize}{!}{$
\begin{array}{|c|l|l|l|l|l|l|l|}
\hline
d & c  & f_{\s\s \e}  & f_{\s\s\e'}& f_{\s\s\e''} \times 10^4 & f_{\sigma\sigma T'} & f_{\sigma\sigma C}  & f_{\sigma\sigma C'} \\
\hline
\mathbf{4} & \mathbf{1} & \mathbf{ 1.4142136} & \mathbf{0} & \mathbf{0} & \mathbf{0} & \mathbf{0.169031} & \mathbf{0} \\
\hline
3.875 & 0.99970(2) & 1.38228(2) & 0.015298(14) & 0.33(10) & 0.003070(2) & 0.1540603(3) & 0.000772(2) \\
3.75  & 0.998594(3) & 1.34586(3) & 0.027517(15) & 1.4(3) & 0.005641(5) & 0.133(8) & 0.00134(10) \\ 
3.5 & 0.9922615(15) & 1.26132(3) & 0.04426(3) & 4.0(2) & 0.00911(10) & 0.105(5) & 0.0021(3) \\
3.25 & 0.976864(6) & 1.16282(4) & 0.05225(3) & 6.0(3) & 0.0106(2) & 0.084(6) & 0.0019(9) \\
\hline\hline
3 &  0.946535(15) & 1.05184(4) &  0.05300(5) & 7.1(4) &  0.010575(15) & 0.065(5)&  0.0020(5) \\
2.75 & 0.893275(15) & 0.92939(4) & 0.04794(8) & 7.0(4) & 0.00901(6) & 0.048(4) & 0.00235(15) \\ 
2.5 & 0.807110(10) & 0.796303(5) & 0.03885(2) & 5.90(9) & 0.00668(3) & 0.033(3) & 0.0029(3) \\
2.25 & 0.677724(2) & 0.65311(2) & 0.02738(4) & 4.27(5) & 0.00394(14) & 0.0195(15) & 0.0035(2) \\
2.2 & 0.64609(7) & 0.62333(6) & 0.0245(5) & 3.76(9) & 0.00352(7) & 0.019(4) & 0.0038(3) \\
2.15 & 0.61243(8) & 0.59313(8) & 0.0225(5) & 3.36(2) & 0.0025(5) & 0.017(3) & 0.00385(15) \\
2.1 & 0.57680(10) & 0.56249(7) & 0.02018(8) & 2.98(7) & 0.00265(5) & 0.016(3) & 0.00395(15)\\
2.05 & 0.53935(15) & 0.53143(8) & 0.01785(5) & 2.58(4) & 0.00230(10) & 0.0135(25) & 0.00390(10)\\
2.01 & 0.5082(3) & 0.5058(6) & 0.01605(5) & 2.246(9)& 0.00193(3) & 0.01550(10) & 0.003920(10)\\
2.00001 & 0.500015(15) & 0.499998(5) & 0.015623(4)& 2.0(2) & 0.0018520(5) & 0.0148235(15) & 0.0039040(10) \\
\hline
\mathbf{2} & \mathbf{0.5} & \mathbf{0.5} & \mathbf{0.0156250} & \mathbf{2.1972656} & \mathbf{0.00185290} & \mathbf{0.0148232} & \mathbf{0.003906} \\
\hline
\end{array}
\nonumber
$}
\ee
\caption{Structure constants of the first few low-lying states for
$4>d>2$. The exact values for $d=2,4$ are given in bold, results for
$3 \ge d > 2$ are taken from~\cite{CMO}.}
\label{tab:structure_const_results}
\end{table}

Tab.~\ref{tab:structure_const_results} reports all data for structure constants: those for $4>d>3$ are new results, the ones for
$3 \ge d > 2$ are taken from~\cite{CMO}. The central
charge $c$ is obtained from the structure constant $f_{\s\s T}$ of the
energy-momentum tensor $T$ by
\be
f_{\s\s T}^2 = \frac{d}{4(d-1)} \, \frac{\D_\s^2}{c}\ .
\ee
For $f_{\s\s \mathcal{O}}$, 
we adopt the by-now standard normalization of~\cite{sd,Henriksson:2022gpa}.
The relation with the earlier normalization
$\wt{f}_{\s\s \mathcal{O}}$ of Ref.~\cite{slava-IM} is
\be\label{f_norm}
f_{\s\s \mathcal{O}}^2=
\frac{\left(\frac{d-2}{2} \right)_\ell}{\left( d-2 \right)_\ell} \,
\wt{f}_{\s\s\mathcal{O}}^2,
\ee
where $(x)_\ell \equiv \Gamma(x+\ell)/\Gamma(x)$ is the Pochhammer symbol.

\begin{figure}[!htb]
\centering
\includegraphics[scale=0.5]{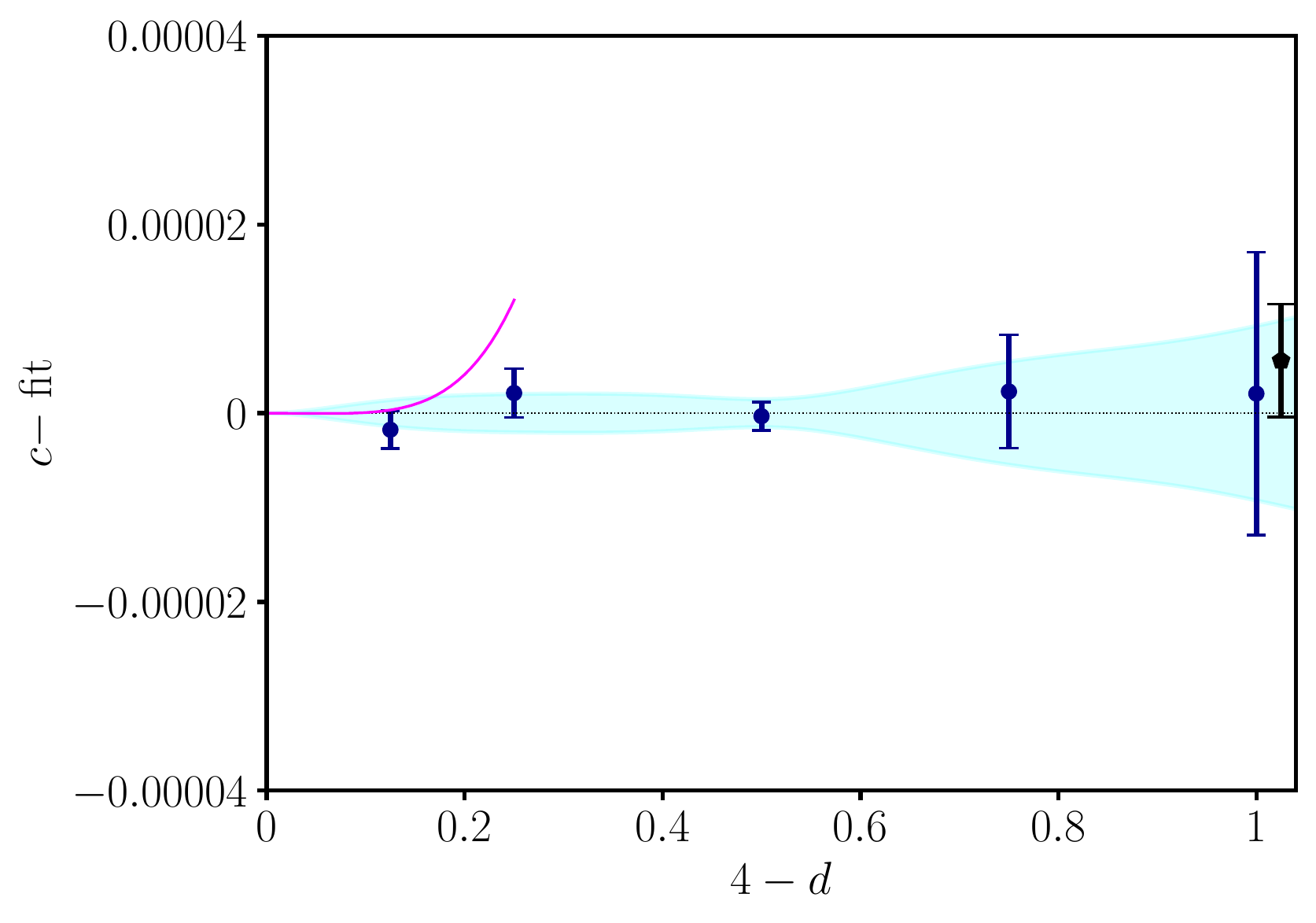}
\caption{Comparison of $c$ data minus best fit: bootstrap (blue circles),
unresummed epsilon expansion~\cite{gopa1,gopa2} (magenta solid curve),
$3$-correlator bootstrap at $d=3$~\cite{sd} (black pentagon).}
\label{fig:f_cc}
\end{figure}

\begin{figure}[!htb]
\centering
\includegraphics[scale=0.5]{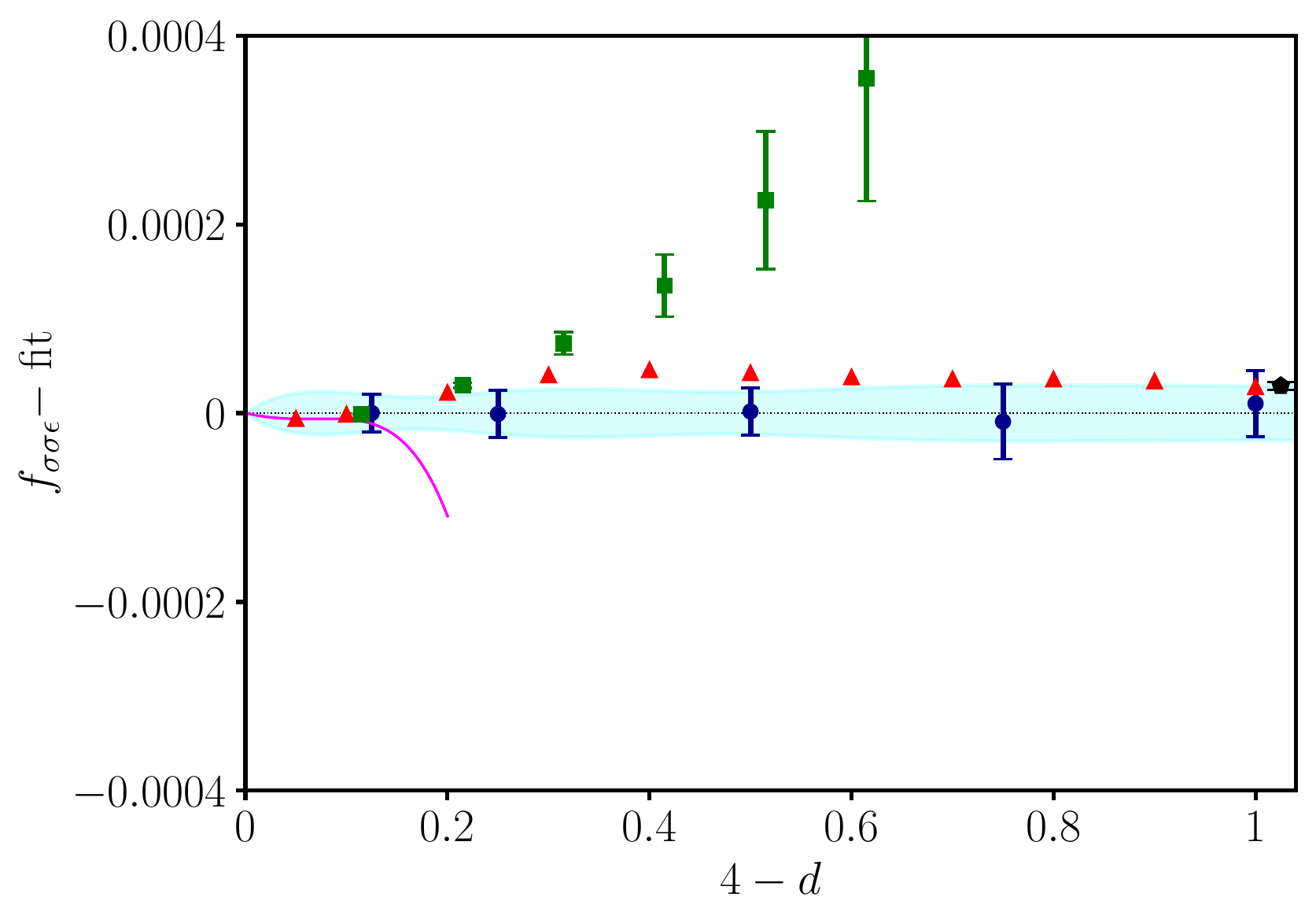}
\includegraphics[scale=0.5]{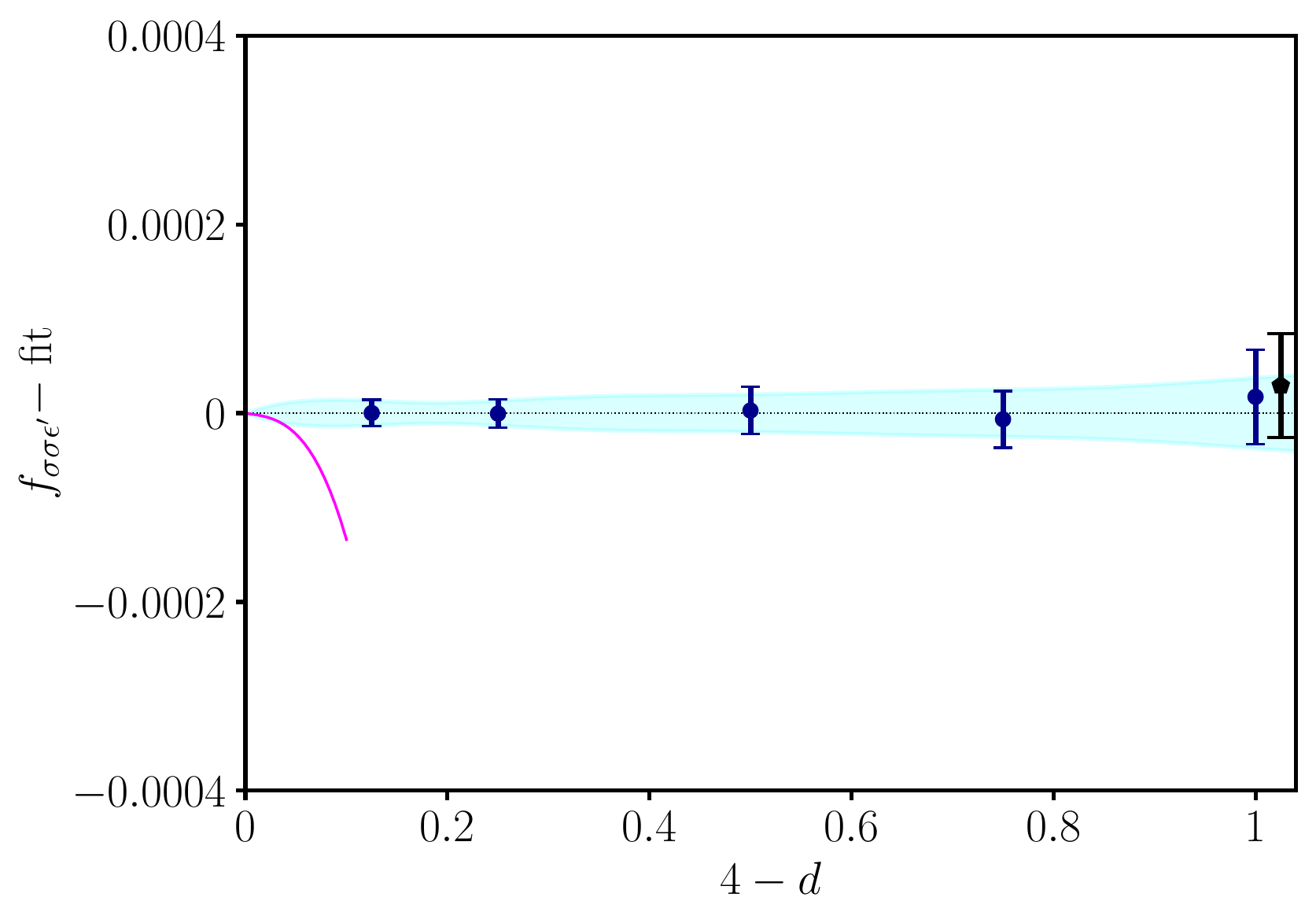}
\caption{Comparison of $f_{\s\s\e}$ and $f_{\s\s\e'}$ minus best
fit: bootstrap (blue circles), unresummed epsilon expansion~\cite{gopa1,gopa2,Henriksson:2022rnm,Bertucci:2022ptt} (magenta
solid curve), $3$-correlator bootstrap at $d=3$~\cite{sd} (black
pentagon). For $f_{\s\s\e}$ we also report the resummed epsilon expansion (green squares) and bootstrap navigator results~\cite{Henriksson:2022gpa} (red triangles).}
\label{fig:f_e_ep}
\end{figure}

The central charge $c$ and the structure constants $f_{\s\s\e}$ and
$f_{\s\s\e'}$ are determined with very high accuracy: their dependence
on $y=4-d$ is obtained with the fit method of
Sec.~\ref{sec:g_s_analysis}, assuming the exact $d=4$ value.
The resulting polynomials are reported  together with
the available epsilon-expansion
series~\cite{gopa1,gopa2,Henriksson:2022rnm,Bertucci:2022ptt}:
\ba
c(y) &=& 1 -0.015415049 y^2 -0.026663929 y^3 -0.004992140 y^4 -0.010357094 y^5
\nl
&& +0.007424814 y^6 -0.004670278 y^7+ 0.001206599 y^8,
\nl
&&  
\qquad \qquad \qquad \qquad\qquad \qquad
\qquad \qquad\qquad ~\mbox{(conformal\ bootstrap)},
\label{fc-cb}
\\
c(y) &=& 1 - 0.0154321 y^2 - 0.0266347 y^3\nl
&& -0.0039608 y^4, \qquad \qquad \qquad \qquad\qquad
\qquad\mbox{(epsilon expansion)},\ \ 
\label{fc-ee}
\ea
\ba
f_{\s\s\e}(y)
&=& \sqrt{2} -0.235465537 y -0.170275458 y^2 + 0.096635030 y^3
- 0.113371408 y^4
\nl
&& + 0.100586943 y^5 -0.054667196 y^6 + 0.016161292 y^7 -0.001992399 y^8,
\nl
&&\qquad \qquad \qquad
\qquad \qquad \qquad \qquad \qquad\qquad \mbox{(conformal\ bootstrap)},
\label{fe-cb}\\
f_{\s\s\e}(y) &=& \sqrt{2} - 0.235702 y - 0.168047 y^2 + 0.103680 y^3
- 0.224776 y^4,
\nl
&&\qquad \qquad \qquad \qquad \qquad \qquad
\qquad \qquad \qquad\mbox{(epsilon expansion)},
\label{fe-ee}
\ea
\ba
f_{\s\s\e'} (y) &=& 0.136221303 y -0.118250195 y^2 +0.067116467 y^3
 -0.058700794 y^4
\nl
&& +0.037159615 y^5 -0.012211017 y^6+ 0.001647332 y^7
\nl
&& \qquad \qquad \qquad \qquad \qquad \qquad \quad
\qquad \qquad~~\mbox{(conformal\ bootstrap)},
\label{fep-cb}
\\
f_{\s\s\e'}(y)&=& 0.1360828 y + 0.11844240525 y^2,
\qquad\qquad \mbox{(epsilon expansion)}.
\label{fep-ee}
\ea
We remark: \textit{i}) the excellent agreement between the first few
terms of the conformal bootstrap and epsilon-expansion series, and
\textit{ii}) the need of a high-order $O(y^7, y^8)$ polynomial for
precise fits.  The corresponding curves are shown in
Figs.~\ref{fig:f_cc} and~\ref{fig:f_e_ep}. Note that $c$, $f_{\s\s\e}$
and $f_{\s\s\e'}$ were determined with strikingly small (relative)
errors, respectively $O( 10^{-5})$, $O( 10^{-4})$ and $O(10^{-4})$
over the entire $d$ range.

The comparison with other conformal bootstrap results is as follows:
The best $3$-correlator determination in $d=3$~\cite{sd} is shown as a
black pentagon in the figures.
Data from the navigator
  method are unfortunately only available
  for $f_{\s\s\e}$ \cite{Henriksson:2022gpa}.
The agreement among different numerical setups is
extremely good. Moreover, as already observed for scaling dimensions,
the unresummed epsilon expansion captures the $d \to 4$ behavior,
and it does it very well, since the lower-order terms of the
respective polynomials~\eqref{fc-cb}--\eqref{fep-ee} are equal within
errors. For $f_{\s\s\e}$, the results of the resummed
epsilon expansion, reported in Tab.~\ref{tab:resummed_f_e}, are also shown, determined by earlier methods: the $4^{\rm th}$-order series~\eqref{fe-ee} only allows for a precise
agreement down to $d\approx 3.6$, given the fine scale of
Fig.~\ref{fig:f_e_ep}. For the remaining quantities, the
epsilon expansion is either too short for a resummation, or not
alternating.

\begin{table}[!htb]
\centering
\begin{center}
\begin{tabular}{|c|l|}
\hline
&\\[-1em]
$d$ & $f_{\s\s\e}$ \\
\hline
&\\[-1em]
3.9 & 1.3890497(2) \\
3.8 & 1.360960(3) \\
3.7 & 1.330222(12) \\
3.6 & 1.29703(3) \\
3.5 & 1.26154(7) \\
3.4 & 1.22386(13) \\
3.3 & 1.1841(2) \\
3.2 & 1.1423(3) \\
3.1 & 1.0986(5) \\
3   & 1.0531(7) \\
\hline
\end{tabular}
\end{center}
\caption{Structure constant $f_{\s\s\e}$ from resummed perturbative expansion, obtained according to the methods of~\cite{Kompaniets:2019zes}.}
\label{tab:resummed_f_e}
\end{table}

\FloatBarrier

% -4.2-------------------------------------------------------
\subsection{Higher fields $T'$ and $C$}\label{sec:higher_op_dim}
\vspace{-\baselineskip}\indent

\begin{figure}[!htb]
\centering
\includegraphics[scale=0.48]{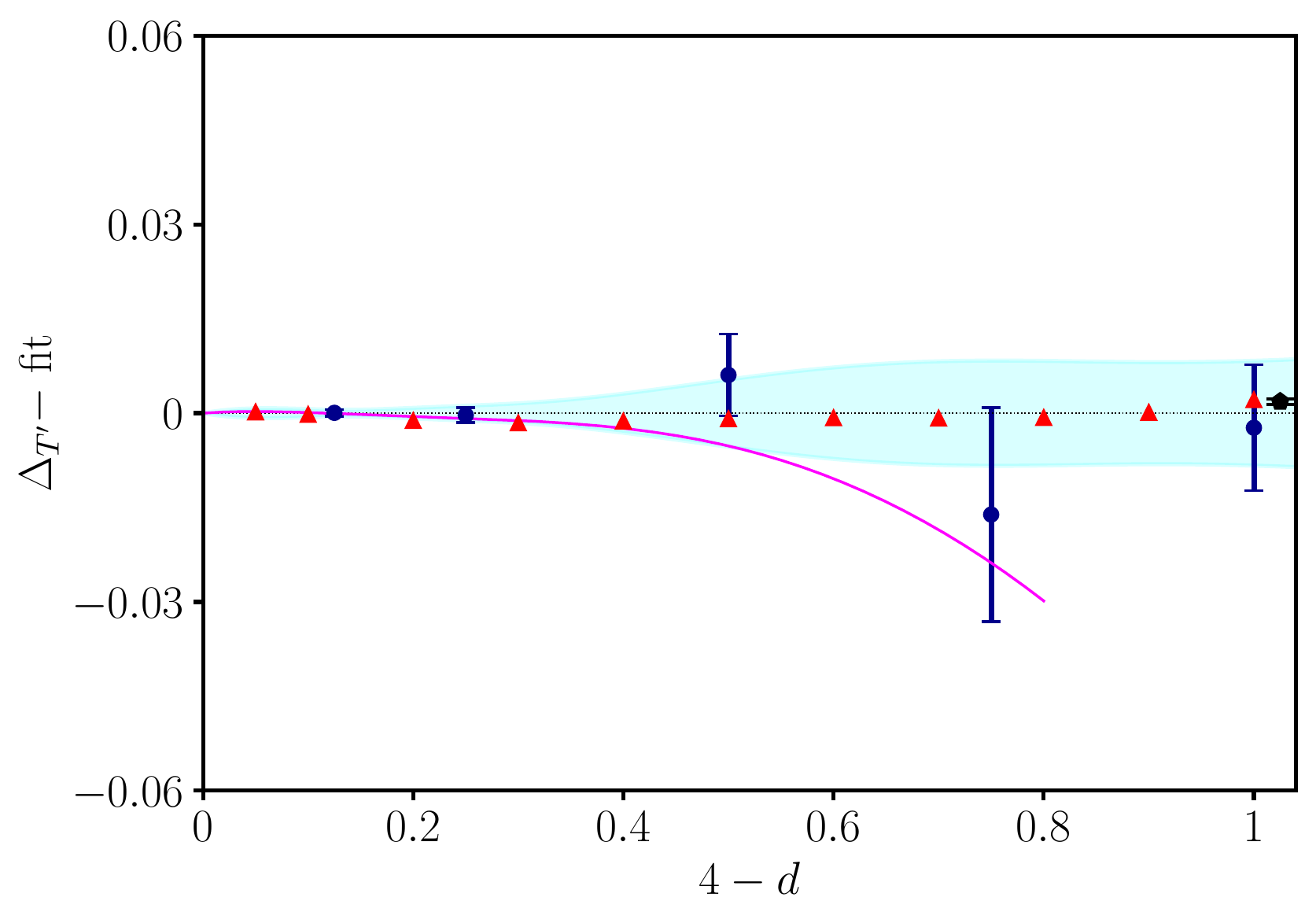}
\hspace{1.5cm}
\includegraphics[scale=0.48]{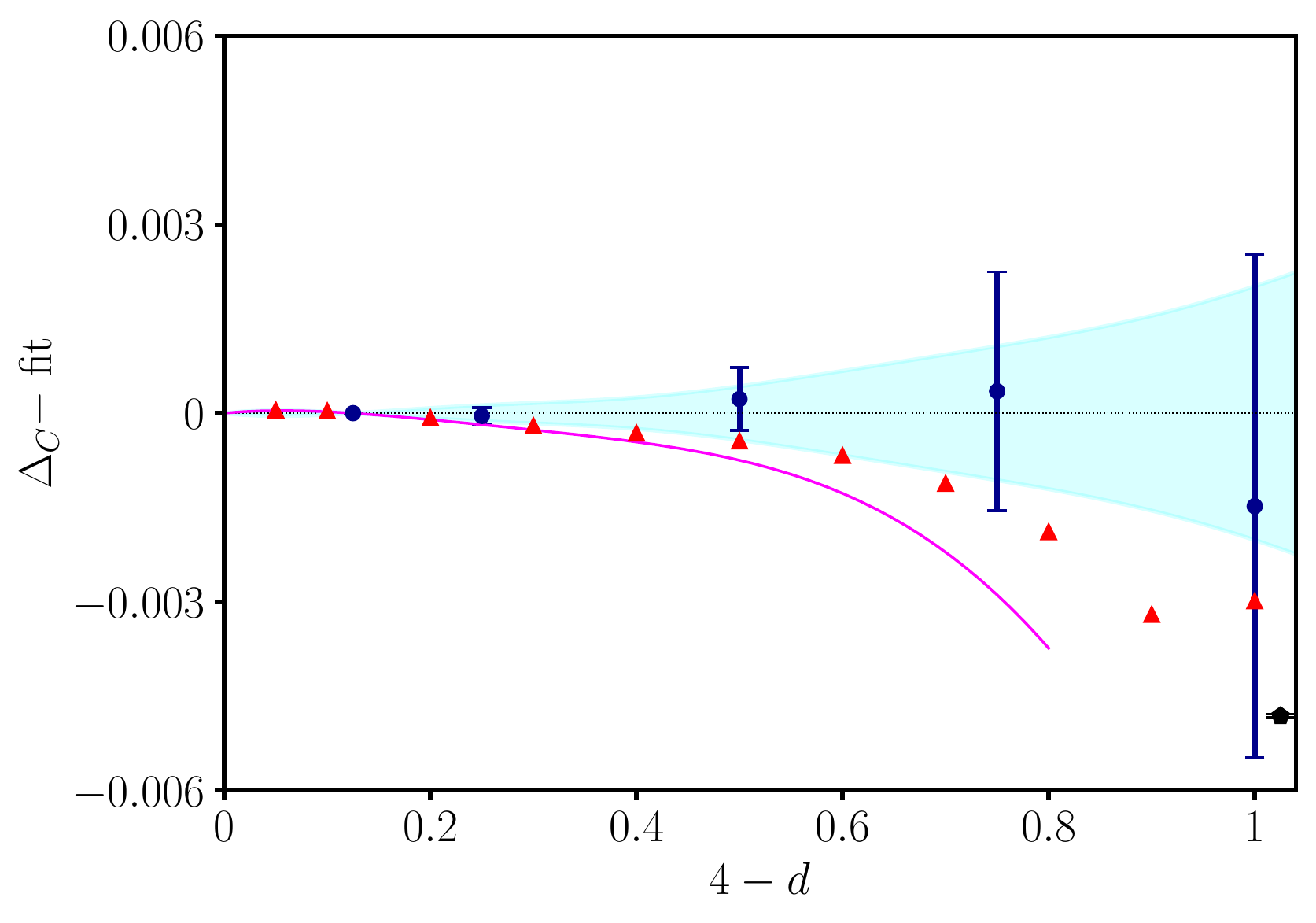}
\caption{Comparison of scaling dimensions minus best fit for $T',C$ fields:
bootstrap (blue round points), navigator method~\cite{Henriksson:2022gpa}
(triangle red points), $3$-correlator bootstrap at $d=3$~\cite{sd} (black
pentagon) and unresummed epsilon expansion~\cite{gopa1,gopa2,4th-1,4th-2} (magenta solid line).}
\label{fig:Tp-C}
\end{figure}

The analysis of the fields $T'$ ($\ell=2$) and $C$ ($\ell=4$) is
done along the same lines. The fit polynomials for $\D_{T'}$ and $\D_{C}$,
obtained as before, are 
\ba
\D_{T'}(y) &=& 6 - 0.567900778 y + 0.1779633663 y^2 -0.806164966 y^3\nl
&& + 1.749534636 y^4 -1.684842086 y^5 + 0.765011179 y^6\nl
&& -0.126284231 y^7, \qquad\qquad\qquad\qquad \mbox{(conformal\ bootstrap)},
\\
\D_C (y)&=& 6 - 1.001598184 y + 0.030791232 y^2\nl
&& -0.033868719 y^3 + 0.041665026 y^4 -0.002907562 y^5\nl
&& -0.006602770 y^6, \qquad\qquad\qquad\qquad \mbox{(conformal\ bootstrap)}.
\ea
They are shown in Fig.~\ref{fig:Tp-C}, along with the
bootstrap results of~\cite{Henriksson:2022gpa} (red triangles)
and the available epsilon-expansion series (magenta solid
lines)~\cite{gopa1,gopa2,4th-1,4th-2}:
\ba
\D_{T'}(y) &=&6 -0.5555556 y, \qquad\qquad\qquad\qquad
~\mbox{(epsilon expansion)},
\\
\D_C(y)&=& 6 - y + 0.01296296 y^2 + 0.01198731 y^3\nl
&&- 0.006591585 y^4, \qquad\qquad\qquad\qquad\mbox{(epsilon expansion)}.
\ea
As shown by the cyan band, representing our fitting error, the scaling
dimensions of these fields are determined with an accuracy comparable
to that achieved for the low-lying $\ell=0$ states:
$\mathrm{Err}(\D_{T'})\approx 10^{-2}$ and
$\mathrm{Err}(\D_{C})\approx 3\times 10^{-3}$, meaning that
$\mathrm{Err}(\D_{T'})/\D_{T'}\approx 10^{-3}$ and
$\mathrm{Err}(\D_{C})/\D_{C}\approx 5 \times 10^{-4}$. Within our
precision, we observe very good agreement with the results
of~\cite{Henriksson:2022gpa} (especially for $T'$). Furthermore, the
unresummed epsilon expansion is again in agreement with the bootstrap
results for $d\to 4$.  Overall, the picture is consistent with the
$\ell=0$ case discussed earlier\footnote{The good behavior of the perturbative expansion for larger values of $y\approx 0.8$ is not stressed, since it may be an artifact of the low order of the series.}.

\begin{figure}[!htb]
\centering
\includegraphics[scale=0.48]{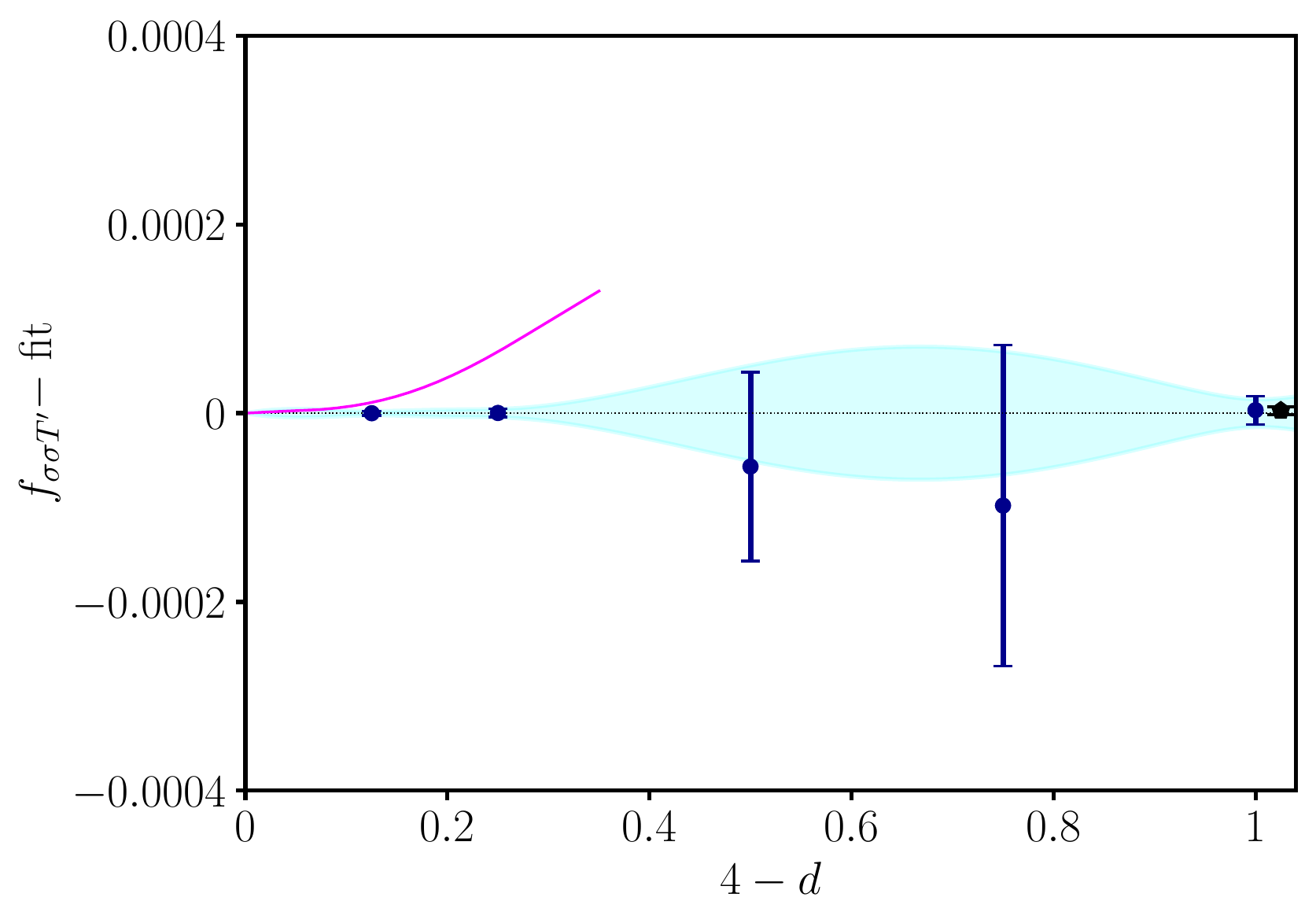}
\includegraphics[scale=0.48]{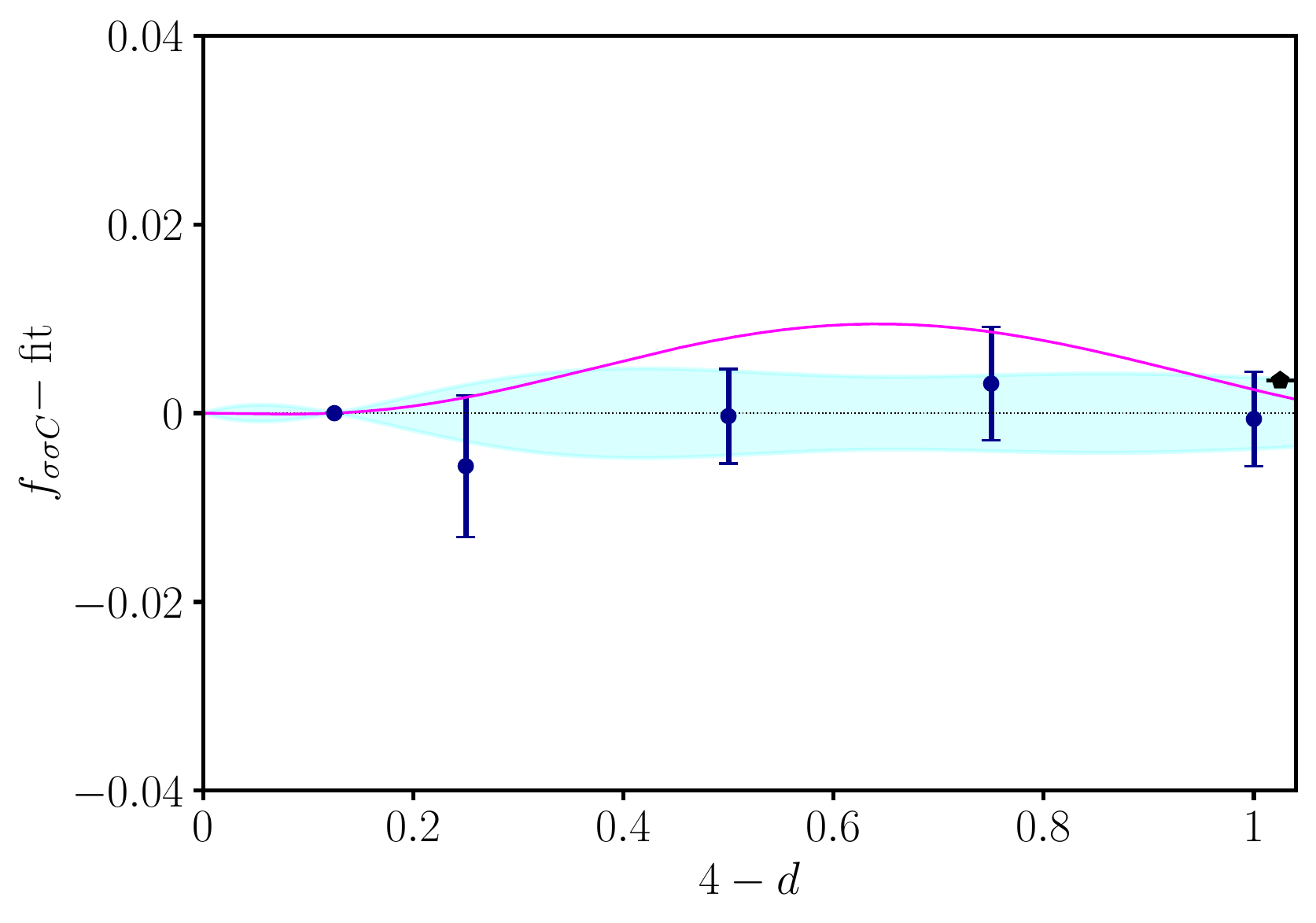}
\caption{Behavior of structure constants $f_{\s\s T'}$ and $f_{\s\s C}$ (round blue points) compared with $3$-correlator bootstrap at $d=3$~\cite{sd} (black pentagon) and epsilon expansion (magenta solid line)~\cite{gopa1,gopa2}.}
\label{fig:f_Tp-C}
\end{figure}

\noindent
The corresponding structure constants are given by the polynomial fits
\ba
f_{\s\s T'}(y)&=& 0.026278214 y -0.012019512 y^2 -0.016779681 y^3
\nl
&& +0.025762223 y^4 -0.018571573 y^5 +0.006902659 y^6
\nl
&& -0.001000504 y^7, \qquad\qquad\qquad\qquad\mbox{(conformal\ bootstrap)},
\\
f_{\s\s C}(y) &=& 0.16903085 -0.122480930 y +0.077087613 y^2-0.591032947 y^3
\nl
&&  +1.331591787 y^4 -1.231373513 y^5+0.512308476 y^6
\nl
&& -0.079520247 y^7,
\qquad\qquad\qquad\qquad\mbox{(conformal\ bootstrap)}.
\ea
They can be compared to the available epsilon expansions~\cite{2loop_eps_sec,gopa1,gopa2,4th-1,4th-2}:
\ba
f_{\s\s T'}(y)&=& 0.02635231 y - 0.013176155 y^2,
\qquad\,\,\mbox{(epsilon expansion)},\ 
\\
\label{pert_fC}
f_{\s\s C}(y)&=& 0.16903085 - 0.12244675 y + 0.02131741 y^2\nl
&& + 0.002168567 y^3 - 0.0019760553 y^4, \quad\,\mbox{(epsilon expansion)}.\ 
\ea
The comparison is shown in Fig.~\ref{fig:f_Tp-C}. Also in this case we observe good agreement between the conformal bootstrap polynomials and the epsilon expansion series up to $ O(y^3)$ terms.

\FloatBarrier

% -4.3----------------------------------------------------
\subsection{Subleading fields $\e''$ and $C'$}\label{high_spectrum}
\vspace{-\baselineskip}\indent

The numerical $1$-correlator bootstrap approach used in this paper is
known to have a limited precision for states higher up in the
conformal spectrum, in particular for our approximation to $190$
components of the truncated bootstrap equations. In this section, we
show that our identification of $\e''$ ($\ell=0$) and $C'$ ($\ell=4$)
has some problems, especially for $d\to 4$. We explain these
difficulties by using the epsilon expansion for conformal dimensions
and structure constants, as well as the $3$-correlator bootstrap
data~\cite{Henriksson:2022gpa} in varying dimensions,
which are definitely more accurate for the higher spectrum than our results.
We think that
these aspects are worth discussing, especially because the $y=4-d$
dependence plays a crucial role.

\begin{figure}[!htb]
\centering
\includegraphics[scale=0.48]{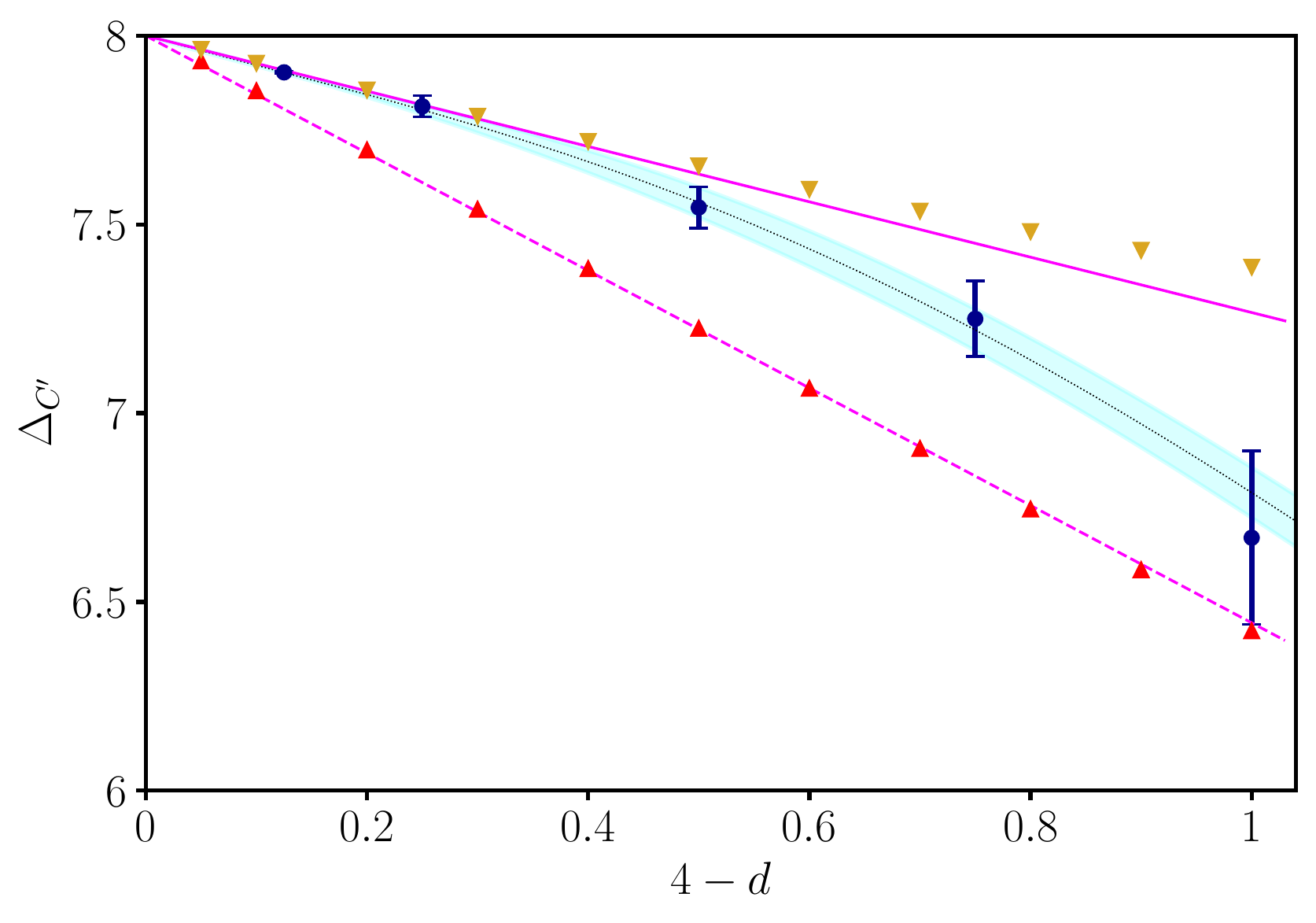}
\includegraphics[scale=0.48]{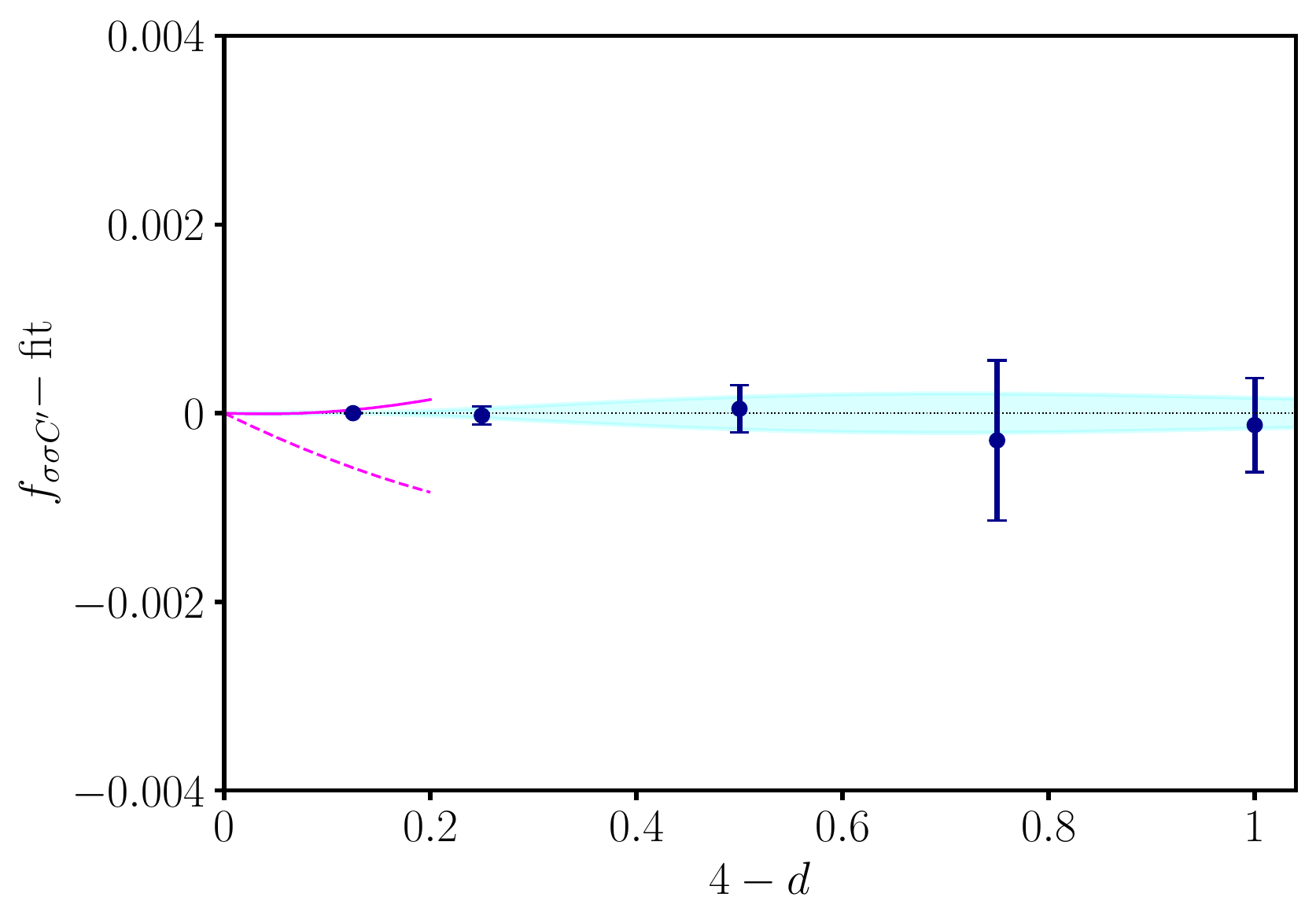}
\caption{Scaling dimension and structure constant of would-be $C'$ operator in our bootstrap spectrum (blue circles). Upward red and downward gold triangles represent navigator results for $C'$ and $C'_2$~\cite{Henriksson:2022gpa}. The dashed and solid magenta lines are the corresponding leading-order epsilon expansion.}
\label{fig:Cp}
\end{figure}

We start our analysis from the subleading twist
$\ell=4$ operator $C'$, for which we find the following best fit
polynomial:
\ba\label{d_Cp-CB}
\D_{C'}(y) &=& 8 - 0.827053961 y -0.055211344 y^2 +0.053430207 y^3\nl
&& +0.010354264  y^4 -0.003205703 y^5, \qquad\mbox{(conformal\ bootstrap)}.\ \ \ 
\ea
These data are shown in Fig.~\ref{fig:Cp} (left part). It turns out that $C'$ is degenerate at $d=4$ with another field with same dimension and spin, called $C'_2$. Their dimensions are
known to leading order in the epsilon expansion,
\ba \label{yCp-ee}
\D_{C'}(y) &=& 8 -1.555556 y, 
\\
\D_{C'_2}(y) &=& 8 -0.833333 y, \qquad\qquad\qquad\mbox{(epsilon expansion)},
\label{Cp2-ee}
\ea
and are plotted in Fig.~\ref{fig:Cp} with magenta dashed and solid lines,
respectively. Near these lines, the navigator bootstrap results~\cite{Henriksson:2022gpa} are plotted with gold and red triangles.

One sees that our results start at $d\to 4$ very close to $C'_2$ (see
first coefficient in polynomials~\eqref{d_Cp-CB} and~\eqref{yCp-ee}) and end up
near $C'$ at $d=3$. Therefore, the state we found is a mixture of $C'$
and $C'_2$: better numerical precision would be needed
for disentangling the two states near $d\to 4$, obtained, e.g., by
increasing the number of components approximating the bootstrap equations.

\noindent
The fit of the structure constant is given by 
\ba\label{f_Cp-CB}
f_{\s\s C'}(y) &=& 0.006871047 y -0.005215834 y^2 -0.003223129 y^3\nl
&& +0.005087571 y^4 -0.001393464 y^5, \qquad \mbox{(conformal\ bootstrap)},
\ \ \ \ 
\ea
and plotted in the right part of Fig.~\ref{fig:Cp}.
The epsilon-expansion results for $C'$ and $C'_2$ read,
\ba
f_{\s\s C'}(y) &=& 0.001543806 y,
\\
f_{\s\s C'_2}(y) &=& 0.006458202 y,
\qquad\qquad\qquad\mbox{(epsilon expansion)},
\ea
and are shown as magenta dashed and solid lines on the right of Fig.~\ref{fig:Cp}.

These perturbative data show a remarkable fact: for $d<4 $ the state
of higher dimension $C'_2$ has a larger structure constant, contrary
to the standard behavior of $f_{\s\s {\cal O}}$ decreasing fast with
$\D_{\cal O}$. It is thus clear that, close to $d=4$, $C'_2$ gives the
dominant contribution to a putative mixed $C'$-$C'_2$ state. This suggests the reason
why our results with limited precision start close to
$C'_2$. The analysis is confirmed by the bootstrap result for the
structure constant in~\eqref{f_Cp-CB}: for $d\to 4$ it fits the perturbative
behavior of $f_{\s\s {C'_2}}$, as seen in the right plot of
Fig.~\ref{fig:Cp}. In conclusion, our subleading $\ell=4$ state is
identified as $C'_2$ for $d\to 4$, but gradually approaches $C'$ in
$d=3$.

\begin{figure}[!htb]
\centering
\includegraphics[scale=0.48]{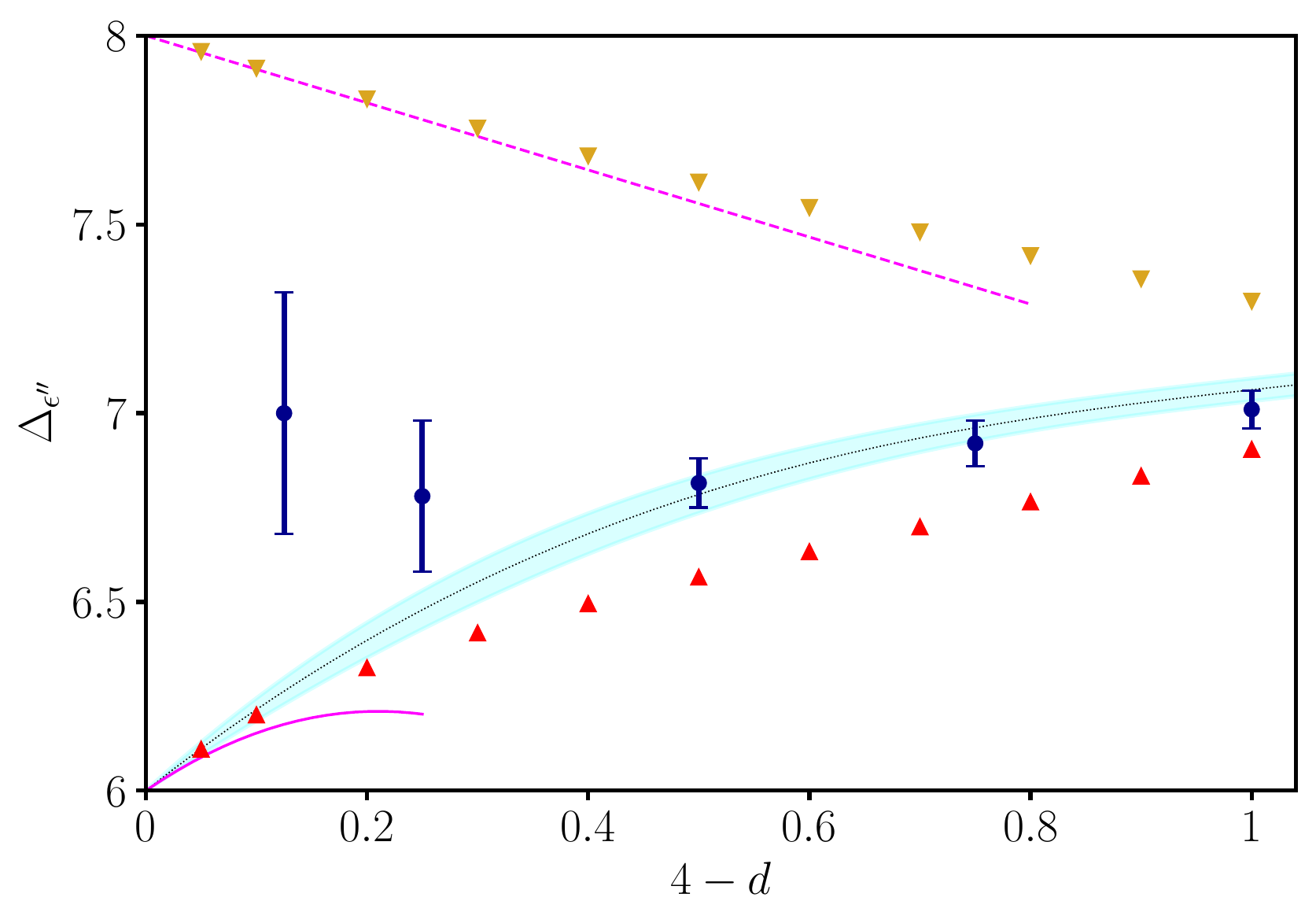}
\includegraphics[scale=0.48]{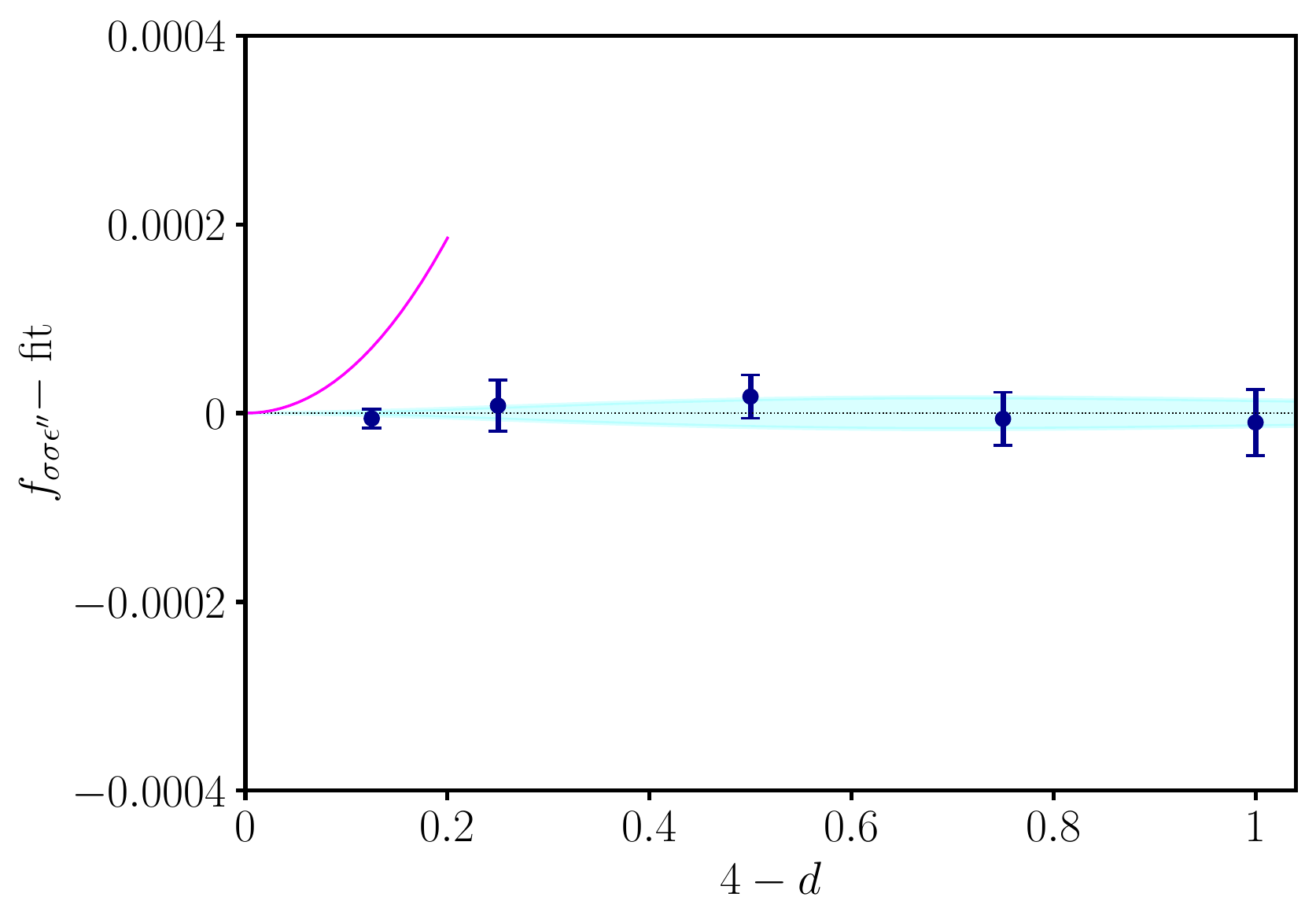}
\caption{Scaling dimension and structure constant of the would-be $\e''$
operator in our bootstrap spectrum (blue circles). Upward red and
downward gold triangles represent navigator results
for $\e''$ and $\e'''$~\cite{Henriksson:2022gpa}. The
solid and dashed magenta lines are the corresponding
leading-order epsilon expansion, which agree with the navigator results, but not ours.}
\label{fig:epp}
\end{figure}

Another problematic identification concerns the $\e''$ field (corresponding to
$\phi^6$ in the $ \phi^4$ theory). The best fit of bootstrap
data gives
\ba
\D_{\e''}(y) &=& 2.313321845 y -1.678645012 y^2+ 0.336440006 y^3
\nl
&&  + 0.090959178 y^4, \qquad\qquad\qquad\mbox{(conformal\ bootstrap)},
\ea
while the leading epsilon-expansion result reads~\cite{null-ref1,null-ref2,2loop_eps_sec}:
\ba\label{eps_exp_epp}
\D_{\e''}(y) = 2y - 4.759259y^2,
\qquad\qquad\qquad\quad\mbox{(epsilon expansion)}.
\ea
For the structure constant we find
\ba
f_{\s\s\e''}(y) &=&  0.002851280 y^2 -0.003188068 y^3+ 0.001218496 y^4
\nl
&&  -0.000161879 y^5, \qquad\qquad\qquad\;\mbox{(conformal\ bootstrap)};
\label{}\\
f_{\s\s\e''}(y) &=& 0.006901444 y^2,
\qquad\qquad\qquad\quad\mbox{(epsilon expansion)}.
\ea
It is apparent that our bootstrap results do not match the leading
perturbative expansion for $d\to 4$. The corresponding plots are shown
in Fig.~\ref{fig:epp}, where the disagreement with bootstrap results
from Ref.~\cite{Henriksson:2022gpa} (red triangles) is also seen.

Let us investigate the possibility of another mixing of states near
$d\to 4$. In this case there is no degenerate field with $\e''$ at
$d=4$. However, the next subleading one $\e'''\sim \Box^2\phi^4$ in the $\phi^4$ theory is present at higher dimension $\D_{\e'''}\le 8$. The epsilon expansion and navigator  
results for this field are also shown in Fig.~\ref{fig:epp} (left part, gold
downward triangles). We remark that a mixing of $\e''$ and $\e'''$
was shown to take place at $d=2.8$, i.e., rather far from $d=4$~\cite{Henriksson:2022gpa}.

We suppose that the limited resolution of our data  
finds a state which is a mixture of $\e''$ and $\e'''$ also for
$d\to 4$, but we cannot be certain of this. As for $C'$ and $C'_2$, support for this argument could come from a comparison of the corresponding structure constants $f_{\s\s\e''}$ and $f_{\s\s\e'''}$. Unfortunately, the epsilon expansion of the latter is not available, so we cannot get a definite explanation of our $\D_{\e''}$ data.

\FloatBarrier

%-5--------------------------------------------------
\section{Conclusions}\label{sec:conclusions}
\vspace{-\baselineskip}\indent

In this paper we obtained the conformal dimensions and structure
constants of the critical Ising CFT as a function of varying dimension
$4 > d \ge 3$ by using the numerical conformal bootstrap approach.

Our main result is the precise determination of the anomalous
dimensions of the $\s,\e,\e'$ fields, which are related to the Ising
critical exponents $\eta$, $\n$, $\w$. Our relatively simple
$1$-correlator bootstrap setup is able to compute the $d$-dependence
of these quantities with up to one-per-thousand relative accuracy;
therefore, our findings can be used as a benchmark for future studies
in non-integer space dimension.
For these low-lying states of the conformal spectrum, our
  results are in very good
  agreement with those of more advanced 3-correlator
  bootstrap techniques~\cite{sd-old,sd,Reehorst:2021ykw,Henriksson:2022gpa},
with a small offset included in the error estimate.

We presented a detailed comparison of available predictions from
different methods. For $d\to 4$, our results agree with those from
unresummed perturbation theory. This shows two things: that
non-perturbative differences, which might effect the bootstrap program
or the resummed series, are negligible for $d\to 4$. The other
non-trivial result is that both approaches agree on the same analytic
continuation in dimension. A possible explanation of this
correspondence is provided by the analytical bootstrap, which on one
hand reproduces the epsilon expansion, and on the other hand uses the
same ingredients as the numerical bootstrap.

For $3\le d<4$, but away from $d=4$, the bootstrap data agree very
well with other results, obtained by resummation techniques of the
perturbative series, Monte Carlo simulations, and other bootstrap
approaches. In the whole $4 > d \ge 3$ range we find overall
consistency among the different approaches; improvements are needed by
adding further terms to the perturbative series in $d=3$, as the
current state of the art still shows a $O(10^{-3})$, $O(10^{-2})$
discrepancy, respectively for $\nu$ and $\omega$, and in general much
larger error bars than bootstrap and Monte Carlo results.

We were able to  compute  bootstrap data for the conformal dimensions of
higher-order fields   in $4>d\ge 3$, including   the lowest-lying
spinful fields $T'$ ($\ell=2$) and $C$ ($\ell=4$), with a precision comparable
 to that of spinless operators. 
The central charge
and OPE coefficients of low-lying fields were obtained with even higher
precision than that of the corresponding anomalous dimensions.
The structure constants agree well with those of the 3-correlator bootstrap,  where
available (mostly in $d=3$),
and with perturbation theory for $d \to 4$.

A possible future development is to improve current
bootstrap results in the region $3 > d \ge 2$, in order to
better understand how the $d=3$ theory approaches
the $d=2$ Virasoro minimal model. To this aim, it is  
important to go beyond the lowest-lying states and precisely probe
higher-dimensional and higher-spin fields. Improved
$3$-correlator bootstrap protocols, such as the recently proposed 
navigator method, may be well suited here.

\vspace*{-0.5\baselineskip}
\section*{Acknowledgements}
\vspace{-\baselineskip}\indent

We are grateful to C.~Bonati, R.~Guida, J.~Henriksson, S.~Kousvos, L.~Maffi, R.~Pisarski, M.~Reehorst, S.~Rychkov, M.~Serone and B.~Sirois for useful discussions. We thank Insitut Pascal for organizing the workshop ``Bootstat'', where this work was initiated. CB acknowledges the support of the Italian Ministry of Education, University and Research
under the project PRIN 2017E44HRF, ``Low dimensional quantum systems:
theory, experiments and simulations''. Numerical computations have
been performed on the \texttt{Zefiro} cluster of the Scientific
Computing Center at INFN Pisa.

%-Appendix-----------------------------------------
\appendix
\section*{Appendix}
\section{Orthogonal polynomial regression}\label{appendix:ortho_poly}
\vspace{-\baselineskip}\indent

Standard polynomial regression of the data set
$S\equiv\left\{x_i, y_i, \Delta y_i\right\}_{i=1}^{N}$ is achieved by
minimizing   
\ba\label{min_chisq}
\chi^2 = \sum_{k=1}^{N} \left( \frac{y_k - f(x_k)}{\Delta y_k} \right)^2,
\ea
with respect to the parameters $\left\{c_i\right\}_{i=0}^{d}$ of
the fit function,
\ba
f_n(x) = \sum_{r=0}^{n} c_r x^r.
\ea
The degree $n$ of the polynomial is not known a priori.
	
A smarter fit   is obtained by changing the basis in
which the polynomial is expressed:
\ba
\mathcal{B}_{\mathrm{naive}} = \left\{1,x,x^2,\dots,x^d\right\}
\rightarrow \mathcal{B}_{\mathrm{ortho}} =
\left\{P_0(x),P_1(x),P_2(x),\dots,P_d(x)\right\},
\ea
where the polynomials $P_k(x)$ (of degree $k$) are chosen to be
\emph{orthogonal} on the independent variables of the dataset $S$, i.e.:
\ba
\braket{P_r(x) P_s(x)}_{S} = \frac{1}{N} \sum_{k=1}^{N} P_r(x_k) P_s(x_k)
= k_r^2 \delta_{rs},
\ea
where $k_r$ are constants. With this choice, the fit function becomes
\ba
f_n(x) = \sum_{r=0}^{n} \alpha_r P_r(x).
\ea
The best fit is obtained by minimizing   $\chi^2$ in
Eq.~\eqref{min_chisq}. The advantage of the orthogonal polynomial
regression is that the  coefficients $\alpha_r$ do not depend on the
$\alpha_s$ with $s>r$, i.e., adding higher-degree polynomials $r>n$ to
$f_n(x)$ does not change the value of $\alpha_r$ with $r\le n$
within the statistical errors~\cite{ortho_poly}. Thus,
this procedure is better suited to assess the optimal degree of the polynomial.
	
The expression of the polynomials $P_r(x)$ is known in the
literature. In this work, we follow the conventions of
Ref.~\cite{ortho_poly}. We start by fixing the $r=0$ and $r=1$
polynomials as
\ba
P_0(x) = 1, \qquad P_1(x) = 2(x-a_1), \qquad a_1 =
\frac{1}{N}\sum_{k=1}^{N}x_k \equiv \overline{x}.
\ea
Higher-order polynomials with $r\ge2$ are obtained
through the recursive relation~\cite{ortho_poly},
\ba
P_{r+1}(x) = 2(x-a_{r+1})P_r(x) - b_r P_{r-1}(x),
\ea
where the coefficients $a_{r+1}$ and $b_r$ are given by
\ba
a_{r+1} = \frac{\sum_{k=1}^{N}x_k P^2_r(x_k)}{\sum_{k=1}^{N}P^2_r(x_k)},
\qquad\qquad\qquad
b_{r}   = \frac{\sum_{k=1}^{N}P^2_r(x_k)}{\sum_{k=1}^{N}P^2_{r-1}(x_k)}.
\ea
In this work, we  find the best fitting
polynomial for $\gamma_{\mathcal{O}}$ and
$f_{\s\s{\cal O}}$ as a function
of $y= 4-d$. We always assume their known analytic value for $d=4$,
for example $\gamma_{\mathcal{O}}(d=4)=0$. To enforce such
constraint, it is sufficient to use as fit function
\ba
h_n(x) = f_n(x) - f_n(0) =
\sum_{r=1}^{n} \tilde{\alpha}_r \left[ P_r(x) - P_r(0) \right].
\ea
Finally, we reconstruct the original expansion in the naive basis  
by summing all equal monomials among every $P_r(x)$ included in the
fit function:
\ba
h_n(x) = \sum_{r=1}^{n} \tilde{\alpha}_r \left[ P_r(x) - P_r(0) \right ]
= \sum_{r=1}^{n} \tilde{c}_r x^r,
\ea
where
\ba
\tilde{c}_r =
\sum_{l=r}^{n} \tilde{\alpha}_l \frac{d^r P_l(x)}{dx^r}\Bigg\vert_{x=0}.
\ea
Once the two expansions are properly matched, the
coefficients obtained from orthogonal polynomials agree with
those obtained using a standard polynomial fit. The advantage
of orthogonal polynomials resides in their improved numerical
stability, which results in an improved precision in the computation of
the $c_i$.

Finally, once the best fitting polynomial is obtained, we
assign an error to our best fit function $h_n(x)$ through standard
error propagation, via the so-called parameter covariance matrix,
\ba
C_{ij} \equiv \mathrm{Cov}(\tilde{\alpha}_i, \tilde{\alpha}_j).
\ea
Let us define $v_i(x)$ as the gradient of the fit function with respect to
the $i^{\text{th}}$ fit parameter,
\ba
v_i(x) = \frac{\partial h_n(x \,\vert\, \vec{\tilde{\alpha}})}{\partial \tilde{\alpha}_i}.
\ea
The error on the best fitting polynomial is
\ba
\mathrm{Err}(h_n)(x) = v^{\mathrm{T}}(x) C v(x) = C_{ij} v_i(x) v_j(x).
\ea
The best fit of $\gamma_{\mathcal{O}}(y)$ via orthogonal polynomial
regression was done by using the \texttt{curve\_fit} routine from the
standard \texttt{Python} library \texttt{scipy}.

%-B--------------------------------------------
\section{Resummation of perturbative series}

\subsection{Toy model example}\label{appendix:baby_integral}
\vspace{-\baselineskip}\indent

\begin{figure}
\begin{center}
\includegraphics[width=6.5cm]{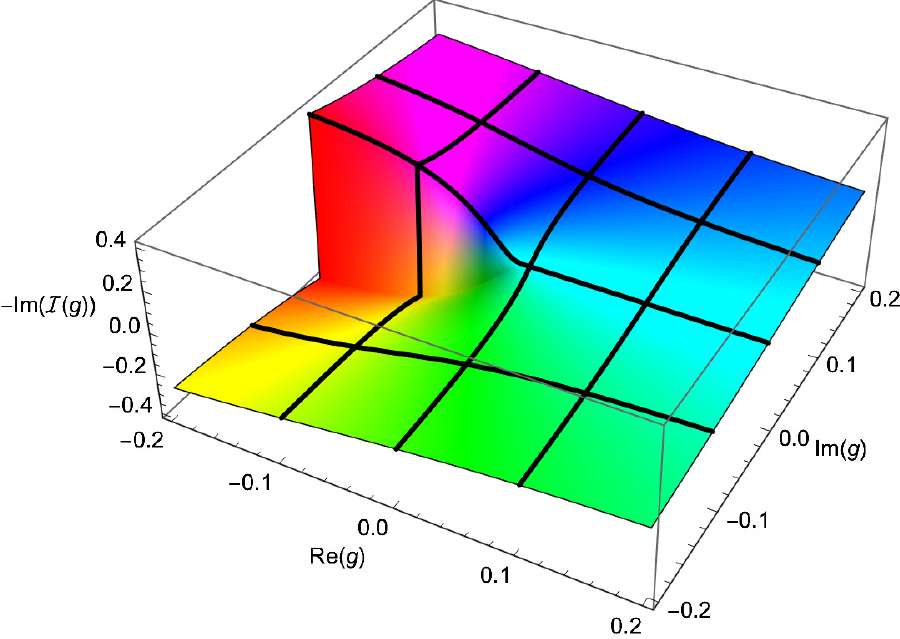} \hfill \includegraphics[width=6.5cm]{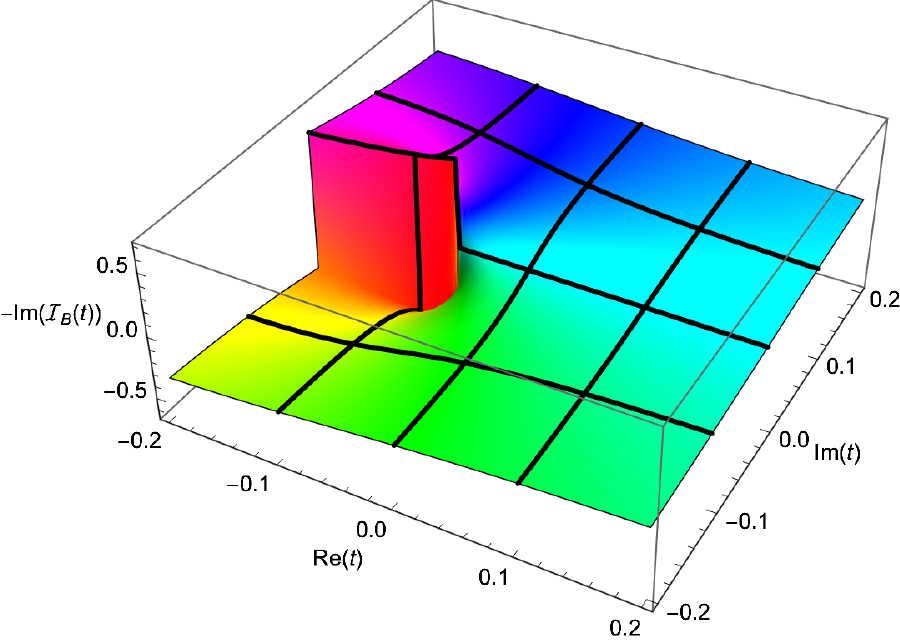}
\end{center}
\vspace*{-1.2cm}
\centerline{\includegraphics[width=6.5cm]{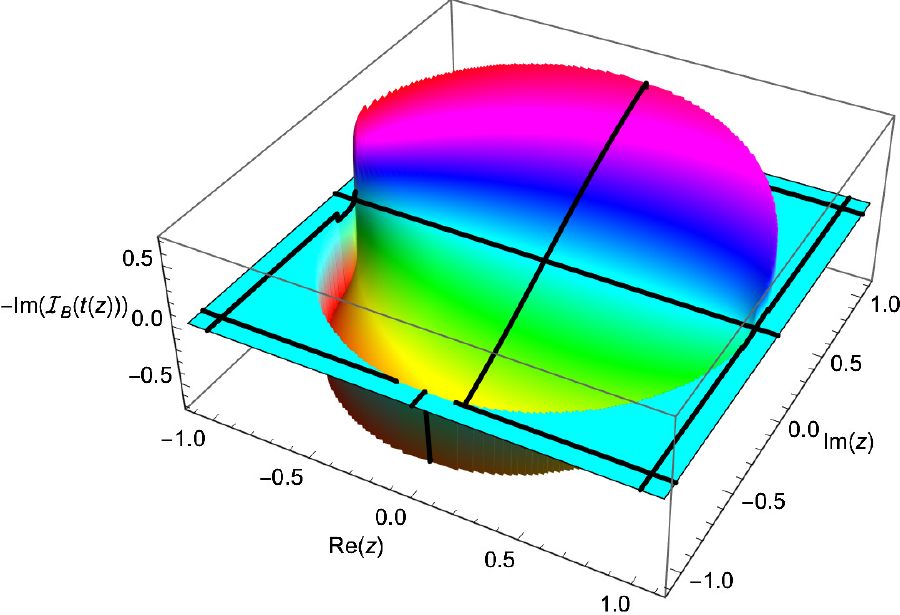}}
\caption{The branch cut in ${\cal I}(g)$ (top left) and ${\cal I}_{\rm B}(t)$ (top right). While the former starts at $g=0$, the latter is moved to $g=-1/16$. The lower plot shows ${\cal I}_{\rm B}(t(z))$, which now has a branch-cut singularity at $|z|=1$. (We set ${\cal I}_{\rm B}(t(z))$ to $0$ outside the disc $|z|\ge1$.)}
\label{f:BC}
\end{figure}

In this appendix, we discuss the perturbative
expansion of a toy model in dimension zero:
\be\label{toy2}
{\cal I}(g) \equiv
\int_{-\infty}^\infty \frac{{\rm d} x}{\sqrt{2\pi}}\, {\rm e}^{-\frac
  {x^2}2 - g x^4}.
\ee
Its perturbative expansion is
\be\label{toy-series}
{\cal I}(g)= \sum_{n=0}^\infty a_n (-g)^n, \qquad a_n =
\frac{(4n)!}{2^{2n}(2n)! n! }\underset{n\to\infty}{\sim}
\frac{2^{4n}}{\sqrt 2 \pi n
}\times n!\ .
\ee
The analytic continuation of the integral~\eqref{toy2}
from ${\rm Re}(g)>0$ to
the full complex plane is given by a second-kind modified
Bessel $K$-function:
\be\label{bess}
{\cal I}(g)=\frac{1}{4\sqrt{\pi g}}e^{\frac{1}{32g}}K_\frac{1}{4}\left(
    \frac{1}{32g}\right) .
\ee
Using the asymptotic behavior of $K_{\frac14}(z)$ for $z\to \infty$, one sees that  the exponential prefactor is canceled, and the series~\eqref{toy-series} recovered. Note that ${\cal I}(g)$ has a cut on the whole negative real axis, see Fig.~\ref{f:BC}.

In field theory, the divergent series is analytically continued
without the knowledge of its exact expression. Let us explain the
strategy on the example of integral~\eqref{toy2}.
The basic idea~\cite{zinn-justin-book} to obtain a convergent
series out of Eq.~\eqref{toy-series}, is to divide each term by
$n!$, defining the \emph{Borel transform} ${\cal I}_{\rm B}(t)$ of the
series.  In a second step, one reconstructs the original series via an
integral transform:
\be\label{inv-Borel}
{\cal I}_{\rm B}(t) \equiv  \sum_{n=0}^\infty \frac{a_n}{n!} (-t)^n, \quad {\cal I}(g) = \int_0^\infty {\rm d}t\, {\rm e}^{-t} \,{\cal I}_{\rm B}(t g).
\ee
In our example we know the analytic expression in terms of the first-kind complete elliptic integral function
\be \label{Iofg-exact}
{\cal I}_{\rm B}(t) = \frac{2 K_{\rm elliptic}\!\left(\frac12 - \frac{1}{2
\sqrt{16 t+1}}\right)}{\pi  \sqrt[4]{16
t+1}}.
\ee
The Borel transform ${\cal I}_{\rm B}(t)$ has a finite radius of convergence,
denoted by $-t_{\rm bc}$ (equal to $1/16$ in our example).
As a consequence, the start of the branch cut is moved from $g=0$ to $t=t_{\rm bc}<0$, see figure~\ref{f:BC}.
Since the radius of convergence of ${\cal I}_{\rm B}(t)$ is still finite, 
the integral transform (\ref{inv-Borel}) does not work as written. One first has to continue ${\cal I}_{\rm B}(t)$ to the domain $0\le t <\infty$. This can be achieved by replacing the known truncated series via a converging Pad\'e approximant, leading to a \emph{Pad\'e-Borel resummation}.

\begin{figure}[t]
\centerline{\includegraphics[width=8.cm]{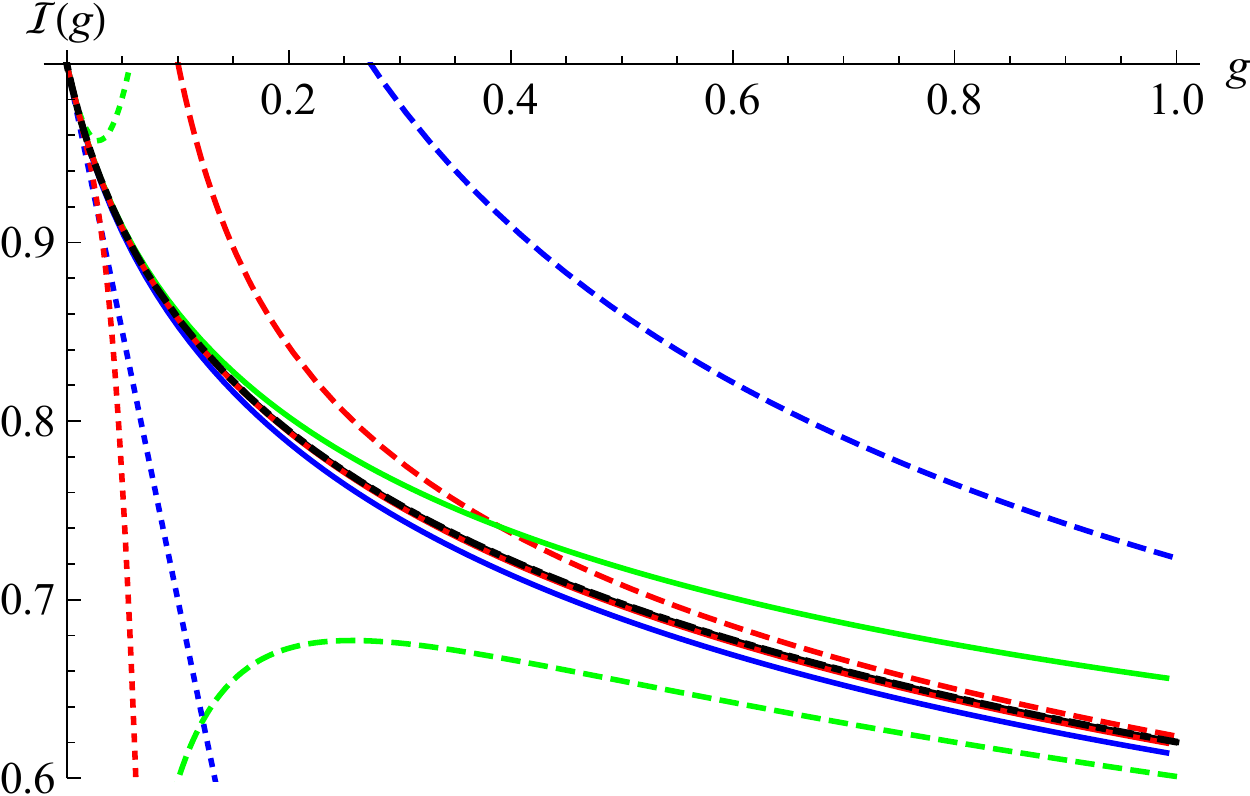}~~~~~~\includegraphics[width=8.cm]{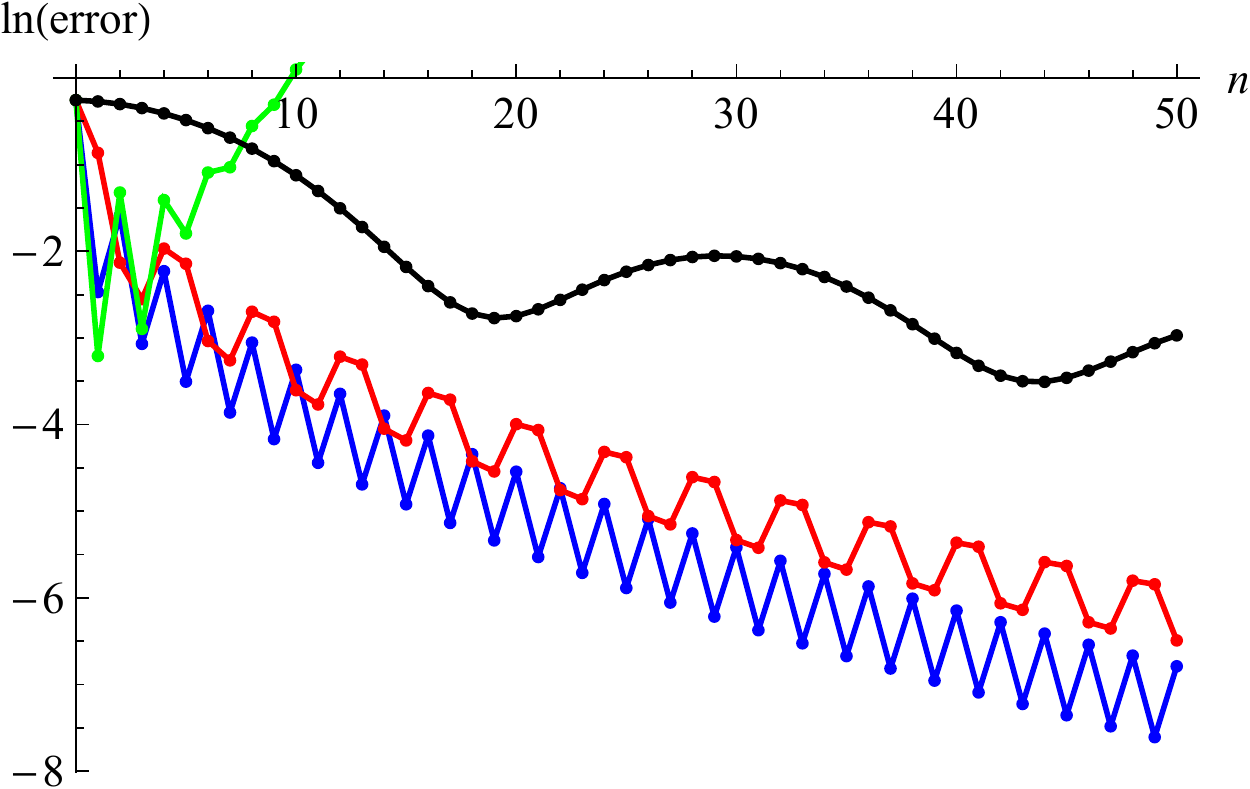}}
\caption{Left: function ${\cal I}(g)$ (black, thick, dot-dashed) and its diverse approximations. Dotted for the series expansion at order 1 (blue), 2 (green), and 3 (red). Solid for the resummed series at the same order. Dashed for the large-$g$ expansion (same color code). Right: deviation of the resummed series~\eqref{Ig-resummed} from the exact result~\eqref{Iofg-exact} for $g=10$ as a function of $n$, assuming one knows $t_{\rm bc}$ only approximately. In blue for $t_{\rm bc}=-1/16$ (the exact result), in red $t_{\rm bc}=-1/32$ (a conservative guess), in black $t_{\rm bc}=-1/1000$ (much too small). Resummation with $t_{\rm bc}=-1/15$ (green) does not work.  We see that conform to expectations, taking a too small value for $-t_{\rm bc}$, the series converges more slowly, while taking a too large value of $-t_{\rm bc}$ the series does not converge.}
\label{f:Iofg}
\end{figure}

A more powerful strategy is to use a conformal mapping.
The most common ansatz is to assume that at $t=t_{\rm bc}<0$ a cut-singularity
starts, which extends on the negative real axis to $t=-\infty$.
One first maps the complex plane with the expected branch cut of ${\cal I}_{\rm B}(t)$ onto the inside of the unit-circle:
\be \label{gbc}
z = \frac{\sqrt{1- t/t_{\rm bc}}-1}{\sqrt{1- t/t_{\rm bc}}+1} \qquad \Longleftrightarrow \qquad t  = \frac{-4 t_{\rm bc} \, z}{(z-1)^2}\ .
\ee
Next one constructs a series in $z$ by expanding both sides in this variable:
\be\label{t-to-z}
f(z) \equiv \sum_{n=0}^\infty c_n z^n  = \sum_{n=0}^\infty   \frac{a_n (-t(z) )^n}{n!} = {\cal I}_{\rm B}(t(z))\ .
\ee
This series is expected to converge for $|z|<1$, a fact we can check for our example (but which is difficult to prove in general):
\be\label{f(z)}
f(z) = 1-\frac{3 z}{4}+\frac{9 z^2}{64}-\frac{51
   z^3}{256}+\frac{1353 z^4}{16384}-\frac{7347
   z^5}{65536}+\frac{61617
   z^6}{1048576}+{\cal O} (z^7 )\ .
\ee
Given $n$ terms in the original series, we know $f(z)$ up to the same order. Using this approximation for $f(z)$, we finally obtain:
\be
{\cal I} (g) = \int_0^\infty {\rm d}  t\, {\rm e}^{-t} {\cal I}_{\rm B}(t g) = \frac1g \int_0^\infty {\rm d} t\, {\rm e}^{-t/g} {\cal I}_{\rm B}(t ) = \frac1g \int_0^1 {\rm d} z\, t'(z)\,{\rm e}^{-t(z)/g} f(z)\ . \label{Ig-resummed}
\ee
The result of this resummation is shown on Fig.~\ref{f:Iofg}. First, in black is the analytic result~\eqref{bess}. Next are the first three orders 
in several expansions, using the same color code for order 1 (blue), 2 (green), and 3 (red): first the direct expansion in $g$ (dotted), then in solid the resummed expansion~\eqref{Ig-resummed}. Dashed, we show a large-$g$ expansion obtained by changing variables $g x^4 \to y$ in the integral~\eqref{toy2}, and then expanding the integrand in powers of $1/\sqrt g$:
\be
\begin{aligned}
{\cal I}(g) &= \frac1{2 \sqrt{2 \pi } \sqrt[4] {g}  }\int_{0}^\infty  {\rmd y}\, \frac{e^{-\frac{\sqrt{y}}{2 \sqrt{g}}-y}}{ 
    y^{\frac{3}{4}}}\\
&= \frac1{2 \sqrt{2 \pi } \sqrt[4] {g}  }    \left[\Gamma \left({\textstyle \frac{1}{4}}\right)-\frac{2}{3}
    \frac{ \Gamma 
   \left({\textstyle\frac{7}{4} }\right)}{\sqrt{{g}}}+\frac{\Gamma
   \left(\frac{5}{4}\right)}{8
   g}+{\cal O} \left( {g}^{-\frac{5}{4}}\right
   ) \right].
\end{aligned}
\ee

\subsection{Details on the resummation method}\label{app:KP17_details}

This appendix aims at providing a ``reader's guide'' to  the analysis in Ref.~\cite{panzer}, which determines the resummed series for the $d=3$
critical exponents $\eta$, $\nu^{-1}$ and $\omega$ (related,
respectively, to $\g_\s$, $\g_\e$ and $\g_{\e'}$). The same methods are
used in our work, by a simple generalization to varying dimension $4>d\ge 3$.
This guide, together with the introduction in the main text and the
example in App.~\ref{appendix:baby_integral}, should provide enough
information to follow the discussion in Ref.~\cite{panzer}. In
particular, we are interested in its Sec.~V. Let us denote the
equations in Ref.~\cite{panzer} by double parentheses, e.g., Eq.~((25)), to
avoid confusion with our numbering.

The resummation procedure with Borel transform and conformal mapping
goes along the lines described in our Sec.~\ref{sec:Borel_transform} and App.~\ref{appendix:baby_integral}.
The perturbative series of a critical exponent $f(\eps)$
in $(-2\eps)=D-4$ (cf.~our $y=4-d$) is defined 
in Eq.~((25)) of~\cite{panzer}:
\be\label{asympt-KP}
f(\eps)=\sum_{k=0}^\infty f_k \; (-2\eps)^k, \qquad
\qquad
f_k \sim C_f\; k!\;  a^k\;  k^{b_f}\quad {\rm as}\quad k\to\infty.
\ee
With respect to our notation (cf.~our Eq.~\eqref{asympt}),
the negative sign of $a$ is included in the power of epsilon, and
the exponent of the power-law behavior earlier denoted by $b$
is now $b_f$.

The values for the parameters $(a,b_f)$ are given in Eq.~((26))
for the $\l\phi^4$ theory with $\mathrm{O}(n)$ symmetry, $n=1$ being the case of
interest, and they are determined by the known asymptotic behavior of the beta function.
With respect to the definition given here in Eq.~\eqref{borel-g},
in~\cite{panzer} the Borel transform is replaced by the more general Borel--Leroy transform, defined as follows (cf.~Eq.~((27)) in~\cite{panzer}):
\be
{\cal B}_f^b(x) = \sum_{k=0}^{\infty} \frac{f_k}{\G(k+b+1)} (-x)^k,
\ee
where $b$ is a free parameter. The function ${\cal B}_f^b(x)$ behaves as ${\cal B}_f^b(x)\sim (1+ax)^{b-b_f-1}$ around $x = -1/a$.

The function ${\cal B}_f^b(x)$ is then modified in three ways in order to
define the inverse transform and improve its convergence. The first step is
the conformal mapping ((29)) already described in App.~\ref{appendix:baby_integral}, involving the known parameter $a$. 
The second step is the addition of the power-law
prefactor in ((30)) with a second free parameter $\l$. The third step is the
``homographic transformation'' $\eps=h_q(\eps')$ defined in ((32))
which introduces a third free parameter $q$.

The resummed epsilon-expansion series $\tilde f(x)$
is finally obtained from the inverse
Borel transform of the modified function ${\cal B}^{b,\l,\ell}_{f\circ h_q}$
reported in ((33)) of~\cite{panzer},  
\be\label{KP-inv}
\tilde f(\eps) =\int_0^\infty t^b e^{-t} {\cal B}^{b,\l,\ell}_{f\circ h_q}
\left(\frac{2\eps t}{1-q\eps}\right) \rmd t\; .
\ee
It   depends on three free
parameters: $b$, $\lambda$ and $q$ ($\ell$ being the perturbative order
considered, $\ell=6$ here).
Let us briefly mention how these are determined.

The behavior of $\tilde f(\eps)\equiv \tilde f^{b,\l,q}_\ell(\eps)$ is
studied in the cubic range
\be\label{cube}
(b, \lambda, q) \in [0, 40] \times [0, 4.5] \times [0, 0.8].~
\ee
The optimal values of the parameters are chosen according to the
principle of ``minimal sensitivity'' (w.r.t.~varying the parameters)
and ``fastest apparent convergence'' (w.r.t.~increasing the
perturbative order by one, $\ell-1\to\ell$).
These dependences are taken into account by a proper definition of the error
function $E^f_\ell(b,\lambda,q)$ that is given in Eq.~((36)).

The global minimum of the error
$\bar{E}^f_\ell=E^f_\ell(\bar{b},\bar{\lambda},\bar{q})$ in the cubic
range~\eqref{cube} identifies the optimal values
$b=\bar{b}, \lambda=\bar{\lambda}$ and $q=\bar{q}$. The final estimate
for the critical exponents is obtained from the inverse Borel transform~\eqref{KP-inv} with these parameters. The optimization
procedure is done independently for each   dimension
$d=4-2\eps$.
The results for $\bar{b}$, $\bar{\lambda}$ and $\bar{q}$ are reported in 
Tab.~\ref{tab:variation_params} for the resummations of 
$\eta$, $\nu^{-1}$ and $\omega$ at the $d$ values considered.
Note the mild dependence of the parameters on $d$.

\begin{table}[!t]
\begin{center}
\begin{tabular}{|c|c|c|c|c|}
\hline
&&&&\\[-1em]
& $d$ & $\bar{b}$ & $\bar{\lambda}$ & $\bar{q}$\\
\hline
\hline
\multirow{5}{*}{$\eta$}
& 3.875 & 11 & 2.56 & 0.20\\
& 3.75  & 11 & 2.56 & 0.20\\
& 3.5   & 11 & 2.56 & 0.20\\
& 3.25  & 11 & 2.56 & 0.20\\
& 3     & 11 & 2.56 & 0.20\\
\hline
\hline
\multirow{5}{*}{$\nu^{-1}$}
& 3.875 & 15   & 1.32 & 0.16\\
& 3.75  & 15   & 1.32 & 0.16\\
& 3.5   & 15   & 1.32 & 0.16\\
& 3.25  & 14   & 1.30 & 0.16\\
& 3     & 13.5 & 1.30 & 0.16\\
\hline
\hline
\multirow{5}{*}{$\omega$}
& 3.875 & 19   & 0.52 & 0.46\\
& 3.75  & 21.5 & 1.02 & 0.40\\
& 3.5   & 21.5 & 1.02 & 0.40\\
& 3.25  & 22   & 1.02 & 0.40\\
& 3     & 22   & 1.02 & 0.40\\
\hline
\end{tabular}
\end{center}
\caption{Optimal variational parameters used here in the resummation procedure
for the critical exponents $\eta$, $\nu^{-1}$ and $\omega$, as a function
of $4> d\ge 3$.}
\label{tab:variation_params}
\end{table}

We remark that this brief outline brushes over
many fine details discussed in Ref.~\cite{panzer}, but which 
are crucial for achieving high-quality results, as well as the comparison
with other methods developed in the extensive literature.
More technical information can be found in Ref.~\cite{panzer} and its supplementary material, available in \href{https://arxiv.org/src/1705.06483v2/anc}{\texttt{arXiv:1705.06483}}.

The Self Consistent resummation of perturbative data used to obtain results shown in Figs.~\ref{fig:e-eps-SC},~\ref{fig:s-lit},~\ref{fig:e-lit},~\ref{fig:ep-lit} and~\ref{fig:f_e_ep}, instead, does not involve the optimization of free parameters introduced before. As explained in Sec.~III of~\cite{Kompaniets:2019zes}, the asymptotic behavior~\eqref{asympt-KP} is fitted from the perturbative series, thus finding the
position of the singularity $x=-1/a$ of the Borel transform. Such fit is done for several values of the free parameter $\alpha$, defined in Eq.~((44)) of~\cite{Kompaniets:2019zes} and analogous to $b_f$
in~\eqref{asympt-KP}, varied in the range $-6\le \alpha<6$ in steps of $0.2$.

For each $\alpha$, the value of $a$ obtained from the fit is used in the conformal
mapping~\eqref{gbc} ($t_{bc}\equiv a$) and the resulting function is Borel
inverted, giving the resummed series.
The best estimate of the resummed quantity with this procedure is obtained through the mean over all the values of $\alpha$ employed, while the error bars represent the maximal and minimal values obtained varying $\alpha$. Since only one parameter is varied, this
error estimate is less reliable than that determined by the methods of
Ref.~\cite{panzer} described earlier.

%-bib--------------------------------------------

\end{document}